\documentclass[11pt,a4paper]{article}
\usepackage{jheppub}
\usepackage{enumitem} 
\usepackage{lmodern}
\usepackage[utf8]{inputenc}
\usepackage{color}
\usepackage[dvipsnames, svgnames, x11names]{xcolor}
\usepackage{float}
\usepackage{amsmath,bm}
\usepackage{amssymb}
\usepackage{graphicx}
\usepackage{setspace}
\usepackage{subfigure}
\usepackage{amsthm}
\usepackage{amsfonts}
\usepackage{braket}
\usepackage{multirow}
\usepackage{ulem}
\usepackage{slashed}
\usepackage{lipsum}
\usepackage{hyperref}
\usepackage{url}
\usepackage{footnote}

\hypersetup{
	colorlinks=true,
	linkcolor=cyan,
	urlcolor=blue,
	citecolor=red,
}

\def\p{\partial}

\def\r{\rho}
\def\bal#1\eal{\begin{align}#1\end{align}}
\def\({\left(}
\def\){\right)}
\def\[{\left[}
\def\]{\right]}
\def\L{\left}
\def\R{\right}

\title{Fluctuations of the AdS$_3$ C-metric}

\author[a]{Shaohua Xue}
\author[b,c]{, Yuxuan Liu}
\author[a]{, and Li-Xin Li}

\affiliation[a]{The Kavli Institute for Astronomy and Astrophysics, Peking University, \\ Beijing 100871, China
}

\affiliation[b]{
Institute of Quantum Physics, School of Physics, Central South University, Changsha 418003, China
}

\affiliation[c]{Kavli Institute for Theoretical Sciences (KITS), University of Chinese Academy of Sciences, Beijing 100190, China}

\emailAdd{xuesh@pku.edu.cn, liuyuxuan93@csu.edu.cn, 
lxl@pku.edu.cn}
\abstract{
We investigate the dilaton fluctuations near the string based on three classes of solutions of the 3D C-metric within the framework of the string-world holography. 
As a setup of holography, we focus on the asymptotic symmetry, recover the Virasoro algebra by central extension and get the central charge of the $\rm AdS_{3}$. Then we reduce the gravity on the brane as a JT gravity model by introducing a fluctuation. As an extension of the braneworld, we also investigate the higher curvature correction to the brane under some conditions. Finally, we make an expansion on generalized entropy of black hole solution with respect to small $l$ and find that the leading term comes from Weyl anomaly, which is different from that in 4-dimensional C-metric.
}

\begin{document}
\maketitle
\section{Introduction}
 C-metric has been studied for a long time\cite{
PhysRevD.2.1359,bonnor1983sources,griffiths2006interpreting} and is used to describe a uniformly accelerating and rotating charged mass in 4D general relativity \cite{plebanski1976rotating,PhysRevD.2.1359}. Then it was generalized to describe a pair of accelerating AdS$_4$ black holes connected by a codimension-two topological defect \cite{dias2003pair}. 
Furthermore, in the Randall-Sundrum scenario, the curvature singularity on the branes of AdS C-metric provides a massive source, driving an acceleration of black holes, which is bound to the brane\cite{emparan2000exact2,emparan2000exact}.
One interesting probe of the quantum nature of black holes is to intersect a brane-like object with a black hole and study the induced black hole localized on the brane~\cite{emparan2000exact,emparan2002quantum,emparan2020quantum}. 
While many of the discussions are about the example of BTZ black holes on the branes in the AdS$_4$ C-metric background, there has been increasing interest on the AdS$_3$ C-metric system~\cite{Astorino:2011mw,Xu:2011vp,Astorino:2016xiy,Arenas-Henriquez:2022www,EslamPanah:2022ihg,Arenas-Henriquez:2023hur,EslamPanah:2023rqw,Cisterna:2023qhh,Fontana:2024odl,Kubiznak:2024ijq,Panella:2024sor}.
In three dimensions, although there is no propagation mode of the graviton, the notion of ``mass'' in the bulk can still be introduced by insertions and identifications of the branes in AdS C-metric\cite{arenas2022acceleration}. On the one hand, the world line of a point mass can be generated by introducing a conical deficit; on the other hand, the bulk geometry can be interpreted as BTZ-like black holes by quotients of the spacetime.


Despite advancements in higher dimensions, the duality remains ambiguous for the case of AdS$_3$ C-metric, especially following the identifications of the strings \cite{arenas2022acceleration}. On the one hand, the induced geometry on the string remains unclear. 

The four dimensional ''C-metric" spacetime describes a scenario where a pair of black holes are pulled to accelerate by conical singularities, or ''cosmic strings"~\cite{PhysRevD.2.1359,bonnor1983sources,Griffiths:2006tk}. 
In this paper we study various properties of C-metric type solution to the Einstein equation $R_{\mu\nu}=-\frac{2}{l^2}G_{\mu\nu}$ in 3d with constant negative cosmological constant, in the presence of  a ``wall" or a ``strut". Configurations with one such object are extensively studied in~\cite{arenas2022acceleration}. In this work, we focus on configurations with two strings, which reveal richer physics in all different classes of the 3d C-metric solutions. 
In the following, we will use the word ''cosmic string" or simply ``string" to denote collectively the ``domain wall" or the ``struct" that causes the acceleration in the C-metric solution if the difference between the strings do not affect the discussion. 
The paper is organized as follows:
In Section 2, we briefly review three classes of solutions for the AdS$_3$ C-metric. Then, in Section 3, as a setup for holography, we examine the asymptotic symmetry and recover the Virasoro algebra from the bulk as a correspondence to the boundary $\rm CFT_{2}$. We also present the reduced model on the boundary based on the Fefferman-Graham (FG) expansion. In Section 4, we discuss the thermodynamic relation of black hole phases.  In Section 5, we derive the effective theory for the three solutions of the C-metric by introducing a fluctuation on the brane to obtain Jackiw–Teitelboim (JT) gravity as the leading term. Additionally, we explore higher curvature corrections in holography as an extension. Finally, in Section 6, we conclude the main results and offer some discussions on future directions. 
 
\section{AdS$_3$ C-metric with two defects}\label{cmetric}

Following the notation in~\cite{arenas2022acceleration}, the line element of 3d C-metric geometries has the following general form 
\begin{equation}\label{2.1}
\mathrm{d}s^2=\frac{1}{A^2(x-y)^2}\left[-P(y)\mathrm{d}\tau^2+\frac{1}{P(y)}\mathrm{d}y^2+ \frac{1}{Q(x)}\mathrm{d}x^2\right],
\end{equation}
which solves  with $A$ a parameter of the acceleration. 
According to the concrete expression of $P(y)$ and $Q(x)$ the general solutions fall into three distinct classes
\begin{description}
    \item[Classe I] $\qquad~~\, Q(x)=1-x^2\,, \quad P(y)=\frac{1}{A^2 l^2}+y^2-1\,,\qquad |x|<1\,,$ 
    \item[Classe II] $\qquad~ Q(x)=x^2-1\,, \quad P(y)=\frac{1}{A^2 l^2}-y^2+1\,,\qquad |x|>1\,,$ 
    \item[Classe III] $\qquad Q(x)=1+x^2\,, \quad P(y)=\frac{1}{A^2 l^2}-y^2-1\,,\qquad \mathbb{R}\,,$ 
\end{description}
where $l$ is AdS radius and the range of allowed $x$ is determined by requiring the spacetime to have Lorentz signature. 
%
The overall conformal factor $\frac{1}{(x-y)^2}$ indicates a boundary of the spacetime at $x=y$. The region $x \ge y$ and $x\leq y$ are related by the reflection $x\to -x$, $y\to -y$~\cite{arenas2022acceleration}, so without loss of generality we consider the region $x>y$.

For now, all these classes merely represent distinct patches of geometries that are locally AdS$_3$. 
More interestingly, these geometries allow simple operation of quotienting, namely cutting the geometries open followed by proper identifications. At the location of the identification, extra localized stress energy, for example in the form of the strings, are needed to be compatible with the Einstein equation. In other words, the localized stress energy tensor should satisfy the Israel junction conditions, or its variance~\cite{Shen:2024dun,Shen:2024itl,SPL2} 
\begin{equation}\label{2.17}
4\pi T_{ij}=K_{ij}-K h_{ij}\,,
\end{equation}
where the indices $i,j$ represent the coordinates localized on the string. The extrinsic curvature is defined as:
    \begin{equation}
    K_{ij}=e^{\alpha}_{i}e^{\beta}_{j}\nabla_{\alpha}n_{\beta},
    \end{equation}
     in which Greek alphabets are the indices of the tensor in three dimension, and Latin alphabets are the indices of the constant-$x$ string. $e^{\alpha}_{i}$ is the Jacobian. $n_{\beta}$ is the norm vector. 
     The nice property of the C-metric spacetime is there exist a special choice of the position of the  string, namely the location with constant $x$, so that the stress energy tensor is simply proportional to the induced metric
\begin{align}
    T_{ij}= \sigma h_{ij}\,,
\end{align}
with $\sigma$ representing the tension on the string. Explicitly, if we cut and glue along the line at $x=X$, we simply compute, via~(\ref{2.17})
\begin{equation}
K_{ij}=\mp A\sqrt{Q(X)}h_{ij}\,,
\end{equation}
where the plus(minus) sign represents the normal vector pointing outward(inward) the region we keep for later construction, and the tension is respectively
\bal
\sigma=\pm \frac{A}{4\pi G_{3}}\sqrt{Q(X)}\ .\label{tension1}
\eal
Therefore as long as $|X|\neq 1$, the tension is nonzero $\sigma \neq 0$. For a wall/strut, we can define a null vector
     \begin{equation}
         N^{\alpha}=\(\pm1,P(y),0 \),
     \end{equation}
     thus we have:
     \begin{equation}
         T_{ij}e^{i}_{\alpha}e^{i}_{\beta}N^{\alpha}N^{\beta}\propto h_{ij}e^{i}_{\alpha}e^{i}_{\beta}N^{\alpha}N^{\beta}\equiv0,
     \end{equation}
     that is to say, both positive-tension wall and negative-tension strut satisfy the null energy condition.

\begin{figure}[htbp]
    \centering
    \begin{minipage}[h]{0.45\textwidth}
        \centering
        \includegraphics[width=\textwidth]{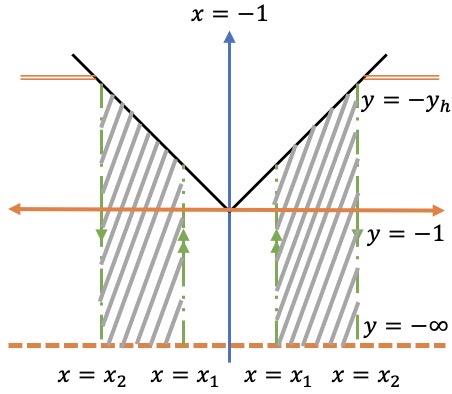}
    \end{minipage}
    \hfill
    \begin{minipage}[h]{0.45\textwidth}
        \centering
        \includegraphics[width=\textwidth]{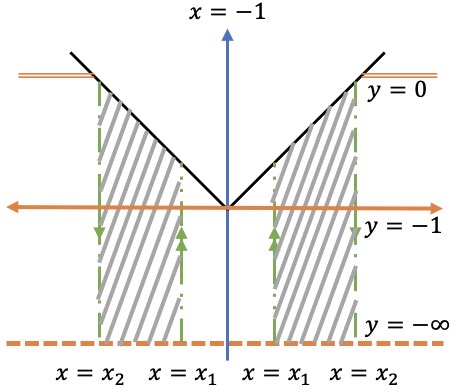}
    \end{minipage}
    
    \vspace{0.7cm} 
    
    \begin{minipage}[h]{0.45\textwidth}
        \hspace{0.18cm}
        \includegraphics[width=\textwidth]{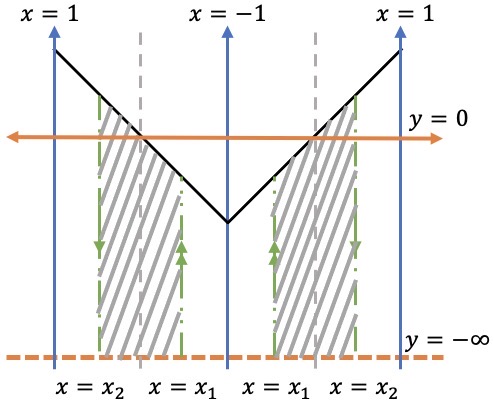}
    \end{minipage}
    \hfill
    \begin{minipage}[h]{0.45\textwidth} 
         \textbf{Upper Left}: Class $\rm I_{p1}$ solutions. \textbf{Upper Right}: Class $\rm I_{p2}$ solutions. \textbf{Lower Left}: Class $\rm I_{p3}$ solutions. The black, green, and blue lines represent the conformal boundary ($x=y$),   the strings, and the constant-$x$ surfaces, respectively. The orange double line represents the event horizon. The grey region is obtained from the entire manifold. The orange dashed lines denote the constant-$y$ lines. The orange dashed lines denote the constant-$y$ lines, and the orange arrows indicate the directions of
         increasing $x$ in the two patches. 
    \end{minipage}
    \caption{Two copies of the geometry with each cut by two  strings in Class I.}
    \label{Class I1}
\end{figure}

Starting with a single spacetime, we first cut it at two different constant-$x$ surfaces, $X=x_1$ and $X=x_2$ ($x_1<x_2$), make two copies of the remaining spacetime within $[x_1,x_2]$, and then glue them along the two cut boundaries head-to-head. This manipulation typically leads to a defect at each of the two gluing surface except for the case where one of the gluing surface is at $|X|=1$ and the gluing is smooth; the latter simply drives us back to the . To avoid cumbersome statements, we will refer to these combined operations simply as CCG (short for ''cut-copy-paste"), and the strings along the gluing junction with positive tension are referred to as (domain) walls, while those with negative tension are called struts.

In the rest of this section, we present different classes of solutions with two  strings after the appropriate CCP. Notice that according to the convention explained above, the two  strings in our later discussion must contain one wall and one strut. 

\subsection{Class I } 
\begin{figure}[htbp]
    \centering
    \begin{minipage}[h]{0.4\textwidth}
        \centering
        \includegraphics[width=\textwidth]{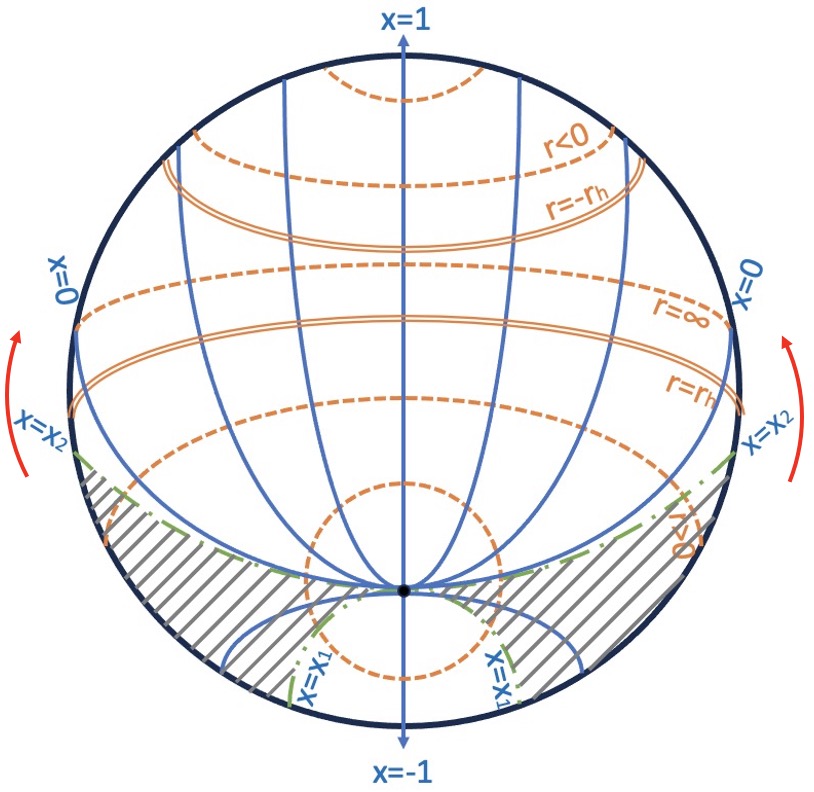}
    \end{minipage}
    \hfill
    \begin{minipage}[h]{0.4\textwidth}
        \centering
        \includegraphics[width=\textwidth]{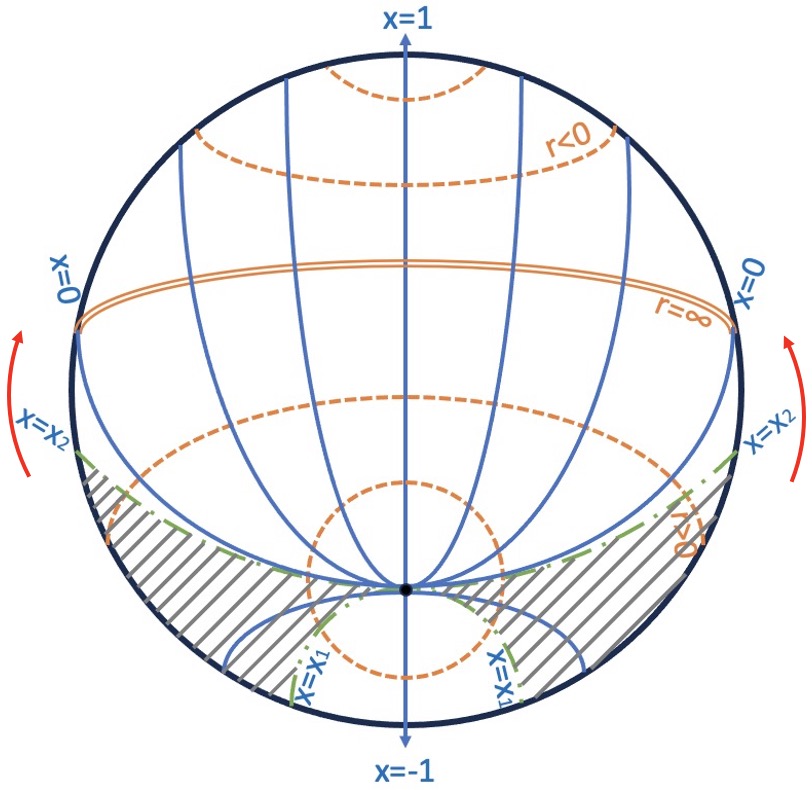}
    \end{minipage}
    
    \vspace{0.7cm} 
    
    \begin{minipage}[h]{0.38\textwidth}
        \hspace{0.1cm}
        \includegraphics[width=\textwidth]{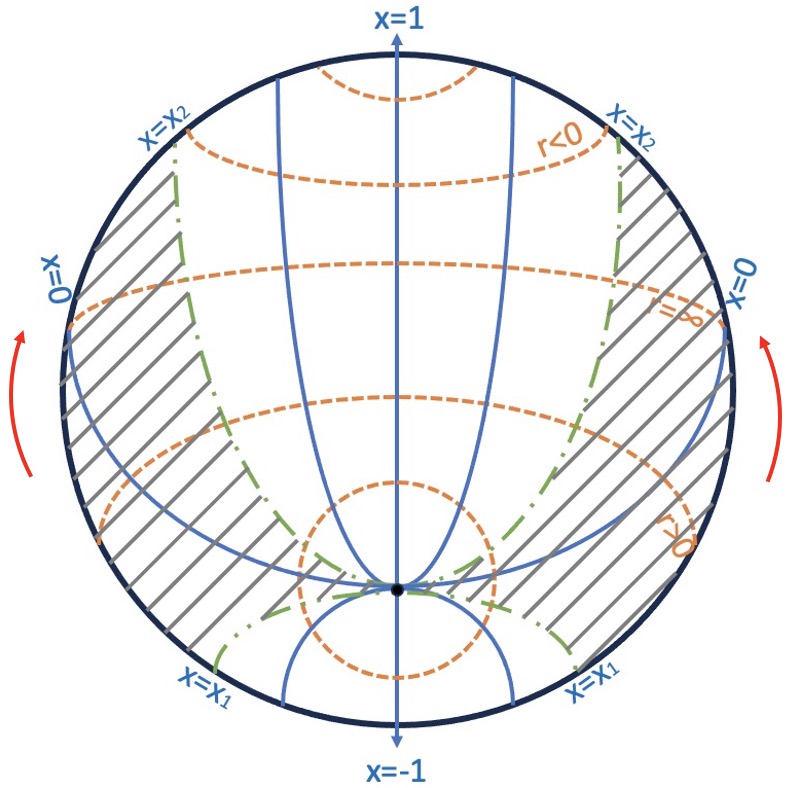}
    \end{minipage}
    \hfill
    \begin{minipage}[h]{0.45\textwidth} 
         The copied spatial slice of the class $\rm I_{p1}$ (\textbf{Upper Left}), class $\rm I_{p2}$ (\textbf{Upper Right}), and class $\rm I_{p3}$ (\textbf{Lower Left}) solutions in $(r,x)$.
         The orange dashed curves are constant-$r$ lines. The Orange double curve denotes the horizon. The blue curves are constant-$x$ lines. The green dot-dash lines are the strings. The black dot is the original point. The black circle is the conformal boundary ($x=y$). The grey shadow region is the part that we cut from the entire manifold as the bulk. The string at $x=x_{2}$($x=x_{1}$) is a strut(wall). The red
        arrows indicate the directions of increasing $x$.
    \end{minipage}
    \caption{The spatial slice of the Class I solutions in $(r,x)$.}
    \label{Class I2}
\end{figure}

 The general line element of the Class I solution is
\begin{equation}\label{2.3}
\mathrm{d}s^2=\frac{1}{A^2(x-y)^2}\left[ -\left(\frac{1}{A^2 l^2}+y^2-1\right)\mathrm{d}\tau^2+\frac{1}{\frac{1}{A^2 l^2}+y^2-1}\mathrm{d}y^2+\frac{1}{1-x^2}\mathrm{d}x^2\right]\,,
\end{equation}
where $|x| \leq 1$. This solution can be further subdivided based on the horizon structure, namely the number of roots $y=y_{\rm h}$ of $\left(\frac{1}{A^2 l^2}+y^2-1\right)=0$ depending on the value of $A$. For example, there is no horizon when $A^2l^2 < 1$; a single horizon at $y_h=0$ when $A^2l^2 = 1$, and two horizons at $y=\pm y_{\rm h}=\pm\frac{\sqrt{A^2l^2-1}}{Al}$ when $A^2l^2 > 1$. They are respectively labelled as ``slow", ``saturated" and ``rapid", and we will refer to them as the $\rm{I_{slow}}, \rm{I_{saturated}}$ and $\rm{I_{rapid}}$ classes. Fig.~\ref{Class I2} shows the spatial slice of all cases in $(x,r)$.

Furthermore, depending on the compactness of the $y$-direction, Class I solution describes either an accelerating particle or an accelerating black hole, thus $\rm I_{rapid}$ and $\rm I_{saturated}$ can be further divided into cases describing a particle or BTZ black hole. 

\subsubsection{An accelerating particle}  
In 3d, a particle is identified as the end point of the string where the angular arc length, in the $x$ direction, shrinks to zero. In the C-metric geometry, such an endpoint with vanishing angular arc length appears at $y\to -\infty$. Therefore, geometries with $y = -\infty$ included have an interpretation of a particle.  There are three different solutions in Class I that describe an accelerating particle, namely the Class $\rm I_{slow}$, the Class $\rm I_{rapid}$ with $y<0$, and $\rm I_{saturated}$ with $y<0$. 
However, we notice that there are subtleties in defining the particle's mass for the latter case due to the presence of the acceleration horizon. For this reason, we sometimes restrains to the cases without horizon, namely require $x_2<-y_h$ for the Class $\rm I_{rapid}$  and $\rm I_{saturated}$.  
 When $\mathcal{A}$ is zero, the spacetime becomes analogous to global AdS with defects.

To gain intuitions of the geometry, we rewrite the line element (\ref{2.3}) in global coordinates, by the following transformation$$y=-\frac{1}{\mathcal{A}r}, \quad \tau=m^2\mathcal{A}t, \quad x=\cos(m\psi+\psi_{0}), \quad  A =m\mathcal{A}\,,$$ with 
\bal
m=[\text{arccos}(x_{2})-\text{arccos}(x_{1})]/\pi\ .
\eal 
Then the metric ($\ref{2.3}$) can be expressed as
\begin{equation}\label{2.5}
\mathrm{d}s^2=\frac{1}{(1+\mathcal{A}r\cos(m\psi+\psi_{0}))^2}\left[-(m^2+\frac{1-\mathcal{A}^2 l^2 m^2}{l^2}r^2)\mathrm{d}t^2+\frac{\mathrm{d}r^2}{m^2+\frac{1-\mathcal{A}^2 l^2 m^2}{l^2}r^2}+r^2\mathrm{d}\psi^2\right],
\end{equation}
where $\psi_{0}=\mathrm{arccos}(x_{1})$.
This metric applies to all the three cases discussed above; they can be distinguished by the range of the parameter $m$.
In particular, we occasionally focus on the cases without a horizon in the physical region, which restricts the range of $x_2<-y_h$ and thus further shrinks the allowed range of $m$.  

To understand the physical meaning of $A$, we examine the covariant velocity of a test particle located at a distant $r$ in the coordinate~\eqref{2.5} $$u=\frac{1}{\sqrt{-G_{tt}}}\p_{t}\,,$$ where $G_{\mu\nu}$ is the metric~\eqref{2.5}. The covariant acceleration is 
\begin{equation}
a^{\mu}=(\nabla_{u}u)^{\mu}=\Gamma^{\mu}_{00}u^{0}u^{0}\ .\label{acceleration}
\end{equation}
Substituting~\eqref{2.5} into the above expression, we obtain the norm of the acceleration
\begin{equation}
|a|=\sqrt{a_{\mu}a^{\mu}}=\sqrt{\frac{((\mathcal{A}^2 l^2 m^2-1)r+\mathcal{A} l^2 m^2\cos(m\psi+\psi_{0}))^2}{l^2(r^2+m^2 l^2(1-\mathcal{A}^2r^2))}+\mathcal{A}^2 m^2\sin(m\psi+\psi_{0})^2}\ .
\end{equation}
Therefore the acceleration of a test particle at the origin accelerates at a rate
$$\lim\limits_{r \to 0}|a|=\mathcal{A} m=A\ .$$ 
 Additionally, the induced metric on a string located at $\psi=\psi_{1}$ reads
\begin{align}\label{Induced I}
\mathrm{d}s^2|_{\psi=\psi_{1}}=&\frac{1}{(1+\mathcal{A}r\cos(\psi_{0}+m\psi_{1}))^2}\left[-(m^2+\frac{1-\mathcal{A}^2 l^2 m^2}{l^2}r^2)\mathrm{d}t^2+\frac{\mathrm{d}r^2}{m^2+\frac{1-\mathcal{A}^2 l^2 m^2}{l^2}r^2}\right]\,,
\end{align}
and therefore the Ricci tensor on the string is 
\begin{equation}\label{induced metric}
R_{ij}(h)=-\frac{(1-\mathcal{A}^2 l^2 m^2\sin(\psi_{0}+m\psi_{1})^2)}{l^2}h_{ij}\equiv -\frac{1}{l_{2}(\psi_{1})^2}h_{ij}\ .
\end{equation}
On the other hand, the tension of the string is, according to~\eqref{tension1},
$$|\sigma|=\frac{\mathcal{A}m|\sin(\psi_{0}+m\psi_{1})|}{4\pi G_{3}}\,,$$ 
This gives a relation between an effective cosmological constant on the string and the ambient 3D cosmological constant 
\begin{equation}
-\frac{2}{l_{2}(\psi_{1})^2}=-\frac{2}{l^2}+32\pi^2 G_{3}^2\sigma^2\ .
\end{equation}

\begin{figure}[h]
\centering
\subfigure[]{
\includegraphics[width=0.43\linewidth]{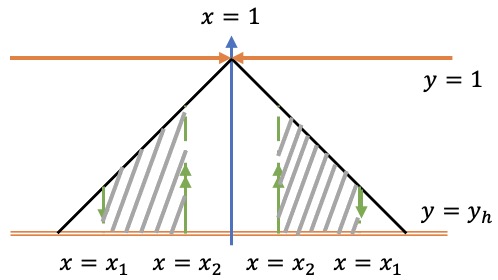}}
\subfigure[]{
\includegraphics[width=0.43\linewidth]{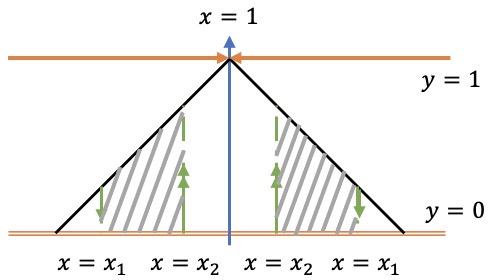}}
\caption{Two copies of the geometry with each cut by two  strings in Class I. \textbf{(a)}: Class $\rm I_{b1}$ solutions; \textbf{(b)}: Class $\rm I_{b2}$ solutions. The black and green 
lines represent the conformal boundary ($x=y$) and the  strings, 
 respectively. The orange double line represents the event horizon. The grey regions are cut from the entire manifold. The orange lines denote the constant-$y$ lines, and the arrows indicate the directions of increasing $x$ in the two patches.
 }
\label{Class Ic1}
\end{figure}

Therefore, for Class I$_{\rm slow}$ and Class I$_{\rm saturated}$, which have $1-\mathcal{A}^2 l^2 m^2\geq0$, the curvature on the string are both negative
since $\sin(\psi_{0}+m\psi_{1})^2\in(0,1)$. For Class $\rm I_{\rm rapid}$ with $1-\mathcal{A}^2 l^2 m^2<0$, the situation is more interesting. When $x_2<-y_h$ the physical region, i.e. CCP of the region between $x_1$ and $x_2$, does not contain horizon, and the induced metric on the  string is locally AdS since $\sin(\psi_{0}+m\psi_{1})^2\in(0,\frac{1}{\mathcal{A}^2 l^2 m^2})$. If we instead consider $x_2 \geq -y_h$, namely the physical region contains horizon, the curvature on the string could be either zero or positive. 


\subsubsection{An accelerating Black Hole}

Contrary to the solutions describing accelerating particles, Class $\rm I_{rapid}$ and $\rm I_{saturated}$ solutions with $y_h < x_1 <x_2 \leq 1\,,~ y\geq y_h$
 does not contain the $y=-\infty $ point due to the existence of horizons at $y_h=\sqrt{1-(A l)^{-2}}$. They describe accelerating black holes. We will call these two types of solutions in Class $\rm I_{rapid}$ and Class $\rm I_{saturated}$  simply as $\rm I_{b1}$ and $\rm I_{b2}$ solutions respectively, which are illustrated in Fig.~\ref{Class Ic1}. 
 
\begin{figure}[]
\centering
\subfigure[]{
\includegraphics[width=0.4\linewidth]{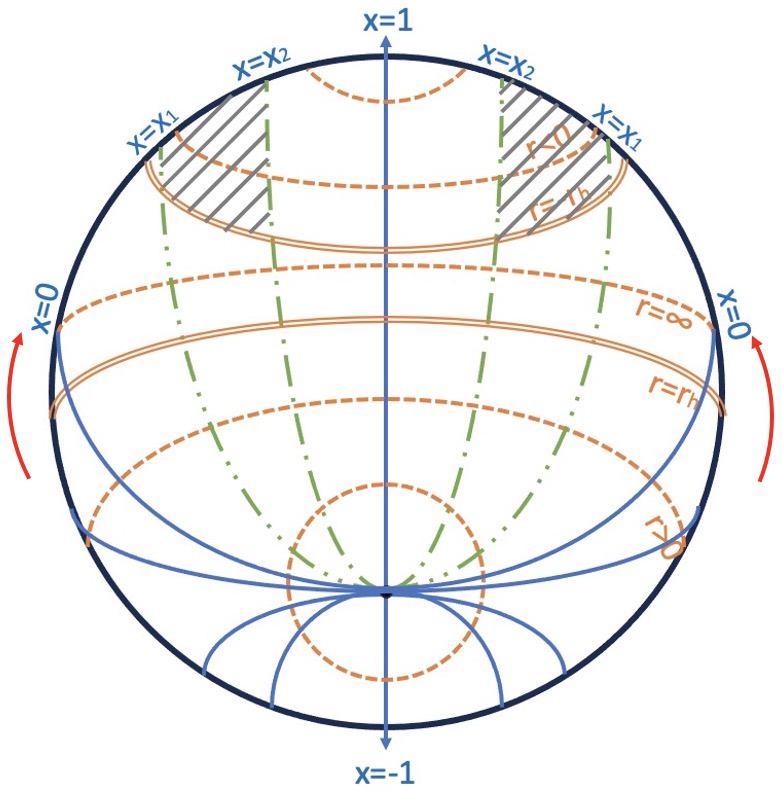}}
\subfigure[]{
\includegraphics[width=0.4\linewidth]{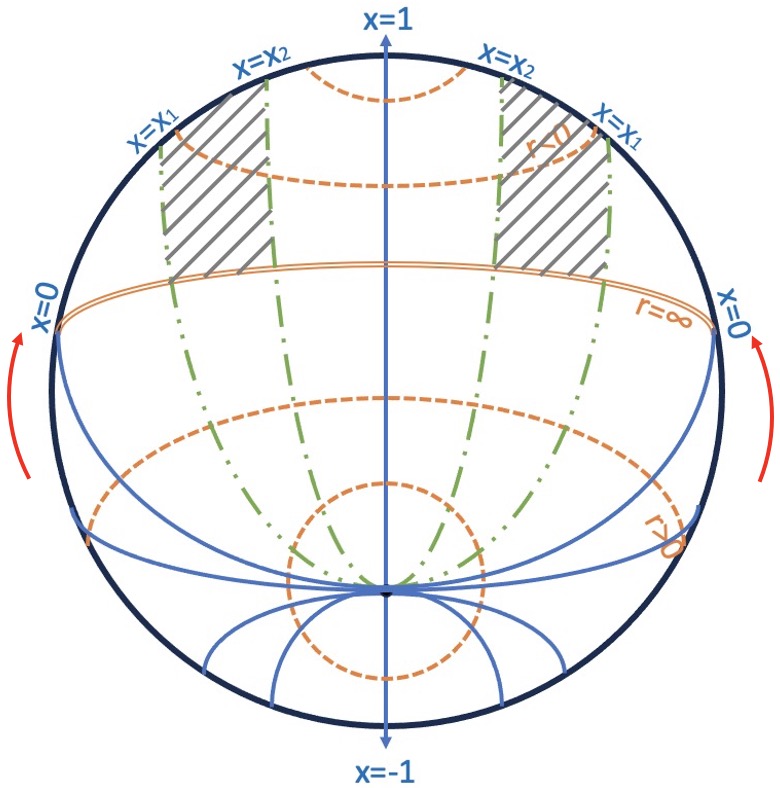}}
\caption{The spatial slice of the \textbf{(a)}: Class $\rm I_{b1}$ solutions, and \textbf{(b)}: Class $\rm I_{b2}$ solutions in the $(r,x)$ coordinates. The orange dashed curves are constant-$r$ lines. The Orange double curve denotes the horizons. The blue curves are constant-$x$ lines. The green dot-dash lines are  strings. The black dot is the original point. The black circle is the conformal boundary ($x=y$). The grey shadow region is the part that we cut from the entire manifold as the bulk. The string at $x=x_{2}$ ($x=x_{1}$) is a strut (wall). The red arrows indicate the directions of increasing $x$.}
\label{Class Ic2}
\end{figure}

 We can similarly rewrite the line element in global coordinate as ($\ref{2.5}$) by the transformation 
 \begin{equation}
\begin{aligned}\label{transformation2.12}
y=\frac{1}{\mathcal{A}r}, \quad \tau=m^2\mathcal{A}t, \quad x=\mathrm{cos}(m\psi+\psi_{0}), \quad  A =m\mathcal{A}\,,
\end{aligned}
\end{equation}
with $m=\left(\arccos(x_{2})-\arccos(x_{1})\right)/\pi$ and the metric becomes
\begin{equation}
    \mathrm{d}s^2=\frac{1}{(1-\mathcal{A}r\cos(m\psi+\psi_{0}))^2}\left[-(m^2-\frac{\mathcal{A}^2 l^2 m^2-1}{l^2}r^2)d t^2+\frac{\mathrm{d}r^2}{m^2-\frac{\mathcal{A}^2 l^2 m^2-1}{l^2}r^2}+r^2\mathrm{d}\psi^2\right],
\end{equation}
with the horizon at $r_{\rm h}=\frac{\ ml}{\ \sqrt{\mathcal{A}^2 l^2 m^2-1}}$. 
The range of $m$ is again determined by the range of $x_1$ and $x_2$, and the radius in this solution is restricted to the region $r \in [\frac{\ 1}{\ \mathcal{A}\cos(m\psi+\psi_{0})},r_{\rm h})$. 
Spatial slices of this solution, after appropriate CCG, are shown in Fig.~\ref{Class Ic2}. Subsequent discussions are parallel to that of the accelerating particle so we will not repeat it here. The  strings, which start from the string horizon and end at the conformal boundary, drag the black hole to accelerate. On the  strings, the induced cosmological constants are negative.

\subsection{Class II}
\begin{figure}[]
\centering
\subfigure[]{
\includegraphics[width=0.43\linewidth]{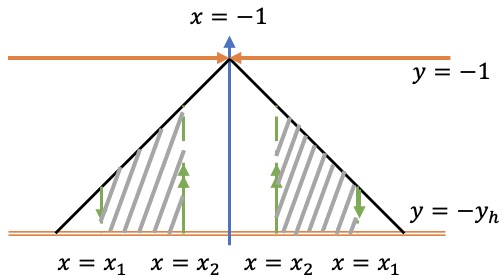}}
\subfigure[]{
\includegraphics[width=0.43\linewidth]{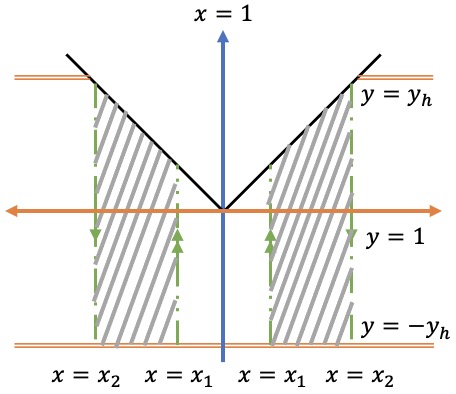}}
\caption{The geometry of  \textbf{(a)} class $\rm II_{b1}$ and \textbf{(b)} class $\rm II_{b2}$ solutions. The dark lines represent the conformal boundaries at $x=y$, the green dash-dotted lines denote  strings, the orange double lines mark the position of the horizon, and the blue lines depict the lines of constant-$x$. The  string at $x=x_{2}$ is a strut, while at $x=x_{1}$ is a wall. The grey-shaded region indicates the remaining manifold after the CCG process. The orange arrows indicate the directions of increasing $x$ in the two patches.}
\label{Class II1}
\end{figure}

The line element of Class II solution  can be expressed as
\begin{equation}\label{2.31}
\mathrm{d}s^2=\frac{1}{A^2(x-y)^2}\left[ -(\frac{1}{A^2l^2}+1-y^2)\mathrm{d}\tau^2+\frac{1}{\frac{1}{A^2l^2}+1-y^2}\mathrm{d}y^2+\frac{1}{x^2-1}\mathrm{d}x^2 \right]\,,
\end{equation}
with $|x|\geq1$. This solution has two horizons at 
\bal
y=\pm y_{\rm h}=\pm\frac{\sqrt{1+A^2l^2}}{Al}\ .
\eal 
After the CCG, the spacetime is compact in the $y$ direction and the solution describes an accelerating black hole. Here we are only interested in the case of $-y_h < x_1 < x_2 <-1$ or $1 < x_1 < x_2 < y_h$, which are bounded by the asymptotic boundary and a horizon. They are referred to as class $\rm II_{b1}$ and $\rm II_{b2}$ respectively. These cases are illustrated in Fig.~\ref{Class II1}.
We can similarly rewrite the line element ($\ref{2.31}$) in global coordinates by the following change of variables 
\bal
y=-\frac{1}{\mathcal{A}r}, \quad \tau=m^2\mathcal{A}t, \quad x=\pm\cosh(m\psi+\psi_{0}), \quad A=m\mathcal{A}\,,\label{changeII}
\eal
in which $m=\frac{\mathrm{arcosh}(|x_{2}|)-\mathrm{arcosh}(|x_{1}|)}{\pi}$ and $\cosh(m\psi+\psi_{0})\in(1,\frac{\sqrt{1+\mathcal{A}^2m^2l^2}}{\mathcal{A}ml})$. The minus (plus) sign corresponds to II$_{\rm b1}$ (II$_{\rm b2}$). Then the metric ($\ref{2.31}$) can be expressed as
\begin{align}\label{2.7}
\mathrm{d}s^2=&\frac{1}{(1\pm\mathcal{A}r\cosh(m\psi+\psi_{0}))^2}\left[-\left(\frac{1+\mathcal{A}^2 l^2 m^2}{l^2}r^2-m^2\right)\mathrm{d}t^2+\frac{\mathrm{d}r^2}{\frac{1+\mathcal{A}^2 l^2 m^2}{l^2}r^2-m^2}+r^2\mathrm{d}\psi^2\right],
\end{align}
where $\psi_{0}=\mathrm{arccosh}(|x_{1}|)$ and $m=\left(\arccos(x_{2})-\arccos(x_{1})\right)/\pi$. If we set $\mathcal{A} = 0$ the geometry~\eqref{2.17} approaches the BTZ black hole. 
The horizon locates at $r_{\rm h}=\frac{ml}{\sqrt{1+\mathcal{A}^2m^2l^2}}$. In the case of II$_{\rm b1}$, $r$ ranges in $(r_{\rm h},\frac{1}{\mathcal{A}\cosh(m\psi+\psi_{0}))})$ and in this region $\mathcal{A}r\cosh(m\psi+\psi_{0}))<1$. However, in II$_{\rm b2}$,  $r$ takes value in two regions, one is $(r_{\rm h},\infty)$ where $\mathcal{A}r\cosh(m\psi+\psi_{0}))+1>0$; the other is $(-\infty,-\frac{1}{\mathcal{A}\cosh(m\psi+\psi_{0}))})$ where $\mathcal{A}r\cosh(m\psi+\psi_{0}))+1<0$. 
Based on the results in the case of particles, it is natural to expect that the parameter $A$ also determines the acceleration of the black hole. However, due to the existence of a horizon, a similar computation as in the case of the point particle does not directly work. We can however show the physical meaning of $A$ indirectly by computing the change of acceleration near the horizon due to a non-zero $A$. The four-acceleration of a static observer $a^{\mu}$ is
\begin{equation}
\begin{aligned}
a^{t}&=0,\\
a^{r}&=\frac{\ (r+\mathcal{A}^2 l^2 m^2 r\pm\mathcal{A} l^2 m^2 \cosh(m\psi+\psi_{0}))(1\pm\mathcal{A}r\cosh(m\psi+\psi_{0}))}{\ l^2},\\
a^{\psi}&=\frac{\ \mathcal{A}m\sinh(m\psi+\psi_{0})(\mp 1-\mathcal{A}r\cosh(m\psi+\psi_{0}))}{\ r}\,,
\end{aligned}
\end{equation}
which becomes
\bal
a^{\mu}&=\left(0,\frac{m}{l} \left(1+A^2 l^2\right)^{-\frac{1}{2}}+\mathcal{A} m^2 \left(\mathcal{A} r_h \left(x^2+1\right)+2 x\right),-\mathcal{A} m \text{sgn}(x) \sqrt{x^2-1}  \left(\mathcal{A} x+\frac{1}{r_h}\right)\right)\,,\label{accII}
\eal
at the horizon $r=r_h$ and $x$ is the same as in~\eqref{changeII}. Notice that although $a^\mu$ is finite its norm diverges at the horizon, as expected.
 
\begin{figure}[]
\centering
\subfigure[]{
\includegraphics[width=0.4\linewidth]{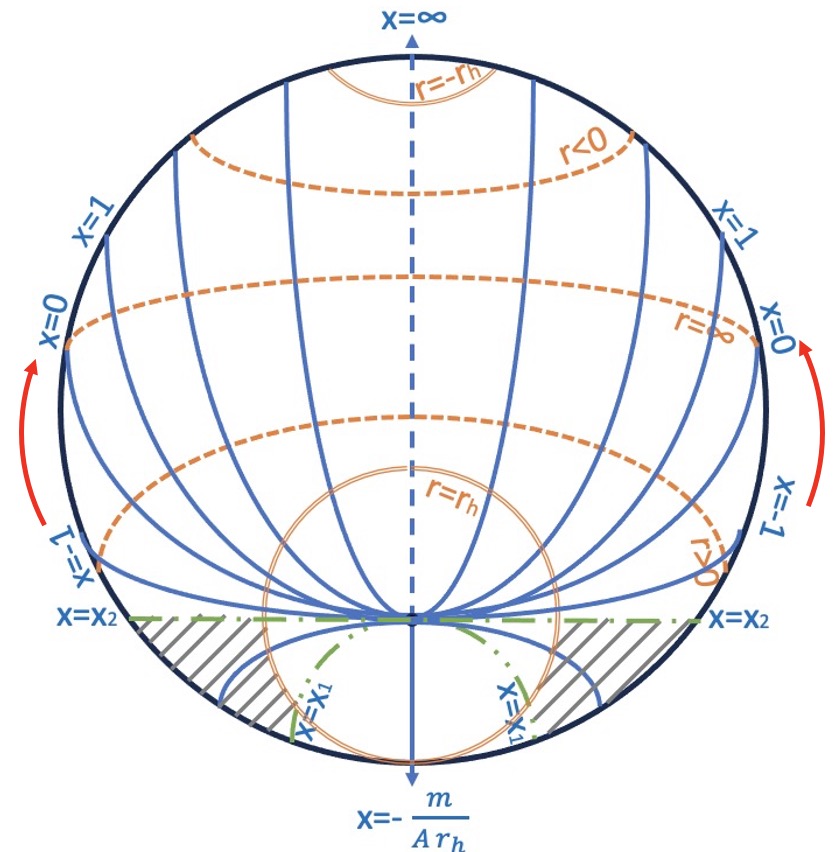}}
\hspace{1cm}
\subfigure[]{
\includegraphics[width=0.4\linewidth]{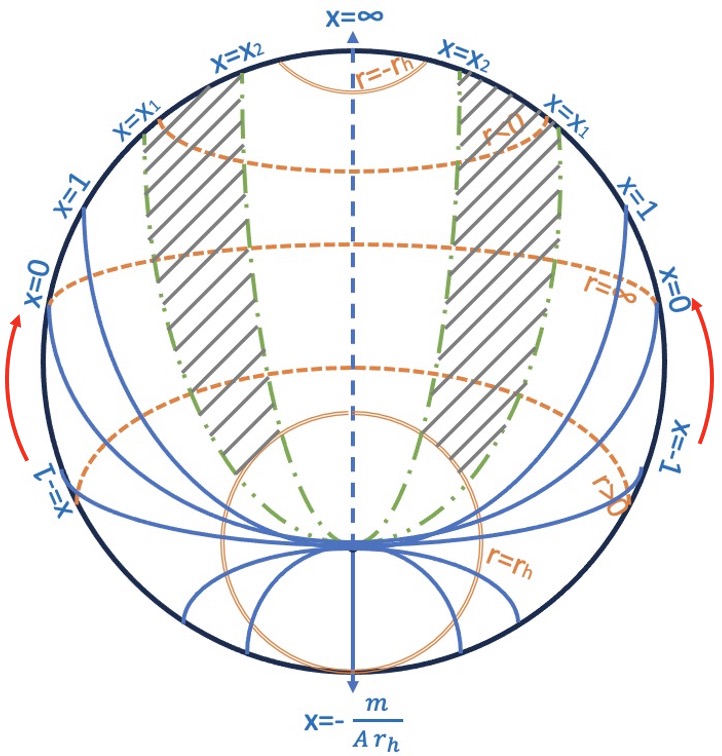}}
\caption{The copied spatial slices of the \textbf{(a)}: class $\rm II_{b1}$ and \textbf{(b)}: class $\rm II_{b2}$ solutions in $(r,x)$. The orange dash curves are lines of constant-$r$. The orange double curves are the horizons. The blue curves are the lines of constant-$x$. The green dash-dotted lines are the positions of the  strings. The black dots are the original points. The black circles are the conformal boundaries ($x=y$). The grey shadow regions are the remaining spacetime after the CCG process.}
\label{ClassII2}
\end{figure}

As a comparison, we calculate the four-acceleration of a static obsever located at the horizon of a static BTZ black hole with mass $\frac{m^2}{8G_{3}}$. The metric of the BTZ black hole is
\begin{equation}
    ds_{\text{BTZ}}^2=G_{\mu\nu (\text{BTZ})}\mathrm{d}x^{\mu}\mathrm{d}x^{\nu}=-(r^2/l^2-m^2)\mathrm{d}t^2+\frac{1}{(r^2/l^2-m^2)}\mathrm{d}r^2+r^2\psi^2\,,
\end{equation}
and the four-velocity of a static observer $(\frac{1}{\sqrt{-G_{tt (\text{BTZ})}}},0,0)$, by definition~\eqref{acceleration}, the four-acceleration of a static observer located at the horizon of the BTZ background is
\begin{equation}\label{acceleration of BTZ}
    a^{\mu}=\left(0,\frac{m}{l},0\right)\ .
\end{equation}
Now we can have a better understanding of~\eqref{accII}. The angular acceleration in~\eqref{accII} is proportional to $A$, which is consistent with the fact that the acceleration of the entire solution due to the pull/push of the  strings induces an acceleration of the static observer in the angular direction. The radial component of four-acceleration is more complicated. the II$_{\rm b1}$ solution is supported in the range $r_{\rm h}< r<\frac{1}{\mathcal{A}\cosh(m\psi+\psi_{0}))}$ which means $1-\mathcal{A}r\cosh(m\psi+\psi_{0}))>0$ and $r>\frac{ml}{\sqrt{1+\mathcal{A}^2m^2l^2}}>\frac{\mathcal{A} l^2 m^2 \cosh(m\psi+\psi_{0})}{1+\mathcal{A}^2m^2l^2}$, thus $a^{r}>0$. For the $r_{\rm h}< r<\infty$ part of the II$_{\rm b2}$ solution, we have $1+\mathcal{A}r\cosh(m\psi+\psi_{0}))>0$, $r>-\frac{\mathcal{A} l^2 m^2 \cosh(m\psi+\psi_{0})}{1+\mathcal{A}^2m^2l^2}$, thus $a^{r}>0$. For the $-\infty< r<-\frac{1}{\mathcal{A}\cosh(m\psi+\psi_{0}))}$ part of the II$_{\rm b2}$ solution, we have $1+\mathcal{A}r\cosh(m\psi+\psi_{0}))<0$ and $r<-\frac{ml}{\sqrt{1+\mathcal{A}^2m^2l^2}}<-\frac{\mathcal{A} l^2 m^2 \cosh(m\psi+\psi_{0})}{1+\mathcal{A}^2m^2l^2}$, thus $a^{r}>0$. Physically, this means the global acceleration of the system due to the drag of the  string does not alter qualitatively the fact that local observers need to accelerate pointing outward from the horizon. In addition, we find as we set $\mathcal{A}=0$, the acceleration~\eqref{accII} reduces to that of the acceleration of a static observable near the horizon of a BTZ black hole; the $\mathcal{A}$ induces additional acceleration on top of the acceleration needed for the test particle to hover at a fixed position near the horizon.
A spatial slice in $(x,r)$ coordinates is illustrated in
Fig.~\ref{ClassII2}.

The induced metric on the string, positioned at $\psi=\psi_{1}$, is given by
\begin{align}
    \mathrm{d}s^2|_{\psi=\psi_{1}}=&\frac{1}{(1\pm\mathcal{A}r\cosh(\psi_{0}+m\psi_{1}))^2}\left[-\left(\frac{1+\mathcal{A}^2 l^2 m^2}{l^2}r^2-m^2\right)\mathrm{d}t^2+\frac{\mathrm{d}r^2}{\frac{1+\mathcal{A}^2 l^2 m^2}{l^2}r^2-m^2}\right]\ .
\end{align}
The tension on the string is $$|\sigma|=\frac{\mathcal{A}m|\sinh(\psi_{0}+m\psi_{1})|}{4\pi G_{3}}\,,$$ and the Ricci tensor is
\begin{equation}\label{induced metric2}
R_{ij}(h)=-\frac{(1-\mathcal{A}^2 l^2 m^2\sinh(\psi_{0}+m\psi_{1})^2)}{l^2}h_{ij}\equiv -\frac{1}{l_{2}(\psi_{1})^2}h_{ij},
\end{equation}
which defines the induced cosmological constants on the  string
\begin{equation}
-\frac{2}{l_{2}(\psi_{1})^2}=-\frac{2}{l^2}+32\pi^2 G_{3}^2\sigma^2 \quad \in \quad \bigg[-\frac{4(1-\mathcal{A}^2 l^2 m^2)}{l^2},0\bigg)\ .
\end{equation}
Moreover, if we set $\mathcal{A} = 0$ the metric on the string reduces to that of a 2d black hole.  
Furthermore, it is easy to observe that the acceleration due to the drag by the  strings are monotonic as a function of $x$ and does not change sign in the entire allowed range, which is consistent with the fact that both  strings at $x_1$ and $x_2$ drag the Blackhole in the same direction. 
\subsection{Class III solution}

\begin{figure}[]
\centering
\includegraphics[width=0.45\linewidth]{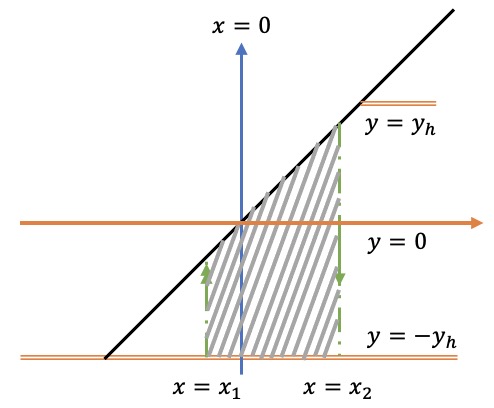}
\caption{The spatial slice of the Class $\rm III_b$ solutions cut by two  strings. The dark line is the conformal boundary at $x=y$, and the green dash dotted lines are two  strings. The orange double lines are the positions of horizons. The blue line is the constant-$x$ line. The grey-shaded region is the remaining spacetime after the cut. The orange arrow indicates the directions of increasing $x$.}
\label{Class III1}
\end{figure}

It is shown in~\cite{arenas2022acceleration} that it is not possible to find a Class III solution with one  string and a compact angular direction, so the simplest solution would involve two  strings. We discuss this two-string solution in this section. 
The line element of the Class III solution can be expressed as
\begin{equation}\label{2.48}
\mathrm{d}s^2=\frac{1}{A^2(x-y)^2}\left[-(\frac{1}{A^2 l^2}-1-y^2)\mathrm{d}\tau^2+\frac{1}{\frac{1}{A^2 l^2}-1-y^2}\mathrm{d}y^2+\frac{1}{x^2+1}\mathrm{d}x^2\right],
\end{equation}
with $x\in\mathbb{R}$. This geometry possesses two horizons at $$y=\pm y_{\rm h}=\pm\frac{\sqrt{1-A^2l^2}}{Al}\ .$$
We are interested in the solution that is compact, bounded by the strings in the $x$-direction and by the horizons and the asymptotic boundary in the $y$-direction. As we will show in the following, the solution describes an accelerating black hole after the CCG. We hereafter denote this solution class $\rm III_b$, and illustrate it in Fig.~\ref{Class III1}.

\begin{figure}[]
\centering
\subfigure[]{
\includegraphics[width=0.4\linewidth]{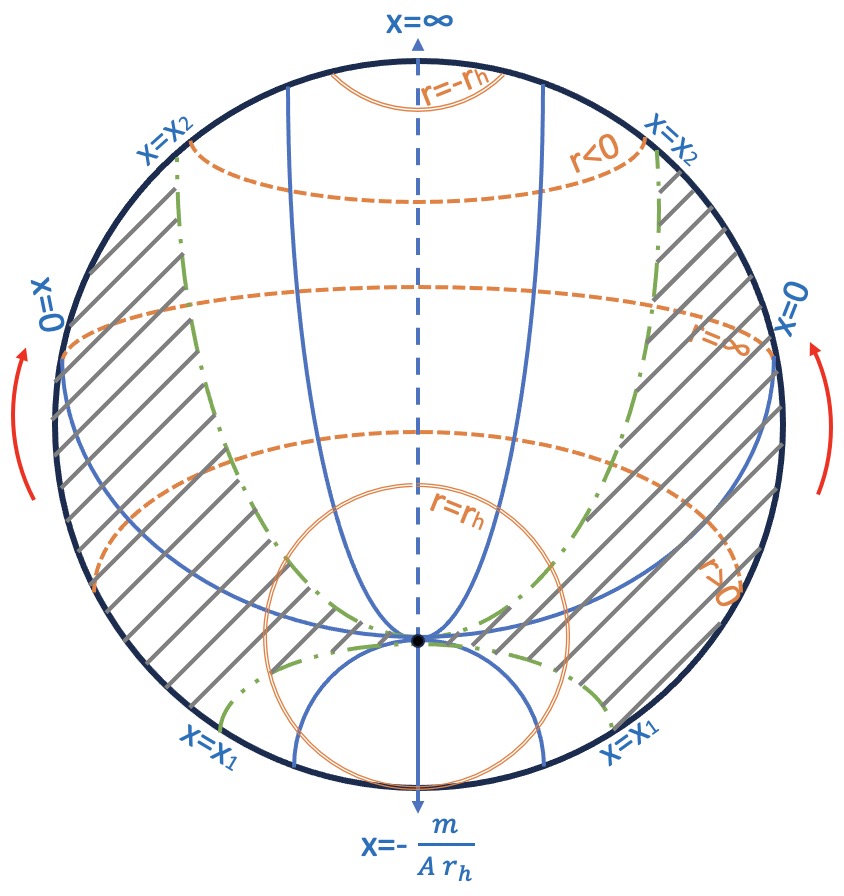}}
\caption{The copied spatial slice of the class $\rm III_b$ solutions in $(r,x)$. The green dash-dotted lines are the  strings. The black dot is the original point. The black circle is the conformal boundary at $x=y$. The grey shadow region is the remaining manifold after the CCG. The red arrows indicate the directions of increasing $x$.}
\label{ClassIII2}
\end{figure}

The line element ($\ref{2.48}$) in global coordinates is
\begin{align}\label{2.9}
\mathrm{d}s^2=&\frac{1}{(1+\mathcal{A}r\sinh(m\psi+\psi_{0}))^2}\left[-(\frac{1-\mathcal{A}^2 l^2 m^2}{l^2}r^2-m^2)\mathrm{d}t^2+\frac{\mathrm{d}r^2}{\frac{1-\mathcal{A}^2 l^2 m^2}{l^2}r^2-m^2}+r^2\mathrm{d}\psi^2\right]\,,
\end{align}
which can be obtained from the following transformation $$y=-\frac{1}{\mathcal{A}r},\quad \tau=m^2\mathcal{A}t,\quad x=\sinh(m\psi+\psi_{0}), \quad A=m\mathcal{A}\,,$$ with  $m=\frac{\mathrm{arsinh}(x_{2})-\mathrm{arsinh}(x_{1})}{\pi}$, 
 $\psi_{0}=\mathrm{arcsinh}(x_{1})$, and the position of horizon at $r_{\rm h}=\frac{ml}{\sqrt{1-\mathcal{A}^2m^2l^2}}$. In this solution the range of the $x$ (or $\psi$) direction is, $-\frac{\sqrt{1-\mathcal{A}^2m^2l^2}}{\mathcal{A}ml}<\sinh(m\psi+\psi_{0})<\frac{\sqrt{1-\mathcal{A}^2m^2l^2}}{\mathcal{A}ml}$. A constant time slice of this geometry is illustrated in Fig.~\ref{ClassIII2}.

Depending on the position of the  strings, there are two different scenarios. 
\begin{itemize}
  \item Class III$_{\rm b1}$ where $0>x_2>x_1$. The geometry is bounded by the horizon $y=-r_h$ and the asymptotic boundary, and there is only one connected region in the global $r$ coordinate; namely $r\in (r_{\rm h},-\frac{1}{\mathcal{A}\sinh(m\psi+\psi_{0})})$ and in this region $\mathcal{A}r\sinh(m\psi+\psi_{0})+1>0$. 
  \item Class III$_{\rm b2}$, where $x_2>0>x_1$. The geometry is bounded by the horizon $y=-r_h$ and the asymptotic boundary, and there are two regions in the global $r$ coordinate; namely $r\in  (r_{\rm h},\infty)$ and $ r \in (-\infty,-\frac{1}{\mathcal{A}\sinh(m\psi+\psi_{0})})$. In the first region $\mathcal{A}r\sinh(m\psi+\psi_{0})+1>0$, while in the second region $\mathcal{A}r\sinh(m\psi+\psi_{0})+1<0$. 
\end{itemize}
The four-acceleration of a static observer at a fixed $r$ is
\begin{equation}
    \begin{aligned}
        a^{t}&=0\\
        a^{r}&=\frac{\ ((1-\mathcal{A}^2m^2l^2)r+\mathcal{A}l^2m^2\sinh(m\psi+\psi_{0}))(1+\mathcal{A}r\sinh(m\psi+\psi_{0}))}{l^2}\\
        a^{\psi}&=-\frac{\mathcal{A}m\cosh(m\psi+\psi_{0})(1+\mathcal{A}r\sinh(m\psi+\psi_{0}))}{r}\ .
    \end{aligned}
\end{equation}
For III$_{\rm b1}$ with $0<\sinh(m\psi+\psi_{0})$ and $r_{\rm h}<r<\infty$, $\mathcal{A}r\sinh(m\psi+\psi_{0})+1>0$ and $r>0>-\frac{\ \mathcal{A}l^2m^2\sinh(m\psi+\psi_{0})}{1-\mathcal{A}^2m^2l^2}$, so $a^{r}>0$.
For the III$_{\rm b2}$ solution with $0>\sinh(m\psi+\psi_{0})$, $\mathcal{A}r\sinh(m\psi+\psi_{0})+1>0$ and $r>\frac{ml}{\sqrt{1-\mathcal{A}^2m^2l^2}}>-\frac{\ \mathcal{A}l^2m^2\sinh(m\psi+\psi_{0})}{1-\mathcal{A}^2m^2l^2}$, so $a^{r}>0$.  
For III$_{\rm b2}$ with $0<\sinh(m\psi+\psi_{0})$ and $-\infty<r<-\frac{1}{\mathcal{A}\sinh(m\psi+\psi_{0})}$, $\mathcal{A}r\sinh(m\psi+\psi_{0})+1<0$ and $r<-\frac{ml}{\sqrt{1-\mathcal{A}^2m^2l^2}}<-\frac{\ \mathcal{A}l^2m^2\sinh(m\psi+\psi_{0})}{1-\mathcal{A}^2m^2l^2}$, so $a^{r}>0$. 
In summary, in all cases of type III, the acceleration $a^r > 0$. 



The induced metric on the string at $\psi=\psi_1$ is
\begin{align}\label{2.14}
\mathrm{d}s^2|_{\psi=\psi_{1}}=&\frac{1}{(1+\mathcal{A}r\sinh(\psi_{0}+m\psi_{1}))^2}\left[-\left(\frac{1-\mathcal{A}^2 l^2 m^2}{l^2}r^2-m^2\right)\mathrm{d}t^2+\frac{\mathrm{d}r^2}{\frac{1-\mathcal{A}^2 l^2 m^2}{l^2}r^2-m^2}\right]\ .
\end{align}
The absolute value of tension on the string is $$|\sigma|=\frac{\mathcal{A}m\cosh(\psi_{0}+m\psi_{1})}{4\pi G_{3}}\,,$$ and the Ricci tensor of the string is
\begin{equation}\label{induced metric3}
R_{ij}(h)=-\frac{(1-\mathcal{A}^2 l^2 m^2\cosh(\psi_{0}+m\psi_{1})^2)}{l^2}h_{ij}\equiv -\frac{1}{l_{2}(\psi_{1})^2}h_{ij}\ .
\end{equation}
This leads to the effective cosmological constant on the string
\begin{equation}
-\frac{2}{l_{2}(\psi_{1})^2} =-   \frac{2}{l^2}+32\pi^2 G_{3}^2\sigma^2\quad \in \quad \bigg[-\frac{2(1-\mathcal{A}^2 l^2 m^2)}{l^2},0\bigg)\,,
\end{equation}
hence the induced theory on the string has a negative cosmological costant. 
Additionally, when $\mathcal{A}$ approaches zero, the geometry~($\ref{2.9}$) approaches a BTZ black hole. 



\section{Asymptotic Symmetry}
It is well known that solutions of 3d Einstein gravity with negative cosmological constants are locally AdS$_3$. Therefore non-trivial dynamics are mainly due to boundary graviton excitations, which are governed by the asymptotic symmetries \`a la Brown-Henneaux~\cite{Brown:1986nw} for any given boundary condition.   
It is shown in~\cite{Brown:1986nw} that the asymptotic symmetry algebra of asymptotic AdS$_3$ contains two copies of the Virasoro algebra, which establish a concrete connection with 2D conformal field theory (CFT$_2$).
In the following, we do a Fefferman-Graham (FG) expansion of the three C-metric solutions near the boundary, 
identify the generators of the asymptotic symmetry algebra, and obtain the central extension.
\subsection{FG expansion}\label{FGsec}
We start with the following FG expansion of 3D metric~\cite{de2001holographic} 
\begin{equation}\label{3.2}
\mathrm{d} s^2=\frac{l^2}{z^2}\mathrm{d} z^2 + \frac{l^2}{z^2}g_{ij}(\tilde{x},z)\mathrm{d} \tilde{x}^{i}\mathrm{d} \tilde{x}^{j}\,,
\end{equation}
where $$g_{ij}(x,z)=g_{(0)}+g_{(2)}z^2+h_{(2)}z^2 \log(z^2)+\mathcal{O}(z^3)\,,$$ and $l$ is the AdS radius. Then near the boundary ($z\ll 1$), we can substitute the metric ($\ref{3.2}$) into the Einstein equations and solve for the coefficients $g_{(d)}(d>0)$ order by order, expressing them in terms of $g_{(0)}$ and the Ricci tensor of $g_{(0)}$. We then integrate out a shell of $z\geq\epsilon$ to get an effective action. In doing this integral, divergences could arise and we need to add appropriate counterterms to cancel these divergences. This renormalizes the effective action to the following form
\begin{equation}\label{3.3}
I_{\rm gr}=I_{\rm gr,reg}+I_{\rm div}
\end{equation}
where $I_{\rm gr,ren}$ represents the regularized on-shell action and $I_{\rm div}$ consists of all divergent term near the boundary and expanded over Ricci scalar.
Specifically, as in e.g.~\cite{chen2020quantum2,emparan1999surface}, by substituting the metric ($\ref{3.2}$) into the action and integrating over $z\geq\epsilon$, we get
\begin{equation}
\begin{aligned}\label{3.4}
I_{\rm div}=&\frac{l}{16\pi G_{3}}\int\mathrm{d}^{2}\tilde{x}\sqrt{\gamma}\\
&\left[\frac{2}{l^2}+\frac{1}{2}R(\gamma)\mathrm{log}\left(-\frac{l^2R(\gamma)}{2}\right)-\frac{1}{2}R(\gamma)+\frac{l^2}{16}R(\gamma)^2+\mathrm{O}(R(\gamma)^3+\tilde{F}(\nabla \Phi)\right],
\end{aligned}
\end{equation}
where $\gamma$ is the induced metric near the conformal boundary and the Ricci scalar $R(\gamma)\propto \epsilon^2$. The detailed derivation of equation (\ref{3.4}) is provided in Appendix \ref{A}. From the divergent part of ($\ref{3.4}$), we obtain the 2D effective theory at the asymptotic boundary with a redefined Newton constant $lG_{2}=G_{3}$. The logarithm term $$R(\gamma)\mathrm{log}\left(-\frac{l^2R(\gamma)}{2}\right)\sim \epsilon^2\log(\epsilon)$$ is convergent and arises from the weyl anomaly. Then the renormalized stress tensor is given by \cite{balasubramanian1999stress}
\begin{equation}\label{mass}
T_{\rm ren}[\gamma]=\frac{l}{8\pi G_{3}}\left(g_{(2)}-g_{(0)}\text{Tr}[g_{(0)}^{-1}g_{(2)}]\right)
\end{equation}
 which allows us to calculate a holographic mass near the boundary.

To apply this method to the cases of C-metric, we need to derive an FG expansion from metric ($\ref{2.1}$) near the conformal boundary. It turns out that it is convenient to change the $(x,y)$ coordinates to $(\xi,z)$ 
\begin{equation}\label{2.74}
y=\xi+\sum^{\infty}_{m=1}F_{m}(\xi)(\frac{z}{l})^{m}, \quad x=\xi+ \sum^{\infty}_{m=1}G_{m}(\xi)(\frac{z}{l})^{m}\ .
\end{equation}
where $z\rightarrow 0$ corresponds to the boundary. Then we substitute ($\ref{2.74}$) into the metric ($\ref{2.1}$) and impose the conditions $G_{zz}=\frac{l^2}{z^2}$ and $G_{z\xi}=0$. In ($\ref{2.74}$) the functions $F_{m}$ and $G_{m}$ can be solved order by order and finally expressed in terms of $F_{1}$, which can be determined to be $$F_1=\frac{(1-A^2l^2Q(\xi))^{\frac{3}{2}}}{Al \omega(\xi)}\ .$$ Although $F_{m}$ and $G_{m}$ can, in principle, be solved to arbitrary order, the coefficients of $g_{ij}(x,z)$ in ($\ref{3.2}$) for terms $z^m$ (with $m>4$) are zero. The only undetermined parameter is a Weyl factor $\omega(\xi)$, which is just a gauge choice. This factor can be fixed arbitrarily, but it will influence the thermodynamic relations \cite{tian2024aspects}. For simplicity, we adopt the ADM gauge here and set $\omega(\xi)=1$ \cite{arnowitt1961coordinate}. The following are the results of all solutions under this gauge choice:
\begin{itemize}
    \item \textbf{Class I}\\
The metric in the FG expansion for the Class I is given by
\begin{equation}\label{2.75}
\mathrm{d}s^2=-\frac{(z^2+l^2(4-A^2 z^2))^2}{16 A^2 l^2 z^2}\mathrm{d}\tau^2+\frac{l^2}{z^2}\mathrm{d}z^2+\frac{l^2(z^2-l^2(4+A^2 z^2))^2}{16z^2(1-\xi^2)(1-A^2 l^2(1-\xi^2))^2}\mathrm{d}\xi^2.
\end{equation}
To compare the metric ($\ref{2.75}$) with that of a Poincare disk, we perform a coordinate transformation 
$$T=\frac{\tau}{A}, \qquad \text{and} \qquad \Xi=f(\xi)-f(x_{2}),$$ with $\xi\in[x_{1},x_{2}]$. The function $f(\xi)$ is defined to be
\begin{equation}
    f(\xi):=
    \begin{cases}
        \frac{l \;\text{arctanh}\left(\sqrt{A^2l^2-1}\sqrt{\frac{1}{\xi^2}-1}\right)}{\sqrt{A^2 l^2-1}} & \rm for \; I_{b1}, \\
        -\frac{l \;\text{arctanh}\left(\sqrt{A^2l^2-1}\sqrt{\frac{1}{\xi^2}-1}\right)}{\sqrt{A^2 l^2-1}} & \rm for \; I_{p1}, \\
        \frac{\sqrt{\frac{1}{\xi^2}-1}}{A^2 l^2}, & \rm for \; I_{b2}, \\
        -\frac{\sqrt{\frac{1}{\xi^2}-1}}{A^2 l^2}, & \rm for \; I_{p2}, \\
        \frac{l \;\text{arctan}\left(\frac{\xi}{\sqrt{1-A^2l^2}\sqrt{1-\xi^2}}\right)}{\sqrt{1-A^2 l^2}} & \rm for \; I_{p3}.
    \end{cases}
\end{equation}
Then the metric becomes
\begin{equation}
\mathrm{d}s^2=-\frac{z^2+l^2 (4-A^2 z^2)}{16 l^2 z^2}\mathrm{d}T^2+\frac{l^2}{z^2}\mathrm{d}z^2+\frac{(4l^2-z^2+A^2 l^2 z^2)^2}{16 l^2 z^2}\mathrm{d}\Xi^2,
\end{equation}
where 
\begin{align}
\Xi \in [0,2\pi \beta], \qquad \text{and } \qquad \beta=\frac{\Xi |_{\xi=x_{1}}}{\pi}. \label{beta1}
\end{align}
    \item \textbf{Class II} \\
The metric in the FG expansion for Class II is given by
\begin{equation}\label{2.82}
\mathrm{d}s^2=-\frac{(z^2+l^2(A^2z^2-4))^2}{16 A^2 l^2 z^2}\mathrm{d}\tau^2 +\frac{l^2}{z^2}\mathrm{d}z^2+\frac{(z^2+l^2(4+A^2 z^2))^2}{16z^2(\xi^2-1)(-1+A^2l^2(-1+\xi^2))^2}\mathrm{d}\xi^2
\end{equation} 
To compare with the Poincare disk, we apply a similar transformation $$T=\frac{\tau}{A}, \qquad \text{and} \qquad \Xi=f(\xi)-f(\xi_{2}),$$ with $\xi \in [x_{1},x_{2}]$. The function $f(\xi)$ is defined as
\begin{equation}
    f(\xi):=
    \begin{cases}
        \frac{l \mathrm{arctanh}\left(\sqrt{A^2l^2+1}\sqrt{1-\frac{1}{\xi^2}}\right)}{\sqrt{A^2 l^2+1}} & \rm for \; II_{b1}, \\
        -\frac{l \mathrm{arctanh}\left(\sqrt{A^2l^2+1}\sqrt{1-\frac{1}{\xi^2}}\right)}{\sqrt{A^2 l^2+1}} & \rm for \; II_{b2}.
    \end{cases}
\end{equation}
Then the metric becomes:
\begin{equation}\label{2.78}
\mathrm{d}s^2=-\frac{(z^2+l^2(A^2z^2-4))^2}{16 l^2 z^2}\mathrm{d}T^2 +\frac{l^2}{z^2}\mathrm{d}z^2+\frac{(z^2+l^2(4+A^2 z^2))^2}{16 l^2 z^2}\mathrm{d}\Xi^2,
\end{equation}
where 
\begin{align}
\Xi \in [0,2\pi \beta], \qquad \text{and } \qquad \beta=\frac{\Xi |_{\xi=x_{1}}}{\pi}. \label{beta2}
\end{align}
    \item \textbf{Class III} \\
The metric in the FG expansion for the Class III is given by
\begin{equation}
\mathrm{d}s^2=-\frac{(z^2-l^2(4+A^2 z^2))^2}{16 A^2 l^2 z^2}\mathrm{d}\tau^2+\frac{l^2}{z^2}\mathrm{d}z^2+\frac{(z^2+l^2(4-A^2z^2))^2}{16z^2(1+\xi^2)(-1+A^2l^2(1+\xi^2))^2}\mathrm{d}\xi^2
\end{equation}
Similarly, we perform the following transformation $$T=\frac{\tau}{A} \qquad \text{and} \qquad \Xi=f(x_{2})-f(\xi),$$ with $\xi \in [x_{1},x_{2}]$, where $f(\xi)$ is defined as

$$f(\xi)=\frac{l \; \text{arctanh}\left(\frac{\ \xi}{\sqrt{1-A^2l^2}\sqrt{1+\xi^2}}\right)}{\sqrt{1-A^2 l^2}} \qquad \rm for \; III_{\rm b}.$$ 
Then the metric becomes
\begin{equation}\label{FG3}
\mathrm{d}s^2=-\frac{(z^2-l^2(4+A^2 z^2))^2}{16 l^2 z^2}\mathrm{d}T^2+\frac{l^2}{z^2}\mathrm{d}z^2+\frac{(z^2+l^2(4-A^2z^2))^2}{16l^2z^2}\mathrm{d}\Xi^2,
\end{equation}
where 
\begin{align}
\Xi\in[0,2\pi\beta] \qquad \text{and} \qquad \beta=\frac{\Xi |_{\xi=x_{1}}}{\pi}. \label{beta3}
\end{align}
\end{itemize}

As we have seen above, the metrics of these three solutions in the FG expansion all exhibit the same asymptotic behavior:
\begin{equation}
    \mathrm{d}s^2=\frac{l^2}{z^2}\mathrm{d}z^2+\frac{l^2(-\mathrm{d}t^2+\mathrm{d}\Xi^2)}{z^2}+\mathcal{O}(1)_{ij}\mathrm{d}x^{i}\mathrm{d}x^{j},
\end{equation}
indicating that these metrics are asymptotically conformally flat. When compared with the Poincare disk in AdS, the metrics in FG expansion include extra constant terms and $z^{2}$ terms. And near the boundary ($z\rightarrow 0$), these terms become subleading, thus the metric ($\ref{2.78}$) is expected to exhibit the same asymptotic algebraic structure as those of the Poincare disk. 
\subsection{Canonical Realization of Asymptotic Symmetry}
This section is largely a review of~\cite{Brown:1986nw}.  We start with a review of the Hamiltonian formalism of gravitational field, which takes a form as
\begin{equation}
H_{0}=\int \mathrm{d}x^2[N\mathcal{H}_{t}+N^{i}\mathcal{H}_{i}],
\end{equation}
 where $\mathcal{H}_{\mu}$ represent the standard constraints, $N$ and $N^{i}$ are lapse and shift functions, respectively, for the spacetime coordinate system under 2+1 decomposition as
\begin{equation}\label{2+1 decomposition}
\mathrm{d}s^2=G_{\mu\nu}\mathrm{d}x^{\mu}\mathrm{d}x^{\nu}=-N^2\mathrm{d}t^2+h_{ij}(N^{i}\mathrm{d}t+\mathrm{d}x^{i})(N^{j}\mathrm{d}t+\mathrm{d}x^{j})\, ,
\end{equation}
and the conjugate variables
\bal
K_{ij}&=\frac{\dot{h}_{ij}-\nabla_{i}N_{j}-\nabla_{j}N_{i}}{2N}\,, \qquad 
\Pi^{ij}=\sqrt{h}(K^{ij}-Kh^{ij})\,,
\eal
where $h$ represents the spatial part of the metric.

To compute the aysmptotic symmetry, we consider all allowed transformations that preserve $SO(2,2)$ isometry and the following Brown-Henneaux type boundary condition~\footnote{It is possible to impose other types of boundary conditions~\cite{Compere:2013bya,Grumiller:2016pqb} and the resulting asymptotic symmetry algebra will be differet. Here we only consider this carnonical boundary condition.}
\begin{equation}\label{3.23}
\delta g={\begin{bmatrix}
\mathcal{O}(1)&\mathcal{O}\(\frac{1}{z}\)&\mathcal{O}(1)\\
\mathcal{O}\(\frac{1}{z}\)&0&\mathcal{O}\(\frac{1}{z}\)\\
\mathcal{O}(1)&\mathcal{O}\(\frac{1}{z}\)&\mathcal{O}(1)\\
\end{bmatrix}}.
\end{equation}
Given ($\ref{3.23}$) and the metric in the FG expansion above, we have:
$$N=\frac{l}{z}+\mathcal{O}(1), \quad N^{i}=0, \quad \Pi^{z}_{z}=\mathcal{O}(z), \quad \Pi^{z}_{\Xi}=\mathcal{O}(z^3).$$
By solving the asymptotic killing equation
\begin{equation}
    \mathcal{L}_{\mathcal{X}^{(3)}}g\approx\delta g\,,
\end{equation}
we get the general form of the asymptotic killing vectors
\begin{equation}
\begin{aligned}\label{asymptotic algebra}
        \mathcal{X}^{(3)t}&=\frac{f_{1}(t+\Xi)+f_{2}(t-\Xi)}{2}+\mathcal{O}(z^2),\\
        \mathcal{X}^{(3)z}&=\frac{(f^{'}_{1}(t+\Xi)+f^{'}_{2}(t-\Xi))z}{2},\\
        \mathcal{X}^{(3)\Xi}&=\frac{f_{1}(t+\Xi)-f_{2}(t-\Xi)}{2}+\mathcal{O}(z^2)\,,
    \end{aligned}
\end{equation}
where $f_{1}(t+\Xi)$ and $f_{2}(t-\Xi)$ are arbitrary functions. The periodicity of $\Xi$ is $\Xi\sim\Xi+2\beta \pi$ where $\beta$ is defined as in section~\ref{FGsec}. 
A set of independent basis that generates the asymptotic symmetry group is 
the left- and right-moving Fourier modes of the Killing vectors
\begin{equation}
\begin{aligned}\label{3.9}
\mathcal{X}^{(3)}_{m(L)}&=\frac{\beta}{2}   e^{-\frac{im(t+\Xi)}{\beta}}\partial_t -\frac{1}{2} \ imz e^{-\frac{im(t+\Xi)}{\beta}}\partial_z+\ \frac{\beta}{2} e^{-\frac{im(t+\Xi)}{\beta}}\partial_\Xi ,\\
\mathcal{X}^{(3)}_{m(R)}&=\frac{\beta}{2}\  e^{-\frac{im(t-\Xi)}{\beta}}\partial_t - \frac{1}{2}\ i mz e^{-\frac{im(t-\Xi)}{\beta}}\partial_z -\frac{\beta}{2} \ e^{-\frac{im(t-\Xi)}{\beta}}\partial_\Xi\,,
\end{aligned}
\end{equation}
 where $n\in\mathbb{Z}$ and the parameter $\beta$ is introduced to ensure that $\frac{\Xi}{\beta}$ ranges within ($-\pi,\pi$). 
 These generators construct a Witt algebra 
 \begin{equation}
 \begin{aligned}
     \[\mathcal{X}^{(3)}_{m(L)}\, ,\mathcal{X}^{(3)}_{n(L)} \]&=i(m-n)\mathcal{X}^{(3)}_{m+n(L)},\\
     \[\mathcal{X}^{(3)}_{m(R)}\, ,\mathcal{X}^{(3)}_{n(R)} \]&=i(m-n)\mathcal{X}^{(3)}_{m+n(R)},\\
     \[\mathcal{X}^{(3)}_{m(L)}\, ,\mathcal{X}^{(3)}_{n(R)} \]&=0\ .
\end{aligned}
 \end{equation}
For convenience, we define a set $\mathbb{W}$ consisting of all generators as:
\begin{equation}
    \mathbb{W}:=\{\lambda_{m}\mathcal{X}^{(3)}_{m(L)}+\lambda_{n}\mathcal{X}^{(3)}_{n(R)}|m,n\in\mathbb{Z},\lambda_{m}\in\mathbb{C},\lambda_{n}\in\mathbb{C}\}.
\end{equation}

In general, we find a Hamiltonian generated by a vector $\mathcal{X}$ in $\rm AdS_{3}$ gravity is given as
\begin{equation}
\begin{aligned}\label{Hamiltonian formalism}
&H(\mathcal{X})=H_{0}(\mathcal{X})+J(\mathcal{X})\\
&=\int \mathrm{d}x^2[-\mathcal{X}^{t}\sqrt{h}\left(R(h)+2\Lambda\right)+\frac{\mathcal{X}^{t}}{\sqrt{h}}\left(\Pi_{ab}\Pi^{ab}-\frac{1}{2}\Pi^2\right)-2\sqrt{h}\mathcal{X}_{\rm b}\nabla_{a}(\Pi^{ab} h^{-1/2})]+J(\mathcal{X}),\\
\end{aligned}
\end{equation}
where $J(\mathcal{X})$ is the surface term that ensures the variation of the Hamiltonian is well-defined, $\Lambda$ is the cosmological constant. 
The surface term $J(\mathcal{X})$ can be defined as a boundary integral
\begin{equation}
\begin{aligned}\label{boundary charge}
&J(\mathcal{X})
=-\frac{1}{16\pi G_{3}}\lim\limits_{z \to 0}\oint \mathrm{d}x_{l}\left[ G^{ijkl}(h)(\mathcal{X}^{t}\nabla_{k}(g_{ij}-h_{ij})-(g_{ij}-h_{ij})\nabla_{k}\mathcal{X}^{t})+2\mathcal{X}^{m}\Pi_{m}^{l}(g)\right],
\end{aligned}
\end{equation}
where $h_{ij}$ is the induced metric in \eqref{2+1 decomposition}, $g_{ij}=h_{ij}+\delta h_{ij}$ is deformed metric compared with $h_{ij}$, and $G^{ijkl}(h)=\sqrt{h}(h^{ik}h^{jl}-h^{ij}h^{kl})$. 
Here $\mathcal{X}$ denotes the direction of deformation corresponding to asymptotic killing vector $\mathcal{X}^{(3)}\in\mathbb{W}$. Then every element in $\mathcal{W}$ corresponds to a direction of deformation, thus we can define a set that consists of all directions of deformation as:
\begin{equation}
\tilde{\mathbb{W}}:=\{P^{\mu}_{\alpha}\mathcal{X}^{(3)\alpha}=\(N\mathcal{X}^{(3)t},\; \mathcal{X}^{(3)z},\;\mathcal{X}^{(3)\Xi}\)|\mathcal{X}^{(3)}\in\mathbb{W}\}.
\end{equation}
Given three deformation vectors $\mathcal{X},\mathcal{Y},\mathcal{Z}\in \tilde{\mathbb{W}}$ corresponding to three generators $\mathcal{X}^{(3)},\mathcal{Y}^{(3)},\mathcal{Z}^{(3)}\in \mathbb{W}$ that satisfy
\begin{equation}
\[\mathcal{X}^{(3)},\mathcal{Y}^{(3)}\]=\mathcal{Z}^{(3)}\,.
\end{equation}
Following \cite{Brown:1986nw} we have
\begin{equation}
\begin{aligned}
\[\mathcal{X},\mathcal{Y}\]^{t}&=N\mathcal{X}^{(3)\mu}\p_{\mu}\mathcal{Y}^{(3)t}+\p_{\mu}(N)\mathcal{X}^{(3)\mu}\mathcal{Y}^{(3)t}-(\mathcal{X}\leftrightarrow\mathcal{Y})=N\mathcal{Z}^{(3)}=\mathcal{Z}^{t},\\
\[\mathcal{X},\mathcal{Y}\]^{i}&=G^{i\mu}N^2\mathcal{X}^{(3)t}\p_{\mu}\mathcal{Y}^{(3)t}+\mathcal{X}^{(3)\mu}\p_{\mu}\mathcal{Y}^{(3)i}-(\mathcal{X}\leftrightarrow\mathcal{Y})=\mathcal{Z}^{(3)i}+\mathcal{O}(z)\approx\mathcal{Z}^{i},
\end{aligned}
\end{equation}
thus the elements in $\tilde{W}$ construct deformation algebra to leading order in $z$.

The Hamiltonian generated by these deformation vectors obeys a central extended  asymptotic symmetry algebra 
\begin{equation}
\{ H(\mathcal{X}) ,H(\mathcal{Y}) \}_{DB}=H(\mathcal{Z})+K(\mathcal{X},\mathcal{Y})\,,
\end{equation}
where the extension term $K(\mathcal{X},\mathcal{Y})$ arises from the algebraic relation of the surface term $J$~\footnote{In our case, the canonical variables are the Killing vector and the induced metric. So the Poisson bracket $\{A, B \}_{PB}$ is defined as $\frac{\delta A}{\delta q^{i}}\frac{\delta B}{\delta p_{i}}+\frac{\delta A}{\delta h^{ij}}\frac{\delta B}{\delta \pi_{ij}}-A\leftrightarrow B $. For standard constraints $\mathcal{H}_{i}$, the Dirac bracket $\{ A,B \}_{DB}$ is defined as $\{ A,B \}_{PB}-\{ A,\mathcal{H}_{i} \}_{PB}\{ \mathcal{H}_{i},\mathcal{H}_{j} \}_{PB}^{-1}\{ B,\mathcal{H}_{j} \}_{PB}$.}. More specifically, the Poisson bracket of $H_{0}$ satisfies
\begin{equation}
\{ H_{0}(\mathcal{X}) ,H_{0}(\mathcal{Y}) \}_{DB}= H_{0}(\mathcal{Z}),
\end{equation}
the Dirac bracket of the surface term can be computed \cite{hanson1976constrained,barnich2007classical,barnich2002covariant}
\begin{equation}
\{J(\mathcal{X}),J(\mathcal{Y}) \}_{DB}=\delta_{\mathcal{Y}} J(\mathcal{X})=J([\mathcal{X},\mathcal{Y}])+K(\mathcal{X},\mathcal{Y})\ .
\end{equation}
According to \eqref{boundary charge}, the deformation of $J(\mathcal{X})$ is given by the variation over $g_{ij}$ and $\mathcal{X}$ as
\begin{equation}
    \begin{aligned}\label{The deformation of J}
       &\delta_{\mathcal{Y}}J(\mathcal{X})\\ 
      =&-\frac{1}{16\pi G_{3}}\lim\limits_{z \to 0}\oint \mathrm{d}x_{l}\left[ G^{ijkl}(h)((\delta_{\mathcal{Y}}\mathcal{X}^{t})\nabla_{k}(g_{ij}-h_{ij})-(g_{ij}-h_{ij})\nabla_{k}\delta_{\mathcal{Y}}(\mathcal{X}^{t}))+2\delta_{\mathcal{Y}}(\mathcal{X}^{m})\Pi_{m}^{l}(g)\right]\\
      &-\frac{1}{16\pi G_{3}}\lim\limits_{z \to 0}\oint \mathrm{d}x_{l}\left[ G^{ijkl}(h)(\mathcal{X}^{t}\nabla_{k}(\delta_{\mathcal{Y}}g_{ij})-(\delta_{\mathcal{Y}}g_{ij})\nabla_{k}\mathcal{X}^{t})+2\mathcal{X}^{m}\delta_{\mathcal{Y}}\Pi_{m}^{l}(g)\right]\\
      =&J([\mathcal{X},\mathcal{Y}])\\
      &-\frac{1}{16\pi G_{3}}\lim\limits_{z \to 0}\oint \mathrm{d}x_{l}\left[ G^{ijkl}(h)(\mathcal{X}^{t}\nabla_{k}(\delta_{\mathcal{Y}}g_{ij})-(\delta_{\mathcal{Y}}g_{ij})\nabla_{k}\mathcal{X}^{t})+2\mathcal{X}^{m}\delta_{\mathcal{Y}}\Pi_{m}^{l}(g)\right],
    \end{aligned}
\end{equation}
where we have used $\delta_{\mathcal{Y}}X=\mathcal{L}_{\mathcal{Y}}\mathcal{X}=[\mathcal{X},\mathcal{Y}]$. The central term thus reads
\begin{equation}
\begin{aligned}
    &K(\mathcal{X},\mathcal{Y})=-\frac{1}{16\pi G_{3}}\lim\limits_{z \to 0}\oint \mathrm{d}x_{l}\left[ G^{ijkl}(h)(\mathcal{X}^{t}\nabla_{k}(\delta_{\mathcal{Y}}g_{ij})-(\delta_{\mathcal{Y}}g_{ij})\nabla_{k}\mathcal{X}^{t})+2\mathcal{X}^{m}\delta_{\mathcal{Y}}\Pi_{m}^{l}(g)\right],
\end{aligned}
\end{equation}
where $\delta{h}_{ij}=\mathcal{L}_{\mathcal{Y}}h_{ij}$ and $\delta_{\mathcal{Y}}$ denotes the deformation along $\mathcal{Y}$.
%
Plugging in the form of the deformation vectors, we find 
\begin{equation}
\begin{aligned}\label{Extended term of three classes}
K\left(\mathcal{X}_{n(L)},\mathcal{X}_{m(L)}\right)=\frac{inl(n^2+\lambda)}{8G_{3}}\delta_{m+n,0}\,,\quad
K\left(\mathcal{X}_{n(R)},\mathcal{X}_{m(R)}\right)=\frac{inl(n^2+\lambda)}{8G_{3}} \delta_{m+n,0}\,,
\end{aligned}
\end{equation}
where $\lambda$ is defined as follows
\begin{equation}
    \lambda:=
    \begin{cases}
        \frac{(A^2l^2-1)\beta^2}{l^2}, & \text{for Class I}, \\
        \frac{(A^2l^2+1)\beta^2}{l^2}, & \text{for Class II},\\
        \frac{(1-A^2l^2)\beta^2}{l^2}, & \text{for Class III},
    \end{cases}
\end{equation}
and $\beta$ takes the corresponding value of the three cases~\eqref{beta1}, \eqref{beta2} and \eqref{beta3} respectively. A further redefinition $J(\mathcal{X}_{0})\to J(\mathcal{X}_{0})-\frac{il(1+\lambda)}{16G_{N}}$ put the result into the standard form of the Virasoro algebra~\footnote{Let us emphasize that the redefinition of the zero mode $J(\mathcal{X}_{0})$ is due to the specific gauge condition, which shift the Casimir energy~\cite{balasubramanian2015entwinement,berenstein2023aspects}. Treatment in a more general gauge and some detailed results are provided in Appendix~\ref{B}. 
}
\begin{equation}
\begin{aligned}\label{3.21}
\{ J(\mathcal{X}_{n(L)}), J(\mathcal{X}_{m(L)})\}_{DB}&=(n-m)J(\mathcal{X}_{n+m(L)})+\frac{i l(n^3-n)}{8G_{3}}\delta_{m+n,0},\\
\{ J(\mathcal{X}_{n(R)}), J(\mathcal{X}_{m(R)})\}_{DB}&=(n-m)J(\mathcal{X}_{n+m(R)})+\frac{i l(n^3-n)}{8G_{3}}\delta_{n+m,0},
\end{aligned}
\end{equation}
and the central charge can be read off from (\ref{3.21}) to be
$$c=\frac{3 l}{2G_{3}}.$$ This result indicates that the C-metric system can be consistently dual to a 2D $\rm CFT$. 
\section{Thermodynamic relation of accelerating BTZ black hole}
In this section we discuss the thermodynamic relation of all phases of accelerating BTZ black hole with two strings.~\footnote{Similar discussion in 4d can be found in, e.g.~\cite{Liu:2024fvq}. }
\begin{itemize}
    \item \textbf{The I$_{\rm b1}$ case}
    
For the I$_{\rm b1}$ case, we set the Weyl factor $\omega(\xi)=1$. Then based on \eqref{mass} and \eqref{2.75}, the holographic mass is
\begin{equation}\label{Mass1}
    M_{\rm I_{b1}}=\int^{x=x_{2}}_{x=x_{1}} \mathrm{d}\xi \sqrt{-g_{(0)}}T^{\tau}_{\tau(\text{ren})}=\frac{\ \sqrt{A^2l^2-1}\mathrm{arctanh}\left(\frac{\ \sqrt{A^2l^2-1}\sqrt{1-x^2}}{\ x}\right)}{\ 8\pi G_{3}}\Bigg|^{x=x_{1}}_{x=x_{2}},
\end{equation}
with a wall at $x=x_{1}$ and a strut at $x=x_{2}$. The temperature and entropy are
\begin{align}
    T_{\rm h}&=\frac{AP'(y_{\rm h})}{4\pi}=\frac{\sqrt{A^2l^2-1}}{2\pi l},\label{Temperature1}\\
    S_{\rm BH}&=\int^{x=x_{2}}_{x=x_{1}}\frac{2}{4G_{3}A(x-y_{\rm h})\sqrt{1-x^2}}\mathrm{d}x=\frac{l}{2G_{3}}\mathrm{arctanh}\(\frac{\sqrt{1-x^2}}{Al-\sqrt{A^2l^2-1}x}\) \Bigg|^{x=x_{1}}_{x=x_{2}}\ .\label{entropy1}
\end{align}
The boundary entropy consisting of the string is
\begin{equation}\label{entropy of boundary}
        S_{\rm boundary}=S_{\rm wall}+S_{\rm strut}\,,
\end{equation}
with \cite{fujita2011aspects,Takayanagi:2011zk}
\begin{equation}
    \begin{aligned}
        S_{\rm wall}&=\frac{2c}{6}\mathrm{arctanh}(4\pi l\sigma)=\frac{l}{2G_{3}}\mathrm{arctanh}\(Al\sqrt{1-x_{1}^2}\),\\
        S_{\rm strut}&=\frac{2c}{6}\mathrm{arctanh}(4\pi l\sigma)=-\frac{l}{2G_{3}}\mathrm{arctanh}\(Al\sqrt{1-x_{2}^2}\),
    \end{aligned}
\end{equation}
in which the factor 2 in front of $c$ comes from two copies.
Then \eqref{entropy of boundary} can be summarized up as 
\begin{equation}
    S_{\rm boundary}=\frac{l}{2G_{3}}\mathrm{arctanh}\(Al\sqrt{1-x^2}\)\Big|^{x=x_{1}}_{x=x_{2}}\ .
\end{equation}
According to the monotonicity of $\sqrt{1-x^2}$, we can see that when $y_{\rm h}<x_{1}<x_{2}<1$, the boundary entropy is negative, and when $-1<x_{1}<x_{2}<-y_{\rm h}$, the boundary entropy is positive.

Then combining \eqref{Mass1}, \eqref{Temperature1} and \eqref{entropy1}, we have:
\begin{equation}
    \begin{aligned}
        S-\frac{2M_{\rm I_{b1}}}{T_{\rm h}}&=\L.\frac{l}{2G_{3}}\left[ \mathrm{arctanh}\(\frac{\sqrt{1-x^2}}{Al-\sqrt{A^2l^2-1}x}\)-\mathrm{arctanh}\(\frac{\ \sqrt{A^2l^2-1}\sqrt{1-x^2}}{\ x}\)\right]\R|^{x=x_{1}}_{x=x_{2}}\\
        &=\L.\frac{l}{4G_{3}}\log\(\frac{1+\frac{\sqrt{1-x^2}}{Al-\sqrt{A^2l^2-1}x}}{1-\frac{\sqrt{1-x^2}}{Al-\sqrt{A^2l^2-1}x}}\times\frac{1-\frac{\ \sqrt{A^2l^2-1}\sqrt{1-x^2}}{\ x}}{1+\frac{\ \sqrt{A^2l^2-1}\sqrt{1-x^2}}{\ x}}\)\R|^{x=x_{1}}_{x=x_{2}}\\
        &=\L.\frac{l}{4G_{3}}\log\(\frac{1+Al\sqrt{1-x^2}}{1-Al\sqrt{1-x^2}}\)\right|^{x=x_{1}}_{x=x_{2}}\\
        &=S_{\rm boundary}\ .
    \end{aligned}
\end{equation}
Thus the Smarr relation of I$_{\rm b1}$ is:
\begin{equation}
    2M_{\rm I_{b1}}=T_{\rm h}(S_{\rm BH}-S_{\rm boundary})\ .
\end{equation}
         \item \textbf{The II$_{\rm b1}$ case}

For the II$_{\rm b1}$ case, we set the Weyl factor $\omega(\xi)=1$. Then based on \eqref{mass} and \eqref{2.82}, the holographic mass is
\begin{equation}\label{Mass2}
    M_{\rm II_{b1}}=\int^{x=x_{2}}_{x=x_{1}} \mathrm{d}\xi \sqrt{-g_{(0)}}T^{\tau}_{\tau(\text{ren})}=\L.\frac{\ \sqrt{A^2l^2+1}\mathrm{arctanh}\(\frac{\sqrt{A^2l^2+1}\sqrt{x^2-1}}{x}\)}{\ 8\pi G_{3}}\R|^{x=x_{2}}_{x=x_{1}}\,,
\end{equation}
with a wall at $x=x_{1}$ and a strut at $x=x_{2}$. The temperature and entropy are
\begin{align}
    T_{\rm h}&=\frac{AP'(y_{\rm h})}{4\pi}=\frac{\sqrt{A^2l^2+1}}{2\pi l},\label{Temperature2}\\
    S_{\rm BH}&=\int^{x=x_{2}}_{x=x_{1}}\frac{2}{4G_{3}A(x-y_{\rm h})\sqrt{x^2-1}}\mathrm{d}x=\frac{l}{2G_{3}}\mathrm{arctanh}\left(\frac{\sqrt{x^2-1}}{Al+\sqrt{A^2l^2+1}x}\right) \Bigg|^{x=x_{2}}_{x=x_{1}}.\label{entropy2}
\end{align}
The boundary entropy consisting of the string is
\begin{equation}\label{entropy of boundary2}
        S_{\rm boundary}=S_{\rm wall}+S_{\rm strut}\,,
\end{equation}
with
\begin{equation}
    \begin{aligned}
        S_{\rm wall}&=\frac{2c}{6}\mathrm{arctanh}\(4\pi l\sigma\)=\frac{l}{2G_{3}}\mathrm{arctanh}\(Al\sqrt{x_{1}^2-1}\),\\
        S_{\rm strut}&=\frac{2c}{6}\mathrm{arctanh}\(4\pi l\sigma\)=-\frac{l}{2G_{3}}\mathrm{arctanh}\(Al\sqrt{x_{2}^2-1}\),
    \end{aligned}
\end{equation}
in which the factor 2 in front of $c$ comes from two copies.
Then \eqref{entropy of boundary2} can be summarized up as 
\begin{equation}
    S_{\rm boundary}=\L.\frac{l}{2G_{3}}\mathrm{arctanh}\(Al\sqrt{x^2-1}\)\R|^{x=x_{1}}_{x=x_{2}}
\end{equation}
Considering that $|x_{1}|>|x_{2}|>1$ in $\rm II_{b1}$, the boundary entropy is positive.

Then combining \eqref{Mass2}, \eqref{Temperature2} and \eqref{entropy2}, we have:
\begin{equation}
    \begin{aligned}
        S-\frac{2M_{\rm II_{b1}}}{T_{\rm h}}&=\L.\frac{l}{2G_{3}}\left[ \mathrm{arctanh}\left(\frac{\sqrt{x^2-1}}{Al+\sqrt{A^2l^2+1}x}\right)-\mathrm{arctanh}\(\frac{\sqrt{A^2l^2+1}\sqrt{x^2-1}}{x}\)\right]\R|^{x=x_{2}}_{x=x_{1}}\\
        &=\L.\frac{l}{4G_{3}}\log\L(\frac{1+\frac{\sqrt{x^2-1}}{Al+\sqrt{A^2l^2+1}x}}{1-\frac{\sqrt{x^2-1}}{Al+\sqrt{A^2l^2+1}x}}\times\frac{1-\frac{\sqrt{A^2l^2+1}\sqrt{x^2-1}}{x}}{1+\frac{\sqrt{A^2l^2+1}\sqrt{x^2-1}}{x}}\R)\R|^{x=x_{2}}_{x=x_{1}}\\
        &=\L.\frac{l}{4G_{3}}\log\L(\frac{1-Al\sqrt{1-x^2}}{1+Al\sqrt{1-x^2}}\R)\R|^{x=x_{2}}_{x=x_{1}}\\
        &=S_{\rm boundary}.
    \end{aligned}
\end{equation}
Thus the Smarr relation of II$_{\rm b1}$ is:
\begin{equation}
    2M_{\rm II_{b1}}=T_{\rm h}(S_{\rm BH}-S_{\rm boundary}).
\end{equation}
 \item \textbf{The II$_{\rm b2}$ case}
 
 For the II$_{\rm b2}$ case, we set the Weyl factor $\omega(\xi)=1$. Then based on \eqref{mass} and \eqref{2.82}, the holographic mass is
\begin{equation}\label{Mass3}
    M_{\rm II_{b2}}=\L.\frac{\ \sqrt{A^2l^2+1}\mathrm{arctanh}\(\frac{\sqrt{A^2l^2+1}\sqrt{x^2-1}}{x}\)}{\ 8\pi G_{3}}\R|^{x=x_{2}}_{x=x_{1}}\,,
\end{equation}
with a wall at $x=x_{1}$  and a strut at $x=x_{2}$. The temperature and entropy are
\begin{align}
    T_{\rm h}&=\frac{AP'(y_{\rm h})}{4\pi}=\frac{\sqrt{A^2l^2+1}}{2\pi l},\label{Temperature3}\\
    S_{\rm BH}&=\int^{x=x_{2}}_{x=x_{1}}\frac{2}{4G_{3}A(x-y_{\rm h})\sqrt{x^2-1}}\mathrm{d}x=\L.\frac{l}{2G_{3}}\mathrm{arctanh}\left(\frac{\sqrt{x^2-1}}{Al+\sqrt{A^2l^2+1}x}\right)\R|^{x=x_{2}}_{x=x_{1}}\ .\label{entropy3}
\end{align}
The boundary entropy consisting of the string  is
\begin{equation}\label{entropy of boundary3}
        S_{\rm boundary}=S_{\rm wall}+S_{\rm strut}\,,
\end{equation}
with
\begin{equation}
    \begin{aligned}
        S_{\rm wall}&=\frac{2c}{6}\mathrm{arctanh}(4\pi l\sigma)=\frac{l}{2G_{3}}\mathrm{arctanh}\(Al\sqrt{x_{1}^2-1}\),\\
        S_{\rm strut}&=\frac{2c}{6}\mathrm{arctanh}(4\pi l\sigma)=-\frac{l}{2G_{3}}\mathrm{arctanh}\(Al\sqrt{x_{2}^2-1}\),
    \end{aligned}
\end{equation}
in which the factor 2 in front of $c$ comes from two copies.
Then \eqref{entropy of boundary3} can be summarized up as 
\begin{equation}
    S_{\rm boundary}=\frac{l}{2G_{3}}\L.\mathrm{arctanh}\(Al\sqrt{x^2-1}\)\R|^{x=x_{1}}_{x=x_{2}}\ .
\end{equation}
Considering that $1<|x_{1}|<|x_{2}|$ in $\rm II_{b2}$, the boundary entropy is negative.

Then combining \eqref{Mass3}, \eqref{Temperature3} and \eqref{entropy3}, we have:
\begin{equation}
    \begin{aligned}
        S-\frac{2M_{\rm II_{b2}}}{T_{\rm h}}&=\L.\frac{l}{2G_{3}}\left[ \mathrm{arctanh}\left(\frac{\sqrt{x^2-1}}{Al+\sqrt{A^2l^2+1}x}\right)-\mathrm{arctanh}\(\frac{\sqrt{A^2l^2+1}\sqrt{x^2-1}}{x}\)\right]\R|^{x=x_{2}}_{x=x_{1}}\\
        &=\L.\frac{l}{4G_{3}}\log\L(\frac{1+\frac{\sqrt{x^2-1}}{Al+\sqrt{A^2l^2+1}x}}{1-\frac{\sqrt{x^2-1}}{Al+\sqrt{A^2l^2+1}x}}\times\frac{1-\frac{\sqrt{A^2l^2+1}\sqrt{x^2-1}}{x}}{1+\frac{\sqrt{A^2l^2+1}\sqrt{x^2-1}}{x}}\R)\R|^{x=x_{2}}_{x=x_{1}}\\
        &=\L.\frac{l}{4G_{3}}\log\(\frac{1-Al\sqrt{1-x^2}}{1+Al\sqrt{1-x^2}}\)\R|^{x=x_{2}}_{x=x_{1}}\\
        &=S_{\rm boundary}.
    \end{aligned}
\end{equation}
Thus the Smarr relation of II$_{\rm b2}$ is:
\begin{equation}
    2M_{\rm II_{b2}}=T_{\rm h}(S_{\rm BH}-S_{\rm boundary}).
\end{equation}
\item \textbf{The III$_{\rm b}$ case}

 For the III$_{\rm b}$ case, we set the Weyl factor $\omega(\xi)=1$. Then based on \eqref{mass} and \eqref{FG3}, the holographic mass is
\begin{equation}\label{Mass4}
    M_{\rm III_b}=\int^{x=x_{2}}_{x=x_{1}} \mathrm{d}\xi \sqrt{-g_{(0)}}T^{\tau}_{\tau(\text{ren})}=\frac{\ \sqrt{1-A^2l^2}\mathrm{arctanh}\left(\frac{\ x}{\sqrt{1-A^2l^2}\sqrt{1+x^2}}\right)}{\ 8\pi G_{3}}\Bigg|^{x=x_{2}}_{x=x_{1}},
\end{equation}
with a wall at $x=x_{1}$ and a strut at $x=x_{2}$. The temperature and entropy are
\begin{align}
    T_{\rm h}&=\frac{AP'(y_{\rm h})}{4\pi}=\frac{\sqrt{1-A^2l^2}}{2\pi l},\label{Temperature4}\\
    S_{\rm BH}&=\int^{x=x_{2}}_{x=x_{1}}\frac{2}{4G_{3}A(x-y_{\rm h})\sqrt{x^2+1}}\mathrm{d}x=\frac{l}{2G_{3}}\mathrm{arctanh}\left( \frac{\sqrt{1-A^2l^2}x-Al}{\sqrt{x^2+1}}\right)\Bigg|^{x=x_{2}}_{x=x_{1}}\ .\label{entropy4}
\end{align}
The boundary entropy consisting of the string is
\begin{equation}\label{entropy of boundary4}
        S_{\rm boundary}=S_{\rm wall}+S_{\rm strut}\,,
\end{equation}
with
\begin{equation}
    \begin{aligned}
        S_{\rm wall}&=\frac{2c}{6}\mathrm{arctanh}(4\pi l\sigma)=\frac{l}{2G_{3}}\mathrm{arctanh}(Al\sqrt{x_{1}^2+1}),\\
        S_{\rm strut}&=\frac{2c}{6}\mathrm{arctanh}(4\pi l\sigma)=-\frac{l}{2G_{3}}\mathrm{arctanh}(Al\sqrt{x_{2}^2+1}),
    \end{aligned}
\end{equation}
in which the factor 2 in front of $c$ comes from two copies.
Then \eqref{entropy of boundary4} can be summarized up as 
\begin{equation}
    S_{\rm boundary}=\L.\frac{l}{2G_{3}}\mathrm{arctanh}\(Al\sqrt{x^2+1}\)\R|^{x=x_{1}}_{x=x_{2}}\ .
\end{equation}
When $|x_{1}|>|x_{2}|$, the boundary entropy is positive, otherwise, the boundary is negative.

Then combining \eqref{Mass4}, \eqref{Temperature4} and \eqref{entropy4}, we have:
\begin{equation}
    \begin{aligned}\label{Smarr relation of Class III}
        S-\frac{2M_{\rm III_b}}{T_{\rm h}}&=\L.\frac{l}{2G_{3}}\left[ \mathrm{arctanh}\left( \frac{\sqrt{1-A^2l^2}x-Al}{\sqrt{x^2+1}}\right)-\mathrm{arctanh}\left(\frac{\ x}{\sqrt{1-A^2l^2}\sqrt{1+x^2}}\right)\right]\R|^{x=x_{2}}_{x=x_{1}}\\
        &=\L.\frac{l}{4G_{3}}\log\(\frac{1+ \frac{\sqrt{1-A^2l^2}x-Al}{\sqrt{x^2+1}}}{1- \frac{\sqrt{1-A^2l^2}x-Al}{\sqrt{x^2+1}}}\times\frac{1-\frac{\ x}{\sqrt{1-A^2l^2}\sqrt{1+x^2}}}{1+\frac{\ x}{\sqrt{1-A^2l^2}\sqrt{1+x^2}}}\)\R|^{x=x_{2}}_{x=x_{1}}\\
        &=\L.\frac{l}{4G_{3}}\log\(\frac{1-Al\sqrt{1+x^2}}{1+Al\sqrt{1+x^2}}\)\R|^{x=x_{2}}_{x=x_{1}}\\
        &=S_{\rm boundary}.
    \end{aligned}
\end{equation}
Thus the Smarr relation of III$_{\rm b}$ is:
\begin{equation}
    2M_{\rm III_b}=T_{\rm h}(S_{\rm BH}-S_{\rm boundary}).
\end{equation}
\end{itemize}
We can see that the entropy of boundary participates in the thermodynamics and all phases of accelerating black hole have the same Smarr relation.

\section{Dynamics on the strings}

The acceleration in the C-metric spacetime is induced by the strings, it is therefore interesting to understand more about their dynamics. In this section we study the reduced gravitational dynamics on the two strings in the system. 
We first discuss the reduction of fluctuations of the strings and get an dilaton-gravity theory. Then we discuss the Karch-Randall-Sundrum construction and introduce higher-curvature corrections to the effective theory. Finally, we study generalized entropy. 
\subsection{Fluctuations of the strings}

The previous discussion assumes the strings that source the acceleration in the solution to be rigid. It is also possible to consider fluctuations of them, and the analysis in~\cite{
Geng:2022slq,geng2022aspects} could be applied to our setup as well. As we will show in this section, the effects of the fluctuations can be equivalently described, from the perspective of the strings themselves, by dynamics of a dilaton field on the strings.



\subsubsection{Strings at $x/y=\rm{constant}$}
\begin{figure}[]
\centering
\includegraphics[width=0.4\linewidth]{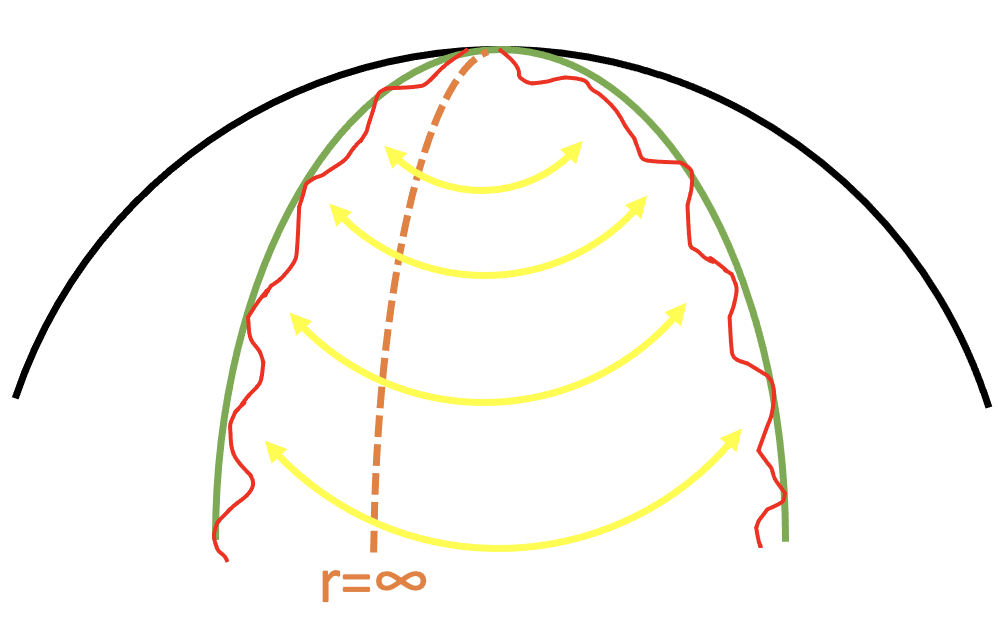}
\caption{
The green curves are two 
strings at different $\r$ = constant lines. 
The red curves denote fluctuations on the string. The black curve represents the aymptotic boundary. The orange dotted curve denotes the position of $y=0$. The yellow curves with double-arrows indicate the wedge of these two strings.}
\label{KRS}
\end{figure}

\noindent $\bullet$ \textbf{Case I$_{\rm p3}$}
We change from $(t,r,\psi)$ to $(T,\rho,\phi)$ using
\begin{equation}
\begin{aligned}\label{KRS 1}
t&=\frac{l T}{\sqrt{1-\mathcal{A}^2 l^2 m^2 }},\\
r&=-\frac{\mathfrak{g}_{s}(\rho)}{\mathcal{A}\cos(m\psi+\psi_{0})},\\
\cos(m\psi+\psi_{0})&=\frac{\mathfrak{g}_{s}(\rho)\sqrt{1-\mathcal{A}^2 l^2 m^2}}{\sqrt{(1-\mathcal{A}^2 l^2 m^2)\cosh(\phi)^2\mathfrak{g}_{s}(\rho)^2+\mathcal{A}^2 l^2 m^2\sinh(\phi)^2}}\,,\\
\mathfrak{g}_{s}(\rho)&=\frac{\mathcal{A}^2 l^2 m^2-\mathcal{A}^4 l^4 m^4+\sqrt{\mathcal{A}^2 l^2 m^2(1-\mathcal{A}^2 l^2 m^2)}\mathrm{sech}(\rho/l)\mathrm{tanh}(\rho/l)}{(1-\mathcal{A}^2 l^2 m^2)(\mathcal{A}^2 l^2 m^2-\mathrm{sech}(\rho/l)^2)}\,,
\end{aligned}
\end{equation}
where $\mathfrak{g}_{s}(\r) \in \mathbb{R}$ is the slope function. Then the line element ($\ref{2.5}$) in global coordinates for I$_{\rm p3}$ reads
\begin{equation}\label{x/y foilation of I3p}
\mathrm{d}s^2=\mathrm{d}\r^2+h_{ij}(\rho,x)\mathrm{d}x^{i}\mathrm{d}x^{j}=\mathrm{d}\r^2+l^2\cosh\(\frac{\r}{l}\)^2g_{ij}(x)\mathrm{d}x^{i}\mathrm{d}x^{j}\ .
\end{equation}
We consider the case with a
strut and at $\r_{1}$ and a wall at $\r_{2}$ ($0<\r_{1}<\r_{2}$), the tension on them can be obtained from Israel's junction condition
\begin{equation}
\mathcal{T}_{1}=\frac{1}{2}K_{1}=\frac{\tanh(\frac{-\r_{1}}{l})}{l},\quad \mathcal{T}_{2}=\frac{1}{2}K_{2}=\frac{\tanh(\frac{\r_{2}}{l})}{l}\ .
\end{equation}
%
Using the metric~\eqref{x/y foilation of I3p}, we get the relation between the bulk Ricci scalar and the 2d Ricci scalar defined by $g_{ij}$
\begin{equation}
R(G_{\rm{bulk}})=\frac{1}{l^2\cosh(\frac{\r}{l})^2}R(g)-\frac{4}{l^2}-\frac{2\tanh(\frac{\r}{l})^2}{l^2}\ .
\end{equation}
The full action in the bulk is
\begin{equation}\label{total action of 3D gravity}
    I_{\rm total}=-\frac{1}{16\pi G_{3}}\int \mathrm{d}^3x\sqrt{-G(x)}\left(\tilde{R}(x)+\frac{6}{l^2}\right)-\frac{1}{8\pi G_{3}}\int \mathrm{d}^2y\sqrt{-h(y)}(K(y)-\mathcal{T})\ .
\end{equation}
Allowing fluctuations of the positions of the strings, namely from constant $\r_{1}$, $\r_{2}$ to $\r_{1}+\delta \r_{1}$, $\r_{2}+\delta \r_{2}$ and integrate out $\rho$-direction, we get an effective action
\bal
\label{reduced action of I3p at x/y=constant}
I_{\rm reduced}=&-\frac{1}{16\pi G_{3}}\int^{\r_{2}+\delta \r_{2}}_{\r_{1}+\delta \phi_{1}}\mathrm{d}\r\int\mathrm{d}^2x\sqrt{-g}\left[ R(g)-4\cosh(\frac{\r}{l})^2-2\sinh(\frac{\r}{l})^2+2\cosh(\frac{\r}{l})^2\right]\nonumber \\
&-\frac{1}{8\pi G_{3}}\int\mathrm{d}^2x\sqrt{-\tilde{g}}l^2\cosh(\frac{\r_{1}+\delta \r_{1}}{l})^2(\frac{2\tanh(\frac{-\r_{1}-\delta \r_{1}}{l})}{l}-\frac{\tanh(\frac{-\r_{1}}{l})}{l})\nonumber \\
&-\frac{1}{8\pi G_{3}}\int\mathrm{d}^2x\sqrt{-\tilde{g}}l^2\cosh(\frac{\r_{2}+\delta \r_{2}}{l})^2(\frac{2\tanh(\frac{\r_{2}+\delta \r_{2}}{l})}{l}-\frac{\tanh(\frac{\r_{2}}{l})}{l})\,,
\eal
where $\tilde{g}$ denotes the metric along the strings at the new fluctuated position. Expanding to quadratic order of the fluctuations $\delta\rho_{1}$ or $\delta\rho_{2}$, it reads
\begin{equation}
   \tilde{g}_{ij}=g_{ij}+\cosh\left(\frac{\rho_{\rm I}}{l}\right)^{-2}\p_{i}\delta\rho_{I}\p_{j}\delta\rho_{I}\,, \quad  I=1,2\ .\label{gtilde}
\end{equation}
Then the effective action up to quadratic order is
\begin{equation}
\begin{split}\label{KRS 4}
I_{\rm reduced}=&-\frac{(\r_{2}-\r_{1})}{16\pi G_{3}}\int \mathrm{d}^2x\sqrt{-g}R(g)-\frac{1}{16\pi G_{3}}\int\mathrm{d}^2x\sqrt{-g}(\delta \r_{2}-\delta\r_{1})\left[ R(g)+2\right]\\
&-\frac{1}{8\pi G_{3}}\int\mathrm{d}^2x\sqrt{-g}[\frac{\tanh(\frac{\r_{2}}{l})}{2l}\nabla_{\mu}\delta \r_{2}\nabla^{\mu}\delta \r_{2}+\frac{\tanh(\frac{\r_{2}}{l})(\delta\r_{2})^2}{l}\\
&-\frac{\tanh(\frac{\r_{1}}{l})}{2l}\nabla_{\mu}\delta \r_{1}\nabla^{\mu}\delta \r_{1}-\frac{\tanh(\frac{\r_{1}}{l})(\delta\r_{1})^2}{l} ]\ .
\end{split}
\end{equation}
Setting $\Phi(x)=\delta\phi_{2}-\delta\phi_{1}$ and $\Psi(x)=\frac{\mathcal{T}_{1}}{\sqrt{\mathcal{T}_{1}+\mathcal{T}_{2}}}\delta\phi_{2}+\frac{\mathcal{T}_{2}}{\sqrt{\mathcal{T}_{1}+\mathcal{T}_{2}}}\delta\phi_{2}$, ($\ref{KRS 4}$) becomes
\begin{equation}
    \begin{aligned}\label{6.5}
        I_{\rm reduced}&=-\frac{(\r_{2}-\r_{1})}{16\pi G_{3}}\int \mathrm{d}^2x\sqrt{-g}R(g)-\frac{1}{16\pi G_{3}}\int\mathrm{d}^2x\sqrt{-g}\Phi(x)\left[ R(g)+2\right]\\
&-\frac{1}{8\pi G_{3}}\int\mathrm{d}^2x\sqrt{-g}\left[\frac{\mathcal{T}_{1}\mathcal{T}_{2}}{2(\mathcal{T}_{1}+\mathcal{T}_{2})}\nabla_{\mu}\Phi(x)\nabla^{\mu}\Phi(x)+\frac{\mathcal{T}_{1}\mathcal{T}_{2}}{(\mathcal{T}_{1}+\mathcal{T}_{2})}\Phi(x)^2\right]\\
&-\frac{1}{8\pi G_{3}}\int\mathrm{d}^2x\sqrt{-g}\left[\frac{1}{2}\nabla_{\mu}\Psi\nabla^{\mu}\Psi+\Psi^2\right]\ .
    \end{aligned}
\end{equation}
Rescaling $g_{ij}\rightarrow g_{ij}e^{-\frac{\mathcal{T}_{1}\mathcal{T}_{2}}{\mathcal{T}_{1}+\mathcal{T}_{2}}}$, ($\ref{6.5}$) becomes:
    \begin{equation}
    \begin{aligned}\label{reduced effective action 1}
        I_{\rm reduced}&=-\frac{(\r_{2}-\r_{1})}{16\pi G_{3}}\int \mathrm{d}^2x\sqrt{-g}R(g)-\frac{1}{16\pi G_{3}}\int\mathrm{d}^2x\sqrt{-g}\Phi(x)\left[ R(g)+2\right]\\
&-\frac{1}{8\pi G_{3}}\int\mathrm{d}^2x\sqrt{-g}\left[\frac{1}{2}\nabla_{\mu}\Psi\nabla^{\mu}\Psi+\Psi^2\right]
    \end{aligned}
\end{equation}
The first term of the effective action is  topological; the second term is the action of  JT gravity; and the third term is the action for a matter field $\Psi$. This result is very similar to that in~\cite{geng2022aspects}. An illustration of the reduction in this section is shown in Fig.~\ref{KRS}. 

To described the dynamical boundary excitations, similar to those governed by the Schwarzian effective action~\cite{Maldacena:2016upp}, we can add cutoffs near the asymptotic boundary in the $\phi$ direction along the strings which will lead to extra boundary terms proportional to the extrinsic curvature. The subsequence derivation is similar to that in~\cite{ Maldacena:2016upp} or~\cite{Geng:2022slq} so we skip further details here.  

\noindent $\bullet$ \textbf{Class I$_{\rm p1}$ and I$_{\rm b1}$}
We apply the following transformation from $(t,r,\psi)$ to $(T,\rho,\phi)$
\begin{equation}
\begin{aligned}
t&=\frac{l T}{\sqrt{\mathcal{A}^2 l^2 m^2 -1}},\\
r&=-\frac{\mathfrak{g}_{s}(\rho)}{\mathcal{A}\cos(m\psi+\psi_{0})},\\
\cos(m\psi+\psi_{0})&=-\frac{\mathfrak{g}_{s}(\rho)\sqrt{\mathcal{A}^2 l^2 m^2-1}}{\sqrt{\mathcal{A}^2 l^2 m^2\cosh(\phi)^2-\mathfrak{g}_{s}(\rho)^2(\mathcal{A}^2 l^2 m^2-1)\sinh(\phi)^2}},\\
\mathfrak{g}_{s}(\rho)&=-\frac{\mathcal{A}^2 l^2 m^2-\mathcal{A}^4 l^4 m^4+\sqrt{\mathcal{A}^2 l^2 m^2(\mathcal{A}^2 l^2 m^2-1)}\sec(\rho/l)\tan(\rho/l)}{(\mathcal{A}^2 l^2 m^2-1)(\ \mathcal{A}^2 l^2 m^2-\frac{1}{\cos(\rho/l)^2})},
\end{aligned}
\end{equation}
where $\mathfrak{g}_{s}(\r) \in \mathbb{R}$.  
Then the line element ($\ref{2.5}$) in global coordinates can be written schematically as
\begin{equation}\label{KRS 2}
\mathrm{d}s^2=-\mathrm{d}\rho^2+h_{ij}(\rho,x)\mathrm{d}x^{i}\mathrm{d}x^{j}=-\mathrm{d}\rho^2+l^2\cos\left(\frac{\rho}{l}\right)^2g_{ij}(x)\mathrm{d}x^{i}\mathrm{d}x^{j}\ .
\end{equation}
The bulk Ricci scalar can be decomposed into 
\begin{equation}\label{2.13}
R[G_{3}]=\frac{R(g)}{l^2 \cos(\frac{\rho}{l})^2}+\frac{2\tan(\frac{\rho}{l})^2}{l^2}-\frac{4}{l^2}\ .
\end{equation}

Denoting the position of the strut and the wall to be $\r_{1}$ and $\r_{2}$ ($\r_{2}>\r_{1}>0$) respectively, we get the tension on them from the  Israel junction condition
\begin{equation}
\mathcal{T}_{1}=\frac{1}{2}K_{1}=-\frac{\tan(\frac{\rho_{1}}{l})}{l}, \quad \mathcal{T}_{2}=\frac{1}{2}K_{2}=+\frac{\tan(\frac{\rho_{2}}{l})}{l}.
\end{equation}
As in the previous case, we substitute Eq.~\eqref{2.13} into the action \eqref{total action of 3D gravity} and allow fluctuations around the string $\r_{1}\rightarrow\r_{1}+\delta \r_{1}$, $\r_{2}\rightarrow\r_{2}+\delta \r_{2}$, then we get
\begin{equation}
\begin{aligned}
I_{\rm reduced}=&-\frac{1}{16\pi G_{3}}\int_{\r_{1}+\delta \r_{1}}^{\r_{2}+\delta \r_{2}}\mathrm{d}\r\int\mathrm{d}^2x\sqrt{-g}[R(g)+2\sin(\frac{\r}{l})^2-2\cos(\frac{\r}{l})^2]\\
&-\frac{1}{8\pi G_{3}}\int\mathrm{d}^2x\sqrt{-\tilde{g}}l^2 \cos(\frac{\r_{1}+\delta\r_{1}}{l})^2[2\frac{\tan(\frac{-\r_{1}-\delta\r_{1}}{l})}{l}-\frac{\tan(\frac{-\r_{1}}{l})}{l}]\\
&-\frac{1}{8\pi G_{3}}\int\mathrm{d}^2x\sqrt{-\tilde{g}}l^2 \cos(\frac{\r_{2}+\delta\r_{2}}{l})^2[2\frac{\tan(\frac{\r_{2}+\delta\r_{2}}{l})}{l}-\frac{\tan(\frac{\r_{2}}{l})}{l}]\,,
\end{aligned}
\end{equation}
where $\tilde{g}$ denotes the deformed $g_{ij}$ similar to the previous case~\eqref{gtilde}.
To the quadratic order, we get
\begin{equation}
\begin{aligned}\label{6.12}
I_{\rm reduced}=&-\frac{\r_{2}-\r_{1}}{16\pi G_{3}}\int\mathrm{d}^2x\sqrt{-g}R(g)-\frac{\delta\r_{2}-\delta\r_{1}}{16\pi G_{3}}\int\mathrm{d}^2x\sqrt{g}(R(g)+2)\\
&-\frac{1}{8\pi G_{3}}\int\mathrm{d}^2x\sqrt{-g}[\frac{\tan(\frac{\r_{1}}{l})}{2l}\nabla_{k}\delta\r_{1}\nabla^{k}\delta\r_{1}+\frac{\tan(\frac{\r_{1}}{l})}{l}\delta\r_{1}^2\\
&-\frac{\tan(\frac{\r_{2}}{l})}{2l}\nabla_{k}\delta\r_{2}\nabla^{k}\delta\r_{2}-\frac{\tan(\frac{\r_{2}}{l})}{l}\delta\r_{2}^2]\ .
\end{aligned}
\end{equation}
Setting $\Phi(x)=\delta\phi_{2}-\delta\phi_{1}$ and $\Psi(x)=\frac{\mathcal{T}_{1}}{\sqrt{\mathcal{T}_{1}+\mathcal{T}_{2}}}\delta\phi_{2}+\frac{\mathcal{T}_{2}}{\sqrt{\mathcal{T}_{1}+\mathcal{T}_{2}}}\delta\phi_{2}$ and $g_{ij}\rightarrow g_{ij}e^{-\frac{\mathcal{T}_{1}\mathcal{T}_{2}}{\mathcal{T}_{1}+\mathcal{T}_{2}}}$, ($\ref{6.12}$) becomes
 \begin{equation}
    \begin{aligned}\label{reduced effective action 2}
        I_{\rm reduced}&=-\frac{(\r_{2}-\r_{1})}{16\pi G_{3}}\int \mathrm{d}^2x\sqrt{-g}R(g)-\frac{1}{16\pi G_{3}}\int\mathrm{d}^2x\sqrt{-g}\Phi(x)\left[ R(g)+2\right]\\
&-\frac{1}{8\pi G_{3}}\int\mathrm{d}^2x\sqrt{-g}\left[\frac{1}{2}\nabla_{\mu}\Psi\nabla^{\mu}\Psi+\Psi^2\right].
    \end{aligned}
\end{equation}

\noindent \textbf{Class III$_{\rm b}$}
In this case, we apply the following transformation from $(t,r,\psi)$ to $(T, \rho,\phi)$ 
\begin{equation}
\begin{aligned}\label{KRS 3}
t&=\frac{l T}{\sqrt{1-\mathcal{A}^2 l^2 m^2 }},\\
r&=-\frac{G_{3}(\rho)}{\mathcal{A}\sinh(m\psi+\psi_{0})},\\
\sinh(m\psi+\psi_{0})&=-\frac{\mathfrak{g}_{s}(\rho)\sqrt{1-\mathcal{A}^2 l^2 m^2}}{\sqrt{\mathcal{A}^2 l^2 m^2\cosh(\phi)^2+\mathfrak{g}_{s}(\rho)^2(1-\mathcal{A}^2 l^2 m^2)\sinh(\phi)^2}},\\
\mathfrak{g}_{s}(\rho)&=\frac{-\mathcal{A}^2 l^2 m^2+\mathcal{A}^4 l^4 m^4+\sqrt{\mathcal{A}^2 l^2 m^2(1-\mathcal{A}^2 l^2 m^2)}\mathrm{sech}(\rho/l)\mathrm{tanh}(\rho/l)}{(1-\mathcal{A}^2 l^2 m^2)(1-\mathcal{A}^2 l^2 m^2-\mathrm{tanh}(\rho/l)^2)}\,,
\end{aligned}
\end{equation}
 where  
 \bal
\mathfrak{g}_{s}(\r) \in\left(-\frac{\mathcal{A}ml}{\sqrt{1-\mathcal{A}^2m^2l^2}},\frac{\mathcal{A}ml}{\sqrt{1-\mathcal{A}^2m^2l^2}}\right)\ .
 \eal
The tension of the strut/wall at the position $\r_{1}$ and $\r_{2}$ ($0<\r_{1}<\r_{2}$) are respectively
\begin{equation}
\mathcal{T}_{1}=\frac{1}{2}K_{1}=\frac{\tanh(\frac{-\r_{1}}{l})}{l},\quad \mathcal{T}_{2}=\frac{1}{2}K_{2}=\frac{\tanh(\frac{\r_{2}}{l})}{l}\ .
\end{equation}
Since the line element~\eqref{2.9} is of the form
\begin{equation}\label{2.57}
\mathrm{d}s^2=\mathrm{d}\rho^2+h_{ij}(\rho,x)\mathrm{d}x^{i}\mathrm{d}x^{j}=\mathrm{d}\rho^2+l^2\cosh(\rho/l)^2g_{ij}(x)\mathrm{d}x^{i}\mathrm{d}x^{j},,
\end{equation}
we can again rewrite the bulk Ricci scalar as
\begin{equation}
R(G_{\rm{bulk}})=\frac{1}{l^2\cosh(\frac{\r}{l})^2}R(g)-\frac{4}{l^2}-\frac{2\tanh(\frac{\r}{l})^2}{l^2}\ .
\end{equation}
Introducing the fluctuations as in the previous cases, integrating out $\rho$ direction gives
\begin{equation}
\begin{split}\label{4.17}
I_{\rm reduced}=&-\frac{1}{16\pi G_{3}}\int^{\r_{2}+\delta \r_{2}}_{\r_{1}+\delta \phi_{1}}\mathrm{d}\r\int\mathrm{d}^2x\sqrt{-g}[R(g)-4\cosh(\frac{\r}{l})^2-2\sinh(\frac{\r}{l})^2+2\cosh(\frac{\r}{l})^2]\\
&-\frac{1}{8\pi G_{3}}\int\mathrm{d}^2x\sqrt{-\tilde{g}}l^2\cosh(\frac{\r_{1}+\delta \r_{1}}{l})^2(\frac{2\tanh(\frac{-\r_{1}-\delta \r_{1}}{l})}{l}-\frac{\tanh(\frac{-\r_{1}}{l})}{l})\\
&-\frac{1}{8\pi G_{3}}\int\mathrm{d}^2x\sqrt{-\tilde{g}}l^2\cosh(\frac{\r_{2}+\delta \r_{2}}{l})^2(\frac{2\tanh(\frac{\r_{2}+\delta \r_{2}}{l})}{l}-\frac{\tanh(\frac{\r_{2}}{l})}{l})\ .
\end{split}
\end{equation}
Expanding to the quadratic order of the perturbation leads to the the effective action 
\begin{equation}
    \begin{aligned}\label{reduced effective action 3}
        I_{\rm reduced}&=-\frac{(\r_{2}-\r_{1})}{16\pi G_{3}}\int \mathrm{d}^2x\sqrt{-g}R(g)-\frac{1}{16\pi G_{3}}\int\mathrm{d}^2x\sqrt{-g}\Phi(x)\left[ R(g)+2\right]\\
&-\frac{1}{8\pi G_{3}}\int\mathrm{d}^2x\sqrt{-g}\left[\frac{1}{2}\nabla_{\mu}\Psi\nabla^{\mu}\Psi+\Psi^2\right]\ .
    \end{aligned}
\end{equation}


\subsubsection{String at $x=\rm{constant}$ }
 In Sec.~\ref{cmetric}, we mainly focus on cases where the strings locate at $x=\text{constant}$, so it is also interesting to consider fluctuations around these type of strings. 
 In this section, we will discuss such fluctuations. It is clear from construction that such reduction can be carried out for each of the strings, so we restrain to the case of a single string at a constant $\psi =\psi_1$ for simplicity. \\
 
\begin{figure}[]
\centering
\includegraphics[width=0.4\linewidth]{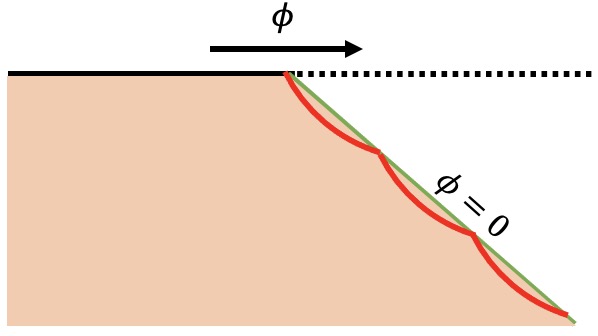}
\caption{
The green line is the  string at $\phi=0$. The red curve denotes the fluctuations on the string. The black line is the asymptotic CFT. The light orange region is the wedge that is cut from the bulk.}
\label{Class I(reduced)}
\end{figure}

\noindent $\bullet$ \textbf{Class I:}
When introducing fluctuations to a string at a fixed $x$ or $\psi$, we need to integrate out $\psi$. It is therefore convenient to go to a coordinate system where $r$ and $\psi$ are decoupled. To achieve this, we use the following transformation to go from the $(t,r,\psi)$ coordinate to the $(t,\rho,\phi)$ coordinate near the wall~\footnote{The computation near the strut is similar.} at $\psi=\psi_{1}$
$$\psi-\psi_{1}=\sum^{\infty}_{i=1}F_{i}(\rho)\phi^{i}, \quad r=\sum^{\infty}_{i=0}G_{i}(\rho)\phi^{i}\,,$$
so that in this coordinate, the line element in ($\ref{2.5}$) becomes
\begin{equation}
\begin{aligned}\label{2.20}
\mathrm{d}s^2=&\mathrm{d}\phi^2+\left(\cosh(\phi/l)+\mathcal{A}lm\sin(\psi_{0})\sinh(\phi/l)\right)^2 g_{ij}\mathrm{d}x^{i}\mathrm{d}x^{j}\,, \\
g_{ij}=& H(\r)\left[\left(-\frac{(l^2m^2+(1-\mathcal{A}^2l^2m^2)G_{0}(\r)^2)^2}{l^4 G_{0}'(\r)^2}\right)\mathrm{d}t^2+\mathrm{d}\r^2 \right]\,,\\
H(\r)=&\frac{G_{0}'^2}{(\mathcal{A}\cos(\psi_{0})G_{0}(\r)+1)^2(m^2+(1/l^2-\mathcal{A}^2m^2)G_{0}(\r)^2)}\ .
\end{aligned}
\end{equation}
The 
coefficients $F_{i}(\rho)$ and $G_{i}(\rho)$ can be solved order by order as functions of the undetermined function $G_{0}(\rho)$, which redefines the radial direction. The simplest choice, which is also the one we adopt in our computation, is $G_{0}(\rho)=\rho$. It is clear that this change of coordinates is valid near $\psi=\psi_1$ or $\phi=0$. 

integrate out $\phi$ from 0 to $\delta\phi$ and get the following induced action that is upto the quadratic order of $\delta \phi$
\begin{equation}
\begin{aligned}\label{4.4}
I_{\rm reduced}=&-\frac{1}{16\pi G_{3}}\int\mathrm{d}^2x\sqrt{-g}\delta \phi\left[R(g)+\frac{2-2\mathcal{A}^2m^2l^2\sin(\psi_{0}+m\psi_{1})^2}{l^2}\right]\\
&-\frac{1}{8\pi G_{3}}\int\mathrm{d}^2x\sqrt{-g}\left[\frac{\mathcal{A}m\sin(\psi_{0}+m\psi_{1})}{2}\nabla_{\mu}\delta \phi\nabla^{\mu}\delta \phi+\mathcal{A}m\sin(\psi_{0}+m\psi_{1})(\delta\phi)^2\right]\\
&+\frac{\mathcal{A}m\sin(\psi_{0}+m\psi_{1})}{8\pi G_{3}}\int \mathrm{d}x^2\sqrt{g}\ .
\end{aligned}
\end{equation}
This is a dilaton gravity with $\delta \phi$ playing the role of the dilaton and an effective cosmological constant  $$\Lambda = \frac{\mathcal{A}^2m^2l^2\sin(\psi_{0}+m\psi_{1})^2-1}{l^2},$$ which is consistent with ($\ref{induced metric}$) at $\psi=\psi_{1}$. 
The reduction process is illustrated in Fig.~\ref{Class I(reduced)}.\\

\noindent $\bullet$ \textbf{Class II:}
Similar to the analysis for Class I, the line element~($\ref{2.7}$) of Class II solution with a wall at $\psi=\psi_1$ has the form
\begin{equation}\label{2.39}
\mathrm{d}s^2=d \phi^2+(\cosh(\phi/l)-\mathcal{A}ml\sinh(\psi_{0}+m\psi_{1})\sinh(\phi/l))^2g_{ij}\mathrm{d}x^{2}\mathrm{d}x^{j}\ .
\end{equation}
The effective action up to quadratic order in $\delta \phi$ is
\begin{equation}
\begin{aligned}\label{4.9}
I_{\rm reduced}=&-\frac{1}{16\pi G_{3}}\int\mathrm{d}^2x\sqrt{-g}\delta \phi\left[R(g)+\frac{2-2\mathcal{A}^2m^2l^2\sinh(\psi_{0}+m\psi_{1})^2}{l^2}\right]\\
&-\frac{1}{8\pi G_{3}}\int\mathrm{d}^2x\sqrt{-g}\left[\frac{\mathcal{A}m\sinh(\psi_{0}+m\psi_{1})}{2}\nabla_{\mu}\delta \phi\nabla^{\mu}\delta \phi+\mathcal{A}m\sinh(\psi_{0}+m\psi_{1})(\delta\phi)^2\right]\\
&+\frac{\mathcal{A}m\sinh(\psi_{0}+m\psi_{1})}{8\pi G_{3}}\int \mathrm{d}x^2\sqrt{g}\,,
\end{aligned}
\end{equation}
with an effective cosmological constant $$\Lambda = -\frac{1-\mathcal{A}^2m^2l^2\sinh(\psi_{0}+m\psi_{1})^2}{l^2}\,,$$ 
which is again consistent with ($\ref{induced metric2}$) at $\psi=\psi_{1}$. \\

\noindent $\bullet$ \textbf{Class III:}
A similar transformation as in the previous cases puts the line element ($\ref{2.9}$) of a Class III solution with a wall at $\psi=\psi_1$ to the following schematic form
\begin{equation}
\mathrm{d}s^2=\mathrm{d}\phi^2+(\cosh(\phi/l)-\mathcal{A}ml\cosh(\psi_{0}+m\psi_{1})\sinh(\phi/l))^2g_{ij}^{\rm AdS}(\rho)\mathrm{d}x^{i}\mathrm{d}x^{j}\ .
\end{equation}
The effective action up to quadratic order in $\delta \phi$ is
\begin{equation}
\begin{aligned}\label{4.12}
I_{\rm reduced}=&-\frac{1}{16\pi G_{3}}\int\mathrm{d}^2x\sqrt{-g}\delta \phi\left[R(g)+\frac{2-2\mathcal{A}^2m^2l^2\cosh(\psi_{0}+m\psi_{1})^2}{l^2}\right]\\
&-\frac{1}{8\pi G_{3}}\int\mathrm{d}^2x\sqrt{-g}\left[\frac{\mathcal{A}m\cosh(\psi_{0}+m\psi_{1})}{2}\nabla_{\mu}\delta \phi\nabla^{\mu}\delta \phi+\mathcal{A}m\cosh(\psi_{0}+m\psi_{1})(\delta\phi)^2\right]\\
&+\frac{\mathcal{A}m\cosh(\psi_{0}+m\psi_{1})}{8\pi G_{3}}\int \mathrm{d}x^2\sqrt{g}\,,
\end{aligned}
\end{equation}
with an effective cosmological constant is $$\Lambda =  -\frac{(1-\mathcal{A}^2m^2l^2\cosh(\psi_{0}+m\psi_{1})^2)}{l^2},$$ which is consistent with ($\ref{induced metric3}$) at $\psi=\psi_{1}$. 

In this subsection, we have decoupled $r$ and $\psi$ in a foliation and reduced the gravity to an AdS$_{2}$ slice at a constant $\phi$ by integrating out $\phi$ from $\phi=0$ (wall) to $\phi=\delta\phi$ (fluctuated string) and introduce a dilaton field. This process results in a dilaton gravity for all three classes of solutions of C-metric. Furthermore, the effective cosmological constant and tension of the string are consistent with the results in Sec. 2.  


\section{A large acceleration limit}

We want to consider the reduction to the string at $x=\rm constant$, and can clearly see how this string approaches the asymptotic boundary within $l\ll 1$ at the same time. Thus a good ideal is to set the string at $x=0$. By using the recipe in Sec.~$3.2$, after integrating out the ultraviolet CFT DOF down to the cutoff energy $A$, the higher-curvature correction will be introduced in effective action \eqref{3.4} and can be expanded over small $l$\footnote{Here we neglect the auxiliary field $\Phi$ that describes the remaining DOF caused by weyl anomaly}:
\begin{equation}
I_{\rm div}=\frac{l}{16\pi G_{3}}\int\mathrm{d}^{2}x\sqrt{-h}
\left[ \frac{2}{l^2}+\frac{1}{2}R(h)\mathrm{log}\left(-\frac{l^2R(h)}{2}\right)-\frac{1}{2}R(h)+\frac{l^2}{16}R(h)^2+\mathrm{O}(l^3)\right].
\end{equation} 
In this framework, we can introduce higher-curvature corrections to string at $x=0$ as an expansion over small $l$. Therefore, the Ricci scalar becomes $R(h)=-\frac{2}{L_{2}^2}$, while the central charge obtained in Sec.~3 is $c=\frac{3l}{2G_{3}}$. Note that there is no contradiction between the requirement of $c\gg 1$ and the small $l$ expansion, since \cite{emparan2020quantum}: $$c\sim \frac{l}{\hbar}\gg 1, \quad \frac{l}{L_{2}}\ll 1.$$
Next, we will incorporate higher-curvature corrections to describe the high-energy DOF of CFT on the string, while introducing some boundary terms and conditions to ensure the convergence of the effective action.

\begin{figure}[]
\centering
\subfigure[]{
\includegraphics[width=0.35\linewidth]{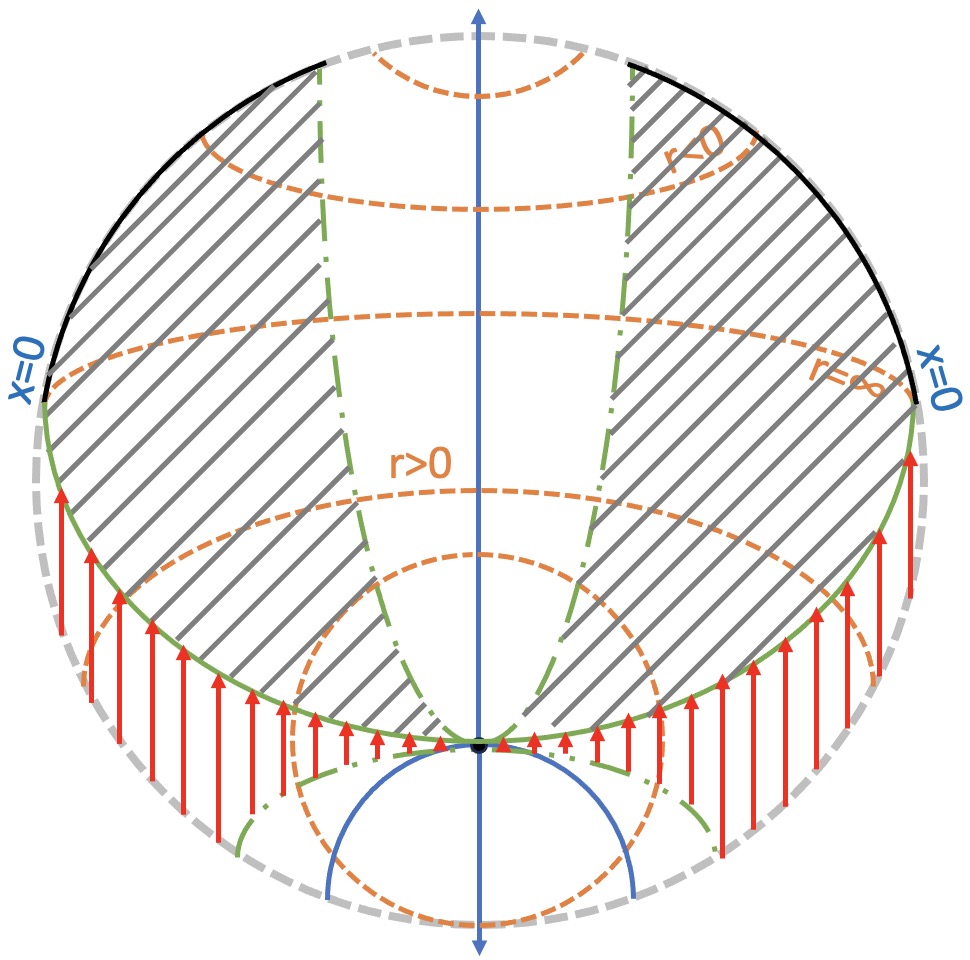}}
\subfigure[]{
\includegraphics[width=0.6\linewidth]{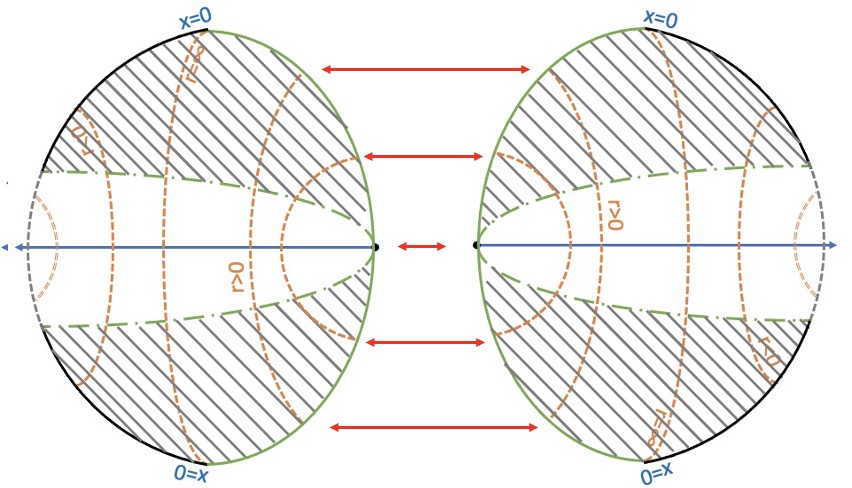}}
\caption{ \textbf{(a)}: The copied AdS$_{3}$ spatial slice with one on the two  strings approaching the conformal boundary. \textbf{(b)}: Gluing the copied geometry along the string near the conformal boundary. The dark curves are conformal boundaries ($x=y$), and the green dash-dotted lines are  strings ($x=x_{1(2)}$). The green line is the string that we make a reduction ($x=0$). The blue lines denote the position at different $x$. The orange dotted curves denote different constant-$r$ lines. The red-shaded region is the portion ($x\leq0$) discarded from the geometry.}
\label{double copies}
\end{figure}

Our analysis begins with the total effective action on the string at $x=0$ of the class $\rm III_b$ solutions. As the acceleration $A$ approaches $\infty$, the tension on the string diverges. Since the Lorentzian signature requires $1>Al$, $l$ must approach zero in this limit, leading to $\frac{x}{y} \to 1$, which is the position of the conformal boundary. In this limit, the two kinds of reduction can be merged, and $I_{\rm div}$ can be introduced as curvature correction on the string. Given ($\ref{3.4}$), the total induced action can be defined as
\begin{equation}\label{total action}
I_{\rm eff}=I_{\rm CFT}+2(I_{\rm div}+I_{\rm ct})+I_{\rm 2D}.
\end{equation}
Here, I$_{\rm CFT}$ is the non-gravitational part and is defined holographically. The factor $2$ arises from the two copies -- Fig.~\ref{double copies}.
The $I_{\rm ct}$ represents the counter-term introduced to cancel out the divergence of the stress tensor near the conformal boundary \cite{balasubramanian1999stress}\footnote{In our paper, the gravitational coupling constant $\kappa$ is negative, leading to a difference in the convention of symbols.}, 
and take the form as
\begin{equation}
I_{\rm ct}=\frac{1}{8\pi G_{3}}\int \mathrm{d}x^2\sqrt{\gamma}\frac{1}{l}=\frac{1}{16\pi G_{2}}\int \mathrm{d}x^2\sqrt{\gamma}\frac{2}{l^2}.
\end{equation}
The $I_{\rm 2D}$ is the action of the two-dimensional gravity with a fluctuation on the wall but without curvature corrections. $I_{\rm 2D}$ consists of two parts:
\begin{equation}
\begin{aligned}\label{4.29}
I_{\rm 2D}=&2I_{\rm reduced}+I_{\rm string}\\
=&-\frac{2}{16\pi G_{3}}\int\mathrm{d}^2x\sqrt{-g}\delta \phi\left[R(g)+\frac{2}{L_{2}^2}\right]\\
&-\frac{2}{16\pi G_{3}}\int\mathrm{d}^2x\sqrt{-g}\left[\mathcal{A}m\nabla_{\mu}\delta \phi\nabla^{\mu}\delta \phi+2\mathcal{A}m(\delta\phi)^2\right]\\
&+\frac{\mathcal{A}m}{4\pi G_{3}}\int \mathrm{d}x^2\sqrt{g}+I_{\rm string},
\end{aligned}
\end{equation}
where the dilaton field $\delta\phi$ in the second and the third lines describe the gravitational DOF since it is coupled with the Ricci curvature. The first term of the last line describes the matter field. With the limit of $\mathcal{A}\rightarrow\infty$, the matter field is divergent. The last term describes the tension on the string as\cite{chen2020quantum2}
\begin{equation}
\begin{aligned}
I_{\rm string}&=-T_{0}\int \mathrm{d}^2x\sqrt{h}\\
T_{0}&=\frac{K}{8\pi G_{3}}=\frac{\mathcal{A}m}{4\pi G_{3}},
\end{aligned}
\end{equation}
which eliminates the former divergence. As a consequence, the dilaton gravitational field is described by the remaining terms as
\begin{equation}
\begin{aligned}\label{4.31}
I_{\rm 2D}=&-\frac{2}{16\pi G_{3}}\int\mathrm{d}^2x\sqrt{-g}\delta \phi\left[ R(g)+\frac{2}{L_{2}^2}\right]\\
&-\frac{2}{16\pi G_{3}}\int\mathrm{d}^2x\sqrt{-g}\left[\mathcal{A}m\nabla_{\mu}\delta \phi\nabla^{\mu}\delta \phi+2\mathcal{A}m(\delta\phi)^2\right].
\end{aligned}
\end{equation}
For the convenience of discussion, this action can be formally rewritten as
\begin{equation}
\begin{aligned}\label{4.40}
I_{\rm 2D}&=k\int\mathrm{d}^2x\sqrt{-g}\left[XR(g)+2X\Lambda+2T(\nabla X)^2+4TX^2\right],\\
k&=-\frac{1}{8\pi G_{3}},\\
X&=\delta\phi ,\\
\Lambda&=\frac{1-\mathcal{A}^2m^2l^2}{l^2} ,\\
T&=\mathcal{A}m.
\end{aligned}
\end{equation}

Different from the cases in \cite{geng2022aspects}, the kinetic term and potential term can not be redefined simultaneously, due to $\Lambda\neq 1$, and the inconsistency of the geometry of the string. To see it clearly, we consider the variation of ($\ref{4.40}$). On the one hand, the variation over $X$ gives
\begin{equation}
\begin{aligned}\label{(1)}
    R+2\Lambda-4T\Box X+8TX&=0,\\
    \Box X=2TX\quad (R+2\Lambda&=0),
\end{aligned}
\end{equation}
where $\Box$ denotes $\nabla_{\mu}\nabla^{\mu}$. On the other hand, the variation of ($\ref{4.40}$) over $g^{ij}$ gives
\begin{equation}
\begin{aligned}\label{4.42}
    \Box Xg_{ij}-\nabla_{i}\nabla_{j}X+X(R_{ij}-\frac{1}{2}(R+2\Lambda)g_{ij})+T\left[2\nabla_{i}X\nabla_{j}X-(\nabla X)^2g_{ij}+2X^2g_{ij}\right]&=0\\
    \Box Xg_{ij}-\nabla_{i}\nabla_{j}X-\Lambda g_{ij}X+T\left[2\nabla_{i}X\nabla_{j}X-(\nabla X)^2g_{ij}+2X^2g_{ij}\right]&=0,
\end{aligned}
\end{equation}
where we have used $R+2\Lambda=0$ and $R_{ij}+\Lambda g_{ij}=0$. 
Taking the trace of ($\ref{4.42}$), we have
\begin{equation}\label{(2)}
    \Box X-2\Lambda X+4TX^2=0.
\end{equation}
Combining ($\ref{(1)}$) and ($\ref{(2)}$), we have
\begin{equation}
    2TX-2\Lambda X+4TX^2=0.
\end{equation}
The solution of the dilaton field $X$ can only be a constant, and the effective theory on the string is still topological, unless the dilaton gravity is weakly coupled with a matter field. At the same time, in the limit of large acceleration, the tension coefficient in kinetic term of dilaton field is divergent, we need to suppress this divergence. If $X=\delta\phi$ is of order $\mathcal{O}(l^2)$, we can neglect the quadratic term because it is of order $\mathcal{O}(l^3)$. For simplification, we set $$\delta\phi=l^2\Phi,\quad \Phi\sim 1,$$ and then the first-order fluctuating term is:
\begin{equation}
I_{\rm 2D}=-\frac{2l}{16\pi G_{2}}\int\mathrm{d}^2x\sqrt{-h}\Phi\left[R(h)+\frac{2}{L_{2}^2}\right],
\end{equation}
which is just a JT gravity. The Ricci scalar $R(h)=-\frac{2}{L_{2}^2}$ is finite, since $$\frac{1}{l^2}=\frac{1}{L_{2}^2}+\mathcal{A}^2m^2\sim \mathcal{A}^2m^2\gg 1, \quad L_{2}\sim 1.$$ Hence, the higher-curvature term of the on-shell action is convergent, and the total on-shell action is convergent except $R(h)\mathrm{log}(l^2)$. But since this term is just topological, the divergence will not affect the dynamics. Finally, the total effective action ($\ref{total action}$) can be obtained as
\begin{equation}
\begin{aligned}\label{5.5}
I_{\rm eff}=&I_{\rm CFT}-\frac{l}{8\pi G_{2}}\int\mathrm{d}^2x\sqrt{-h}\Phi\left[R(h)+\frac{2}{L_{2}^2}\right]\\
&-\frac{1}{16\pi G_{2}}\int\mathrm{d}^{2}x\sqrt{h}\left[ R(h)\mathrm{log}\left(-\frac{l^2R(h)}{2}\right)-R(h)+\frac{l^2}{8}R(h)^2+\mathrm{O}(l^3)+\tilde{F}(\nabla \tilde{\Phi})\right],\\
=&I_{\rm CFT}+I_{\rm{JT}}+I_{\rm c}+I_{\rm af},
\end{aligned}
\end{equation}
in which each part is defined respectively as
\begin{equation}
    \begin{aligned}\label{effective action}
        I_{\rm{JT}}&:=-\frac{l}{8\pi G_{2}}\int\mathrm{d}^2x\sqrt{-h}\Phi\left[ R(h)+\frac{2}{L_{2}^2}\right],\\
        I_{\rm c}&:=-\frac{1}{16\pi G_{2}}\int\mathrm{d}^{2}x\sqrt{h}\left[ R(h)\mathrm{log}\left(-\frac{l^2R(h)}{2}\right)-R(h)+\frac{l^2}{8}R(h)^2+\mathrm{O}(l^3)\right],\\
        I_{\rm af}&:=-\frac{1}{16\pi G_{2}}\int\mathrm{d}^{2}x\sqrt{h}\tilde{F}(\nabla \tilde{\Phi}).
    \end{aligned}
\end{equation}
In higher-curvature correction terms, we do not couple the dilaton field to the action, because the variation of the action over the dilaton field $\delta\phi$ usually gives $R=-\frac{2}{L_{2}^2}$. Lack of information of matter field on the string, in principle, we can't solve out the dilaton field. We consider a simplest situation that JT gravity is not coupled with a matter field. After the analysis, we can find the dynamics of the string by taking a variation as
\begin{equation}
\begin{aligned}
&R=-\frac{2}{L_{2}^2},\\
&R_{\mu\nu}=-\frac{2}{L_{2}^2}g_{\mu\nu},\\
&\nabla_{\mu}\nabla_{\nu}\Phi=\frac{1}{L_{2}^2}g_{\mu\nu}\Phi.
\end{aligned}
\end{equation}
The solution of the last equation is
\begin{equation}
\Phi(\tilde{r},T)=\frac{A_{1}\tilde{r}+A_{2}\sqrt{\tilde{r}^2-L_{2}^2m^2}\mathrm{cosh}(mT)+A_{3}\sqrt{\tilde{r}^2-L_{2}^2m^2}\mathrm{sinh}(mT)}{L_{2}m},
\end{equation}
where $A_{1}$, $A_{2}$ and $A_{3}$ are undetermined constants. After reducing the bulk gravity on the string with higher curvature corrections, a matter field is introduced to ensure that the solution of JT gravity satisfies the effective action.

Different from the 4D cases, a logarithmic term $R(h)\mathrm{log}\left(-\frac{l^2R(h)}{2}\right)$ caused by Weyl anomaly, exists in the curvature correction terms. The contribution of this term can be considered by the non-local Polyakov action as \cite{skenderis2000quantum}
\begin{equation}\label{4.41}
I=\frac{\chi}{8\pi G_{3}}\int \mathrm{d}x^2\sqrt{-h}\left[-\frac{1}{2}(\nabla \tilde{\phi})^2+\tilde{\phi} R(h)+\lambda e^{-\tilde{\phi}}\right],
\end{equation}
where $\chi$ and $\lambda$ are constants. Assuming a constant Ricci scalar, the simple solution of the scalar field is $\tilde{\phi}_{0}=\log (\lambda/R)$. Then the Polyakov action is reduced to
\begin{equation}
I_{\rm Poly}|_{\tilde{\phi}=\tilde{\phi}_{0}}=-\frac{\chi}{8\pi G_{3}}\int \mathrm{d}x^2\sqrt{-h}\left[R(h)\log(R(h)/\lambda)-R(h)\right].
\end{equation}
By setting $\chi=\frac{1}{4}$ and  $\lambda=-\frac{2}{l^2}$, the action $I_{\rm Poly}|_{\tilde{\phi}=\tilde{\phi}_{0}}$ will correspond to the logarithmic term. Evaluate the stress tensor at $\tilde{\phi}=\tilde{\phi}_{0}$, and we have
\begin{equation}
T_{(\text{Poly}) ij}|_{\tilde{\phi}=\tilde{\phi}_{0}}=\frac{l}{32\pi G_{3}}h_{ij}R(h).
\end{equation}
Taking the trace will recover the anomaly as $$T_{(\text{Poly})}(h) =\frac{c}{24\pi}R(h).$$ 
In general, the curvature correction terms can be divided into two parts:
\begin{equation}
I_{c}=I_{\rm Poly}+I_{\rm hc},
\end{equation}
where $I_{\rm hc}$ corresponds to higher-curvature corrections of orders larger than $1$.

Moreover, introducing the higher-curvature term enables the discussion of the Wald-Dong entropy. Formally, Wald-Dong entropy is defined by the area term on the co-dimension $2$ surface. In 3D, the area term has zero contribution. Therefore, the leading contribution comes from the logarithmic term in the action ($\ref{5.5}$):
\begin{equation}
    S_{\rm WD}\sim\frac{l}{4G_{3}}\log\left(-\frac{l^2R(h)}{2}\right),
\end{equation}
which can be derived from the Polyakov action. 

In addition to the perturbative contribution of the Polyakov term, it is also necessary to construct the stress tensor corresponding to other higher-curvature terms defined by the induced metric. The stress tensor is obtained as
\begin{equation}
T_{\rm (hc)\mu\nu}=-\frac{2}{\sqrt{-h}}\frac{\delta I_{\rm hc}}{\delta h^{ij}}.
\end{equation}
For the cases of the $f(R)$ action, taking the variation with respect to the induced metric gives
\begin{equation}
\begin{aligned}
\delta \sqrt{-h}f(R(h))&=\sqrt{-h}(f'(R)R_{\mu\nu}-\frac{1}{2}f(R)g_{\mu\nu}-\nabla_{\mu}\nabla_{\nu}f'(R)+g_{\mu\nu}\nabla_{\mu}\nabla^{\mu}f'(R))\delta h^{\mu\nu},\\
&=-\sqrt{-h}\left[2\frac{f'\left(-\frac{2}{L_{2}^2}\right)}{L_{2}^2}+\frac{1}{2}f\left(-\frac{2}{L_{2}^2}\right)\right]h_{\mu\nu}\delta h^{\mu\nu}.
\end{aligned}
\end{equation}
In above calculation, the covariant derivative of $f(R)$ vanishes for Ricci scalar is a constant.
Then we have the order expansion over $l$ as
\begin{equation}
\begin{aligned}\label{4.37}
T_{\mu\nu}&=T_{(2)\mu\nu}l^2+...\\
T_{(2)\mu\nu}&=-\frac{3}{4L_{2}^{4}}h_{\mu\nu},\\
&...
\end{aligned}
\end{equation}
By introducing this kind of matter field, we have aligned the induced metric with the solution of the effective theory on the string. 
In our case, the classical Hawking entropy defined at $\tilde{r}_{\rm h}=L_{2}m$ is
\begin{equation}\label{4.37}
S_{\rm 2D}=2\frac{\delta\phi(\tilde{r}_{\rm h})}{4G_{3}}=\frac{A_{1}l }{2 G_{2}}.
\end{equation}
With the temperature $T_{\rm h}=\frac{m}{2\pi L_{2}}$, we have
\begin{equation}\label{4.52}
T_{\rm h}S_{\rm 2D}=\frac{A_{1}ml}{4\pi G_{2}L_{2}}.
\end{equation}
Here the entropy is doubled due to the two copies of string at $x=0$.

In the same way, we can consider the modified model on the string at $x=0$ in Class I$_{\rm p3}$. Recalling the transformation \eqref{KRS 1}, conditions such as $l\ll 1$, $\mathcal{A}\gg1$ and $\mathcal{A}^2m^2l^2<1$ lead to that the string (constant-$x$) approaching the conformal boundary. The effective cosmological constant is:$$\Lambda=\frac{1-\mathcal{A}^2m^2l^2}{l^2}=\frac{1}{L_{2}^2},$$ and the total action is
\begin{equation}
\begin{aligned}\label{Effective action of I3p}
I_{\rm eff}=&I_{\rm CFT}+2(I_{\rm div}+I_{\rm ct})+I_{\rm 2D}\\
=&I_{\rm CFT}-\frac{l}{8\pi G_{2}}\int\mathrm{d}^2x\sqrt{-h}\Phi\left[R(h)+\frac{2}{L_{2}^2}\right]\\
&-\frac{1}{16\pi G_{2}}\int\mathrm{d}^{2}x\sqrt{-h}\left[R(h)\mathrm{log}\left(-\frac{l^2R(h)}{2}\right)-R(h)+\frac{l^2}{8}R(h)^2+\mathrm{O}(l^3)+\tilde{F}(\nabla \tilde{\Phi})\right].
\end{aligned}
\end{equation}
In parallel, we can construct the stress tensor in a manner similar to the cases of Class $\rm III_b$ solutions; hence, we will not restate the process here. The final solution for the dilaton field is obtained as follows
\begin{equation}
\Phi(\tilde{r},T)=\frac{A_{1}\tilde{r}+A_{2}\sqrt{\tilde{r}^2+L_{2}^2m^2}\mathrm{cos}(mT)+A_{3}\sqrt{\tilde{r}^2+L_{2}^2m^2}\mathrm{sin}(mT)}{L_{2}m},
\end{equation}
where $A_{1}$, $A_{2}$ and $A_{3}$ are undetermined constants. From ($\ref{4.37}$), the stress tensor is influenced by higher curvature corrections and is not traceless. This phenomenon can be explained by the breaking of the conformal symmetry, after the introduction of a cutoff in the effective theory.
\subsection{The Expension of Generalized Entropy}
In the previous section, the introduction of higher-curvature causes the Weyl anomaly $$R(h)\log\left(-\frac{l^2R(h)}{2}\right),$$ which dominates in the expansion of small $l$. However, in semi-classical approximation, the leading term of action effects directly on the string and dominates the geometry of string, that is to say, the leading term should be JT term. To explain this issue, we can start with the 3D generalized entropy. According to \eqref{Smarr relation of Class III}, the generalized entropy is made up by two parts:
\begin{equation}\label{generalized entropy}
    S_{\rm BH}=\frac{2M_{\rm III_b}}{T_{\rm h}}+S_{\rm  boundary}.
\end{equation}
Then we expand these two parts over small $l\sim\nu$ respectively as follows:
 \begin{equation}
\begin{aligned}\label{2D entropy}
    \frac{2M_{\rm III_b}}{T_{\rm h}}&=\L.\frac{l}{2G_{3}}\mathrm{arctanh}\(\frac{\ x}{\sqrt{1-A^2l^2}\sqrt{1+x^2}}\)\R|^{x=x_{\rm b}}_{x=0}\\
    &=\frac{\sqrt{\frac{\ 1+x_{\rm b}^2}{x_{\rm b}^2}}}{2G_{2}}\nu+\frac{\(\frac{\ 1+x_{\rm b}^2}{x_{\rm b}^2}\)^{3/2}\nu^{3}}{6G_{2}}+\mathcal{O}(\nu)^4+...\; ,\\
    S_{\rm boundary}&=\L.\frac{l}{2G_{3}}\mathrm{arctanh}\(Al\sqrt{x^2+1}\)\R|^{x=x_{\rm b}}_{x=0}\\
 &=\frac{1}{2G_{2}}\mathrm{arctanh}\(\frac{1}{\sqrt{1+x_{\rm b}^2}}\)+\frac{\sqrt{1+x_{\rm b}^2}}{4G_{2}x_{\rm b}^2}\nu^2+\frac{\log(\nu/2)}{2G_{2}}-\frac{\nu^2}{8G_{2}}+\mathcal{O}(\nu)^3+...\; ,
\end{aligned}
\end{equation}
 with $\nu=\frac{l}{L_{2}},\quad L_{2}=\frac{l}{\sqrt{1-\mathcal{A}^2m^2l^2}}\text{ and }\frac{G_{3}}{l}=G_{2}$.
 Obviously, in the second equation, we find that the third term $$\frac{\log(\nu/2)}{2G_{2}}$$ is the leading term that corresponds to the Weyl anomaly and the first term of the first equation $$\frac{1}{2G_{2}}\frac{\sqrt{1+x_{\rm b}^2}}{x_{\rm b}}\nu\sim S_{\rm JT}$$ corresponds to the contribution of the JT gravity. They come from the value of the integral at the lower limit of integration. In summary, the contributions can be attributed as $$S_{\rm BH}=S_{\rm WD}+S_{\rm JT}+\mathcal{O}(\nu^2).$$ Considering the need to exclude the component of boundary entropy from the generalized entropy in thermodynamic relationships, the true classic term that dominates the geometry of string is JT.

 As a comparison, in I$_{\rm p3}$, given the effective action \eqref{Effective action of I3p}, we can also discuss the Wald-Dong entropy in small $l$. Similar to III$_{\rm b}$, the Wald-Dong entropy arising from the Weyl anomaly has the same form as $$R(h)\log\left(-\frac{l^2R(h)}{2}\right)\sim\log(\nu),$$with $\nu=\frac{l}{L_{2}},\quad L_{2}=\frac{l}{\sqrt{1-\mathcal{A}^2m^2l^2}}$. With no generalized entropy in III$_{\rm b}$, the only source that arises the Wald-Dong entropy is the boundary entropy, which is defined as:
 \begin{equation}
 \begin{aligned}
     S_{\rm boundary}&=\frac{2c}{6}\mathrm{arctanh}\(4\pi l\sigma\)|_{x=x_{2}}^{x=0}\\
     &=\frac{l}{2G_{3}}\mathrm{arctanh}(Al\sqrt{1-x^2})|_{x=x_{2}}^{x=0}\\
     &\sim\mathrm{arctanh}(Al)-\mathrm{arctanh}\(Al\sqrt{1-x_{2}^2}\)\\
     &\sim\log(\nu/2)+\mathcal{O}(1)+....
\end{aligned}
 \end{equation}
 As we can see, the Wald-Dong entropy also comes from the boundary entropy.

 On the contrary, in the limit of small acceleration, the higher-curvature correction is vanished, for the  string is no longer narrowly stuck to the conformal boundary. As is shown in Sec~$5.1.1$, the effective actions of all phases are summed up as:
 \begin{equation}
     I_{\rm reduced}=I_{\rm JT}+I_{\rm CFT}.
 \end{equation}
 Furthermore, for accelerating particle, the entropy of JT gravity is zero for its dependence of horizon. Thus the boundary entropy arises from the boundary CFT. 
\section{Conclusion} 
 In this study, we primarily focused on the reduction of the three-dimensional C-metric to the  string. As a holographic setup, we first analyzed the asymptotic symmetry of the 3D C-metric. In the FG framework, we managed to recover the Virasoro algebra through classical central extension and determine the central charge of $\rm AdS_{3}$. Subsequently, we derived a JT gravity model by introducing a fluctuation on the string.  
 otably, we found that the leading contribution of the effective action stems from the Weyl anomaly, unlike in the four-dimensional case. By constructing a proper stress tensor, we get the effective on-shell equation on the string in semi-classical approximation. Moreover, by expanding the 3D generalized entropy of C-metric with respect to small $l$, we prove the validity of the effective action we obtain. However, several questions remain open for future investigation.

\textbf{Future Directions}
\begin{itemize}
    \item \textbf{dS solution:}\\ Following the foundational work by Strominger on dS/CFT \cite{strominger2001ds}, we can also consider to solve the C-metric ansatz in dS background. However, we will face a problem: the conformal boundary will be completely obscured by the event horizon, and the mass will be hard to define. As a promote for our discussion in this paper, we can investigate this case in the future.
    \item \textbf{More symmetries:}\\
        It is interesting to embeding the geometry discussion in the current paper theories with more symmetry, such as supersymmetry and higher-spin symmetry in AdS$_3$. One can compute the asymptotic symmetry and thermodynamic relations in the these theories following previous works such as~\cite{Campoleoni:2010zq,Henneaux:2012ny,Hanaki:2012yf,Peng:2012ae}. 
    \item \textbf{Weyl Factor:}\\ Unlike typical $\rm AdS_{3}$ metrics, the 3D C-metric incorporates an undefined Weyl factor in the FG expansion. For simplicity, we chose a special gauge where the Weyl factor $\omega(\xi)$ is constant. This choice simplifies the calculation of asymptotic algebra and central charge. However, different gauges could shift the level of the zero mode, warranting further investigation into how these factors influence physical definitions near the boundary.
    \item \textbf{Geometric Defect:}\\ As we have discussed above, the accelerating particle and accelerating BTZ are both constructed by gluing two copies of patch that is cut from the entire geometry of solution. Naturally, we have to deal with a defective manifold. We have notice that \cite{balasubramanian2015entwinement} has discussed AdS$_{3}/\mathbb{Z}_{n}$ with conical defects. It can been investigated by lifting it to covering space and symmetrization, and the conical defect results in the shift of spectrum of Virasoro algebra. In the dual CFT, this geometry corresponds to a heavy operator that creates a highly excited state at the boundary \cite{berenstein2023aspects,grabovsky2024heavy}.  It provides us a new perspective to study C-metric. Further studies are necessary to explore these effects more comprehensively.
     \item \textbf{Matter Field:}\\ Out of simplifying our discussion, we don't consider the matter field coupled with JT gravity on the string. In principle, the introduction of fluctuation makes it possible to couple the 2D string gravity with a matter field. As shown in e.g.~\cite{Peng:2021vhs} adding matter fields could make the theory significantly richer, it is thus very intereting to further investigate this question in the future.
    \item \textbf{Volume Complexity:}\\ We notice that in 4D C-metric, the conformal factor functions as a delta function in the integral of the extremal surface with the limit of large acceleration, and the volume complexity recovers the thermodynamic of the quantum BTZ on the string at the leading order\cite{Emparan:2021hyr,Chen:2023tpi}. In parallel, similar operation can be exercised in our 3D version.
    \item \textbf{AdS/BCFT and entanglement entropy:}\\ In AdS/BCFT, g-theorem can recover the boundary entropy. Thus the g-function bear some information of the corresponding boundary state. In the future, we can combine the C-metric with boundary and investigate the RG flow. We can also compute entanglement entropy with defects such as discussed in~\cite{Liu:2023ggg} or the entanglement among different boundaries/defects similar to the discussion in~\cite{Liu:2022pan}.
\end{itemize}
\section*{Acknowledgement}
We thank Cheng Peng, Jia Tian and Gabriel Arenas-Henriquez for helpful discussion. LYX is supported by the NSFC under Grant NO. 12175237 and No.~12405079. SX and LL are supported by
NSFC NO. 12473038.

\appendix
\section{Holographic Renormalization}\label{A}
 In this part, we give the details to get the effective action at the boundary. First of all, we rewrite the line element ($\ref{3.2}$) for AdS$_{d+1}$ by a transformation $z^2=\rho$ as
\begin{equation}
\mathrm{d} s^2=\frac{l^2}{4\rho^2}\mathrm{d} \rho^2 + \frac{l^2}{\rho}g_{ij}(x,\rho)\mathrm{d} x^{i}\mathrm{d} x^{j}\ .\label{A1}
\end{equation}
For spacetime with vanishing Weyl tensor, the Einstein equation in the coordinate~\eqref{A1} reduces to~\cite{skenderis2000quantum}
\begin{equation}
\begin{aligned}\label{A.2}
\rho [2g''-2g'g^{-1}g'+\text{Tr}(g^{-1}g')g']+\text{Ric}(g)-(d-2)g'-\text{Tr}(g^{-1}g')g&=0\\
\nabla_{i}\text{Tr}(g^{-1}g')-\nabla^{j}g'_{ij}&=0\\
\text{Tr}(g^{-1}g'')-\frac{1}{2}\text{Tr}(g^{-1}g'g^{-1}g')&=0\ .
\end{aligned}
\end{equation}
Then we can make a expansion of $g_{ij}(x,\rho)$ over $\rho$ as 
\begin{equation}
g_{ij}(x,\rho)=g_{(0)}+g_{(2)}\rho+...+g_{(d)}\rho^{\frac{d}{2}}+h_{(d)}\rho^{\frac{d}{2}}log(\rho)+...
\end{equation}
Next, we only consider the leading term of the the equation over $\rho$. As $\rho$ approaches 0, from the first line of ($\ref{A.2}$), we get:
\begin{equation}
\begin{aligned}
R_{ij}(g)&=(d-2)g'+\text{Tr}(g^{-1}g')g\\
R(g_{(0)})&=2(d-1)\text{Tr}(g_{(2)})\\
\text{Tr}(g_{(2)})&=\frac{1}{2(d-1)}R(g_{(0)})\\
g_{(2)ij}&=\frac{1}{d-2}\[R(g_{(0)})_{ij}-\frac{1}{2(d-1)}R(g_{(0)})g_{(0)ij})\]
\end{aligned}
\end{equation}
After introducing a cut-off near the boundary at $\rho=\epsilon$, the regulated action can be obtained by integrate out $\rho$ from the $\epsilon$ to $\infty$:
\begin{equation}
\begin{aligned}
I_{\rm gr,reg}=&\frac{1}{16\pi G_{N}}\[\int_{\rho \geq \epsilon}\mathrm{d}^{d+1}x\sqrt{G}(R(G)+2\Lambda)-\int_{\rho = \epsilon}\mathrm{d}^{d} x\sqrt{\gamma}2K\]\\
=&\frac{1}{16\pi G_{N}}\int \mathrm{d}^{d} x\[\int_{\epsilon}\mathrm{d}\rho\frac{d}{\rho^{\frac{d}{2}+1}}\sqrt{\mathrm{det}g(x,\rho)}+\frac{1}{\rho^{\frac{d}{2}}}(-2d\sqrt{\mathrm{det}g(x,\rho)}\R.\\
&\L.+4\rho\partial_{\rho}\sqrt{\mathrm{det}g(x,\rho)})|_{\rho=\epsilon}\]
\end{aligned}
\end{equation}
When $\epsilon$ approaches 0, the counter term is made up by those divergent terms. Meanwhile, we set $d=2$, counter term is given by 
\begin{equation}
\begin{aligned}\label{A.6}
I_{\rm ct}&=-\frac{l}{16\pi G_{N}}\int\mathrm{d}^{2}x\sqrt{\mathrm{det}g_{0}}\(-2\epsilon^{-1}-\text{Tr}(g_{(2)})\mathrm{log}\epsilon\)\\
&=\frac{l}{16\pi G_{N}}\int\mathrm{d}^{2}x\sqrt{\mathrm{det}g_{0}}\(2\epsilon^{-1}+\text{Tr}(g_{(2)})\mathrm{log}\epsilon\),
\end{aligned}
\end{equation}
where
\begin{equation}
\begin{aligned}
\sqrt{g_{(0)}}&=\epsilon\(1-\frac{1}{2}\epsilon \text{Tr}(g_{(2)})+\frac{1}{8}\epsilon^2\((\text{Tr}(g_{(2)}))^2+\text{Tr}(g_{(2)}^2)\)+\mathrm{O}(\epsilon^3)\)\sqrt{\gamma}\\
\text{Tr}(g_{(2)})&=\frac{1}{2}R(g_{(0)})\\
&=\frac{1}{2\epsilon}R_{ij}(\gamma)\(\gamma^{ij}+g^{ij}_{(2)}\)\\
&=\frac{1}{2\epsilon}\(R(\gamma)+g^{ij}_{(2)}R_{ij}(\gamma)\)
\end{aligned}
\end{equation}
We can't work out the $g_{(2)ij}$, but we can construct the $g_{(2)ij}$ by introducing a auxiliary field $\phi$ and corresponding energy-stress tensor $T_{ij}$, which meet:
\begin{equation}
\begin{aligned}
g_{(2)ij}&=\frac{1}{2}\(R(g_{(0)})g_{(0)}+T_{ij}\)\\
\nabla^{i}T_{ij}&=0,\text{Tr}(g^{ij}_{(0)}T_{ij})=-R(g_{(0)})\\
T_{ij}&=\frac{1}{2}\nabla_{i}\phi \nabla_{j}\phi+\nabla_{i}\nabla_{j}\phi-\frac{1}{2}g_{(0)ij}\(\frac{1}{2}(\nabla\phi)^2+2\Box \phi\)\\
&=\frac{1}{2}\nabla_{i}\phi \nabla_{j}\phi+\nabla_{i}\nabla_{j}\phi-\frac{1}{2}g_{(0)ij}\(\frac{1}{2}(\nabla\phi)^2+2R(g_{(0)})\)\\
\text{Tr}(g_{(2)ij}^2)&=\frac{1}{4}g^{lm}_{(0)}(R(g_{(0)})g_{mp}+T_{mp})g^{pq}_{(0)}(R(g_{(0)})g_{ql}+T_{ql})\\
&=\frac{1}{4}(2R(g_{(0)})^2+2\text{Tr}(T)R(g_{(0)})+\text{Tr}(T^2) )\\
&=\frac{1}{4}\text{Tr}(T^2)\\
&=f(\nabla \phi)
\end{aligned}
\end{equation}
Then ($\ref{A.6}$) becomes
\begin{equation}
\begin{aligned}
I_{\rm ct}&=\frac{1}{16\pi G_{N}}\int\mathrm{d}^{2}x\sqrt{\gamma}\left[1-\frac{1}{2}\epsilon \text{Tr}(g_{(2)})+\frac{1}{8}\epsilon^2[(\text{Tr}(g_{(2)}))^2+\text{Tr}(g_{(2)}^2)]+\mathrm{O}(\epsilon^3)\right]\\
&\times(2+\epsilon \text{Tr}(g_{(2)})\mathrm{log}\epsilon)\\
&=\frac{1}{16\pi G_{N}}\int\mathrm{d}^{2}x\sqrt{\gamma}\left[1-\frac{1}{2}\epsilon \text{Tr}(g_{(2)})+\frac{1}{8}\epsilon^2(\text{Tr}(g_{(2)}))^2+\mathrm{O}(\epsilon^3)\right]\\
&\times(2+\epsilon \text{Tr}(g_{(2)})\mathrm{log}\epsilon+F(\nabla \phi))\\
&=\frac{1}{16\pi G_{N}}\int\mathrm{d}^{2}x\sqrt{\gamma}\left[1-\frac{1}{4}(R(\gamma)+g^{ij}_{(2)}R_{ij}(\gamma))+\frac{1}{32}(R(\gamma)+g^{ij}_{(2)}R_{ij}(\gamma))^2+\mathrm{O}(\epsilon^3)\right]\\
&\times\left[ 2+\frac{1}{2}(R(\gamma)+g^{ij}_{(2)}R_{ij}(\gamma))\mathrm{log}\epsilon \right]\\
&=\frac{1}{16\pi G_{N}}\int\mathrm{d}^{2}x\sqrt{\gamma}\[(2+\frac{1}{2}R(\gamma)\mathrm{log}\epsilon-\frac{1}{2}R(\gamma)+\frac{1}{16}R(\gamma)^2-\frac{1}{2}g^{ij}_{(2)}R_{ij}\R.\\
&\L.+\frac{1}{8}R(\gamma)g^{ij}_{(2)}R_{ij}+\frac{1}{16}(g^{ij}_{(2)}R_{ij})^2)+F(\nabla \phi) \],
\end{aligned}
\end{equation}
where 
\begin{align}
g^{ij}_{(2)}R_{ij}&=\frac{1}{4}R_{ij}\nabla^{i}\phi\nabla^{j}\phi+\frac{1}{2}R_{ij}\nabla^{i}\nabla^{j}\phi-\frac{1}{8}R(g_{(0)})(\nabla \phi)^2.
\end{align}
Finally, we get the counter term as an expansion of Ricci scalar
\begin{equation}
\begin{aligned}
I_{\rm ct}&=\frac{1}{16\pi G_{N}}\int\mathrm{d}^{2}x\sqrt{\gamma}\[2+\frac{1}{2}R(\gamma)\mathrm{log}\epsilon-\frac{1}{2}R(\gamma)+\frac{1}{16}R(\gamma)^2+\mathrm{O}(R(\gamma)^3)\R.\\
&\L.+p(\nabla \phi)+F(\nabla \phi)\]\\
&=\frac{1}{16\pi G_{N}}\int\mathrm{d}^{2}x\sqrt{\gamma}\[2+\frac{1}{2}R(\gamma)\mathrm{log}\epsilon-\frac{1}{2}R(\gamma)+\frac{1}{16}R(\gamma)^2+\mathrm{O}(R(\gamma)^3)+\tilde{F}(\nabla \phi)\].
\end{aligned}
\end{equation}


\subsection{Higher-curvature corrections}
 In this part, we introduce high curvature corrections. In above, we have transformed the coordinates in a different foliation to decouple $r$ and $\psi$ as $(t,r,\psi)\rightarrow(T,\rho,\phi)$. Since the transformations are well-defined for $\rm I_{p3}$ and $\rm III_{\rm b}$ and these two phases both have good asymptotic behaviors in limit of $l\ll 1$, we only focus on these phases in this subsection. 

For I$_{\rm p3}$, to compare with the foliation in Sec.~$5.2$, we add a substitute to the transformation \eqref{KRS 1} as $$\sinh(\phi)^2=\frac{\tilde{r}^2}{L_{2}^2m^2},$$
with $L_{2}=\frac{l}{\sqrt{1-\mathcal{A}^2m^2l^2}}$ the radius of AdS$_{2}$ slice. Then the line element ($\ref{2.5}$) in global coordinates for I$_{\rm p3}$ is represented as
\begin{equation}
\begin{aligned}
\mathrm{d}s^2&=\mathrm{d}\rho^2+h_{ij}(\rho)\mathrm{d}x^{i}\mathrm{d}x^{j},\\
&=\mathrm{d}\rho^2+\frac{l^2}{L_{2}^2}\cosh\left(\frac{\rho}{l}\right)^2g_{ij}(\rho)\mathrm{d}x^{i}\mathrm{d}x^{j},\\
&=\mathrm{d}\rho^2+\frac{l^2}{L_{2}^2}\cosh\left(\frac{\rho}{l}\right)^2\left[ -\left(\frac{\tilde{r}^2}{L_{2}^2}+m^2\right)\mathrm{d}T^2+\frac{1}{\frac{\tilde{r}^2}{L_{2}^2}+m^2}\mathrm{d}\tilde{r}^2\right],
\end{aligned}
\end{equation}
in which $g_{ij}$ is the metric of the AdS$_{2}$ slice. The induced metric $h_{ij}(\r)$ at constant $\r$ satisfies the Israel junction condition and fulfills $$\partial_{\r}h_{ij}(\r)\propto h_{ij}(\r).$$ Accordingly, the absolute value of the tension at $\r=\r_{\rm b}$ is obtained as
\begin{equation}
    |\mathcal{T}_{\rm b}|=\frac{1}{2}|K_{\rm b}|=\frac{\tanh(\frac{\r_{\rm b}}{l})}{l},
\end{equation}
where $K_{\rm b}$ is the extrinsic curvature of the string. To compare with $I_{\rm reduced}$ at $x=\text{constant}$, we make a substitute to \eqref{reduced action of I3p at x/y=constant} as $$g_{ij}=\frac{g_{ij}(\r)}{L_{2}^2},\quad \r_{1}=\r_{2}=\r_{\rm b},\quad \delta\r_{2}=\delta\r,\quad \delta\r_{1}=0,$$ and keep \eqref{reduced action of I3p at x/y=constant} to second order of $\delta\r$, then we get:
\begin{equation}
\begin{aligned}
I_{\rm reduced}&=-\frac{1}{16\pi G_{3}}\int\mathrm{d}^2x\sqrt{-g}\delta \r\left[ R(g)+\frac{2}{L_{2}^2}\right]\\
&-\frac{1}{8\pi G_{3}}\int\mathrm{d}^2x\sqrt{-g}\left[\frac{\tanh(\frac{\r_{\rm b}}{l})}{2l}\nabla_{\mu}\delta \r\nabla^{\mu}\delta \r+\frac{\tanh(\frac{\r_{\rm b}}{l})(\delta\r)^2}{l}\right]\\
&+\frac{\mathcal{A}m}{8\pi G_{3}}\int \mathrm{d}^2x\sqrt{-g},
\end{aligned}
\end{equation}
Given that the string at $\frac{x}{y}=0$ in \eqref{KRS 1} and \eqref{KRS 3} corresponds to the one at $x=0$ in Sec.~$5.1.2$, we set $\mathfrak{g}_{s}(\r_{\rm b})=0$ and solve for $\rho_{\rm b}$, the absolute value of tension on the string can be obtained as
\begin{equation}
    |\mathcal{T}_{\rm b}|=\mathcal{A}m,
\end{equation}
which is consistent with \eqref{4.4} with $\psi_{0}+m\psi_{1}=\frac{\pi}{2}(x=0)$.

For III$_{\rm b}$, in parallel, we add a substitute to the transformation \eqref{KRS 3} as $$\cosh(\phi)^2=\frac{\tilde{r}^2}{L_{2}^2m^2},$$
with $L_{2}=\frac{l}{\sqrt{1-\mathcal{A}^2m^2l^2}}$ the radius of the AdS$_{2}$ slice. Then the line element ($\ref{2.9}$) in global coordinates becomes
\begin{equation}
\begin{aligned}
\mathrm{d}s^2&=\mathrm{d}\rho^2+h_{ij}(\rho)\mathrm{d}x^{i}\mathrm{d}x^{j},\\
&=\mathrm{d}\rho^2+\frac{l^2}{L_{2}^2}\cosh\left(\frac{\rho}{l}\right)^2g_{ij}(\rho)\mathrm{d}x^{i}\mathrm{d}x^{j},\\
&=\mathrm{d}\rho^2+\frac{l^2}{L_{2}^2}\cosh(\frac{\rho}{l})^2\left[ -(\frac{\tilde{r}^2}{L_{2}^2}-m^2)\mathrm{d}T^2+\frac{1}{\frac{\tilde{r}^2}{L_{2}^2}-m^2}\mathrm{d}\tilde{r}^2\right].
\end{aligned}
\end{equation}
The induced metric $h_{ij}(\rho)$ also satisfies the Israel junction condition, and the absolute value of tension at $\r=\r_{\rm b}$ is given by
\begin{equation}
    |\mathcal{T}_{\rm b}|=\frac{1}{2}|K_{\rm b}|=\frac{\tanh(\frac{\r_{\rm b}}{l})}{l}.
\end{equation}
Similarly,  we make a substitute to \eqref{4.17} as $$g_{ij}=\frac{g_{ij}(\r)}{L_{2}^2},\quad \r_{1}=\r_{2}=\r,\quad \delta\r_{2}=\delta\r,\quad \delta\r_{1}=0,$$ and keep \eqref{4.17} to second order of $\delta\r$, then we can also get:
\begin{equation}
\begin{aligned}
I_{\rm reduced}&=-\frac{1}{16\pi G_{3}}\int\mathrm{d}^2x\sqrt{-g}\delta \r\left[ R(g)+\frac{2}{L_{2}^2}\right]\\
&-\frac{1}{8\pi G_{3}}\int\mathrm{d}^2x\sqrt{-g}\left[\frac{\tanh(\frac{\r}{l})}{2l}\nabla_{\mu}\delta \r\nabla^{\mu}\delta \r+\frac{\tanh(\frac{\r}{l})(\delta\r)^2}{l}\right]\\
&+\frac{\mathcal{A}m}{8\pi G_{3}}\int \mathrm{d}^2x\sqrt{-g}.
\end{aligned}
\end{equation}
Likewise, we set $\mathfrak{g}_{s}(\r_{\rm b})=0$ and solve for $\rho_{\rm b}$, the absolute value of tension on the string can be obtained as
\begin{equation}
    |\mathcal{T}_{\rm b}|=\mathcal{A}m,
\end{equation}
which is consistent with \eqref{4.12} with $\psi_{0}+m\psi_{1}=0(x=0)$.

As we can see, here the acceleration measures the tension on the string at $x=0$ and the result is consistent with the tension at $Q(x)=1$ in Class I and III solutions. Furthermore, the line element of the AdS$_{2}$ slice is consistent with that of the string at $x=0$, and $L_{2}$ corresponds to $l_{2}(\psi_{0}+m\psi_{1})$ at $x=0$. When fixing $L_{2}$, the relation between $l$ and $A$ becomes
\begin{equation}
    \frac{1}{l^2}=\frac{1}{L_{2}^2}+A^2,
\end{equation}
indicating that $A\gg 1$ when $l\ll 1$, the position of the  string at $x=0$ is closer to the AdS$_{3}$ asymptotic boundary.

\section{Central Extension of 3D C-metric in Generic Gauge}\label{B}
In this part, we show the results of the FG expansion of 3D C-metric in a generic Weyl gauge. Following the FG expansion in Sec. 3 and transfer $\xi$ into $\Xi$ and set $\Omega(\Xi)=\omega(\xi)$, we can get:\\
\textbf{Class I}:\\
\begin{equation}
\begin{aligned}
    &\mathrm{d}s^2=\\
    &-\frac{((1-A^2l^2)z^2\Omega(\Xi)^2+4l^2\Omega(\Xi)^4+z^2l^2\Omega'(\Xi)^2)^2}{16l^2z^2\Omega(\Xi)^6}\mathrm{d}t^2\\
    &+\frac{l^2}{z^2}\mathrm{d}z^2\\
    &+\frac{((A^2l^2-1)\Omega(\Xi)^2z^2+4l^2\Omega(\Xi)^4-3l^2\Omega'(\Xi)^2z^2+2l^2\Omega(\Xi)\Omega''(\Xi)z^2)^2}{16l^2\Omega(\Xi)^6z^2}\mathrm{d}\Xi^2
\end{aligned}
\end{equation}
\textbf{Class II}:\\
\begin{equation}
\begin{aligned}
    \mathrm{d}s^2=&-\frac{((1+A^2l^2)z^2\Omega(\Xi)^2-4l^2\Omega(\Xi)^4-z^2l^2\Omega'(\Xi)^2)^2}{16l^2z^2\Omega(\Xi)^6}\mathrm{d}t^2\\
    &+\frac{l^2}{z^2}\mathrm{d}z^2\\
    &+\frac{((1+A^2l^2)\Omega(\Xi)^2z^2+4l^2\Omega(\Xi)^4-3l^2\Omega'(\Xi)^2z^2+2l^2\Omega(\Xi)\Omega''(\Xi)z^2)^2}{16l^2\Omega(\Xi)^6z^2}\mathrm{d}\Xi^2,
\end{aligned}
\end{equation}
\textbf{Class III}:\\
\begin{equation}
    \begin{aligned}
        &\mathrm{d}s^2=\\
    &-\frac{((1-A^2l^2)z^2\Omega(\Xi)^2-4l^2\Omega(\Xi)^4-z^2l^2\Omega'(\Xi)^2)^2}{16l^2z^2\Omega(\Xi)^6}\mathrm{d}t^2\\
    &+\frac{l^2}{z^2}\mathrm{d}z^2\\
    &+\frac{((1-A^2l^2)\Omega(\Xi)^2z^2+4l^2\Omega(\Xi)^4-3l^2\Omega'(\Xi)^2z^2+2l^2\Omega(\Xi)\Omega''(\Xi)z^2)^2}{16l^2\Omega(\Xi)^6z^2}\mathrm{d}\Xi^2.
    \end{aligned}
\end{equation}
Near the boundary ($z\rightarrow 0$), three solutions are all like:
\begin{equation}
    \mathrm{d}s^2=\frac{l^2}{z^2}\mathrm{d}z^2+ \frac{l^2\Omega(\Xi)(-\mathrm{d}t^2+\mathrm{d}\Xi^2)}{z^2}+\mathcal{O}_{ij}(1)\mathrm{d}x^{i}\mathrm{d}x^{j},
\end{equation}
we can see in generic gauge this metric is asymptotically flat and has similar asymptotic symmetric and algebra with the case in ADM gauge.

Given the boundary condition:
\begin{equation}
\delta g={\begin{bmatrix}
\mathcal{O}(1)&\mathcal{O}\(\frac{1}{z}\)&\mathcal{O}(1)\\
\mathcal{O}\(\frac{1}{z}\)&0&\mathcal{O}\(\frac{1}{z}\)\\
\mathcal{O}(1)&\mathcal{O}\(\frac{1}{z}\)&\mathcal{O}(1)\\
\end{bmatrix}}.
\end{equation}
Here we set that $\delta g_{tt}$, $\delta g_{t\Xi}$ and $\delta g_{\Xi\Xi}$ is of $\mathcal{O}(1)$ in order to find a non-trivial asymptotic killing vector and construct the structure of algebra. The asymptotic killing vector is:
\begin{equation}
    \begin{aligned}\label{B.7}
       \mathcal{X}^{(3)}=&\(\frac{T(t+\Xi)+M(t-\Xi)}{2}+\mathcal{O}(z)\)\partial_{t}\\
       &+\(\frac{z\partial_{\Xi}(\Omega(\Xi)(T(t+\Xi)-M(t-\Xi)))}{2\Omega(\Xi)}+O(z^2)\)\partial_{z}\\
       &+\(\frac{T(t+\Xi)-M(t-\Xi)}{2}+\mathcal{O}(z)\)\partial_{\Xi},
    \end{aligned}
\end{equation}
$T(t+\Xi)$ and $M(t-\Xi)$ are arbitrary functions and correspond to two modes. ($\ref{B.7}$) holds for all of three solutions. Considering that only leading term in \eqref{B.7} make a contribution to the non-trivial central term, we neglect the sub-leading term. Similar to Sec~$3$, we introduce two Fourier modes $\beta e^{-i n (t+\Xi)/\beta}$ and $\beta e^{-i n (t-\Xi)/\beta}$ and define $\mathcal{X}_{(R)n}$ and $\mathcal{X}_{(L)n}$ as:
\begin{equation}
    \begin{aligned}
        \mathcal{X}^{(3)}_{n(R)}&=\frac{\beta e^{-\frac{i n (t+\Xi)}{\beta}}}{2}\partial_{t}+\frac{z\beta \partial_{\Xi}\(\Omega(\Xi)e^{-\frac{i n (t+\Xi)}{\beta}}\)}{2\Omega(\Xi)}\partial_{z}+\frac{\beta e^{-\frac{i n (t+\Xi)}{\beta}}}{2}\partial_{\Xi}\\
        \mathcal{X}^{(3)}_{n(L)}&=\frac{\beta e^{-\frac{i n (t-\Xi)}{\beta}}}{2}\partial_{t}-\frac{z\beta \partial_{\Xi}\(\Omega(\Xi)e^{-\frac{i n (t-\Xi)}{\beta}}\)}{2\Omega(\Xi)}\partial_{z}-\frac{\beta e^{-\frac{i n (t-\Xi)}{\beta}}}{2}\partial_{\Xi},
    \end{aligned}
\end{equation}
in which $\beta$ is an arbitrary non-zero constant. These generators meet the classical commutation relation:
\begin{equation}
    \begin{aligned}
        [ \mathcal{X}^{(3)}_{n(R)},\mathcal{X}^{(3)}_{m(R)}]&=i(n-m)\mathcal{X}^{(3)}_{n+m(R)}\\
        [ \mathcal{X}^{(3)}_{n(L)},\mathcal{X}^{(3)}_{m(L)}]&=i(n-m)\mathcal{X}^{(3)}_{n+m(L)}\\
        [ \mathcal{X}^{(3)}_{n(R)},\mathcal{X}^{(3)}_{m(L)}]&=0\\
    \end{aligned}
\end{equation}
After using \eqref{The deformation of J}, the results of three classes are as follows:

For Class I, the central term in generic gauge is:
\begin{equation}
\begin{aligned}\label{Central term of I in generic gauge}
        &K(\mathcal{X}_{n(R/L)},\mathcal{X}_{m(R/L)})\\
        =&\delta_{m+n}\int^{\beta\pi}_{-\beta\pi}\(-\frac{iln^3}{16G_{3}\pi\beta}\R.\\
        &\L.-\frac{in\beta(2(A^2l^2-1)\Omega(\Xi)^2-l^2\Omega'(\Xi)^2+l^2\Omega(\Xi)\Omega''(\Xi))}{32G_{3}l\pi\Omega(\Xi)^2}\R.\\
        &\L.-\frac{l\beta^2(2\Omega'(\Xi)^3-3\Omega(\Xi)\Omega'(\Xi)\Omega''(\Xi)+\Omega(\Xi)^2\Omega'''(\Xi))}{32G_{3}\pi\Omega(\Xi)^3}\)\mathrm{d}\Xi, 
\end{aligned}
\end{equation}
in which $\delta_{m+n}=1$ with $m+n=0$ or $\delta_{m+n}=0$. As we can see above, the coefficient of $n^{3}$ tell us that the central charge is still $\frac{\ 3l}{\ 2G_{2}}$. To recover the Virasoro algebra, the third term of right hand side in \eqref{Central term of I in generic gauge} is required to be 0, that is to say, we require:
\begin{equation}
    \begin{aligned}\label{Equation1}
         2\Omega'(\Xi)^3-3\Omega(\Xi)\Omega'(\Xi)\Omega''(\Xi)+\Omega(\Xi)^2\Omega'''(\Xi)&=0,\\
         \Omega(\Xi)(\ln(\Omega(\Xi)))'''&=0,\\
         (\ln(\Omega(\Xi)))'''&=0(\Omega(\Xi)\neq 0).
    \end{aligned}
\end{equation}
After calculating, the general form of $\Omega(\Xi)$ that meets \eqref{Equation1} is:
\begin{equation}\label{Solution1}
    \Omega(\Xi)=e^{\alpha+\mu\Xi+\gamma \Xi^2},
\end{equation}
with $\alpha$, $\mu$ and $\gamma$ the free constants. After applying \eqref{Solution1} to \eqref{Central term of I in generic gauge} we can get:
\begin{equation}
\begin{aligned}\label{Extended term in generic gauge of Class I}
    &K(\mathcal{X}_{n(R/L)},\mathcal{X}_{m(R/L)})\\
    =&-\frac{il}{8\pi G_{3}}\(n^3+ n\beta^2\(\frac{A^2l^2-1}{l^2}+\gamma\)\)\delta_{m+n}=-i\frac{c}{12}\(n^3+ n\beta^2\(\frac{A^2l^2-1}{l^2}+\gamma\)\)\delta_{m+n},
\end{aligned}
\end{equation}
with $c=\frac{3l}{2G_{3}}$.

For Class II, the central term in generic gauge is:
\begin{equation}
\begin{aligned}\label{Central term of II in generic gauge}
    &K(\mathcal{X}_{n(R/L)},\mathcal{X}_{m(R/L)})\\
    =&\delta_{m+n}\int^{\beta\pi}_{-\beta\pi}\(-\frac{iln^3}{16G_{3}\pi\beta}\R.\\
        &\L.-\frac{in\beta(2(1+A^2l^2)\Omega(\Xi)^2-l^2\Omega'(\Xi)^2+l^2\Omega(\Xi)\Omega''(\Xi))}{32G_{3}l\pi\Omega(\Xi)^2}\R.\\
        &\L.-\frac{l\beta^2(2\Omega'(\Xi)^3-3\Omega(\Xi)\Omega'(\Xi)\Omega''(\Xi)+\Omega(\Xi)^2\Omega'''(\Xi))}{32G_{3}\pi\Omega(\Xi)^3}\)\mathrm{d}\Xi,
\end{aligned}
\end{equation}
in which $\delta_{m+n}=1$ with $m+n=0$ or $\delta_{m+n}=0$. Similarly, the central charge can be read off from the coefficient of $n^{3}$ as $\frac{\ 3l}{\ 2G_{3}}$. Then we require:
\begin{equation}\label{Equation2}
        2\Omega'(\Xi)^3-3\Omega(\Xi)\Omega'(\Xi)\Omega''(\Xi)+\Omega(\Xi)^2\Omega'''(\Xi)=0,
\end{equation}
the general form of $\Omega(\Xi)$ that meets \eqref{Equation2} is also:
\begin{equation}\label{Solution2}
    \Omega(\Xi)=e^{\alpha+\mu\Xi+\gamma \Xi^2},
\end{equation}
with $\alpha$, $\mu$ and $\gamma$ the free constants. After applying \eqref{Solution2} to \eqref{Central term of II in generic gauge}, we can also get:
\begin{equation}
\begin{aligned}\label{Extended term in generic gauge of Class II}
   &K(\mathcal{X}_{n(R/L)},\mathcal{X}_{m(R/L)})\\
    =&-\frac{il}{8\pi G_{3}}\(n^3+ n\beta^2\(\frac{A^2l^2+1}{l^2}+\gamma\)\)\delta_{m+n}=-i\frac{c}{12}\(n^3+ n\beta^2\(\frac{A^2l^2+1}{l^2}+\gamma\)\)\delta_{m+n},
\end{aligned}
\end{equation}
with $c=\frac{3l}{2G_{3}}$.

For Class III, the central term in generic gauge is:
\begin{equation}
\begin{aligned}\label{Central term of III in generic gauge}
        &K(\mathcal{X}_{n(R/L)},\mathcal{X}_{m(R/L)})\\
        =&\delta_{m+n}\int^{\beta\pi}_{-\beta\pi}\(-\frac{iln^3}{16G_{3}\pi\beta}\R.\\
        &\L.-\frac{in\beta(2(1-A^2l^2)\Omega(\Xi)^2-l^2\Omega'(\Xi)^2+l^2\Omega(\Xi)\Omega''(\Xi))}{32G_{3}l\pi\Omega(\Xi)^2}\R.\\
        &\L.-\frac{l\beta^2(2\Omega'(\Xi)^3-3\Omega(\Xi)\Omega'(\Xi)\Omega''(\Xi)+\Omega(\Xi)^2\Omega'''(\Xi))}{32G_{3}\pi\Omega(\Xi)^3}\)\mathrm{d}\Xi,
\end{aligned}
\end{equation}
in which $\delta_{m+n}=1$ with $m+n=0$ or $\delta_{m+n}=0$. Likewise, the central charge is $\frac{\ 3l}{\ 2G_{3}}$. Then we require:
\begin{equation}\label{Equation3}
        2\Omega'(\Xi)^3-3\Omega(\Xi)\Omega'(\Xi)\Omega''(\Xi)+\Omega(\Xi)^2\Omega'''(\Xi)=0,
\end{equation}
the general form of $\Omega(\Xi)$ that meets \eqref{Equation2} is also:
\begin{equation}\label{Solution3}
    \Omega(\Xi)=e^{\alpha+\mu\Xi+\gamma \Xi^2},
\end{equation}
with $\alpha$, $\mu$ and $\gamma$ the free constants. After applying \eqref{Solution3} to \eqref{Central term of III in generic gauge}, we can also get:
\begin{equation}
\begin{aligned}\label{Extended term in generic gauge of Class III}
    &K(\mathcal{X}_{n(R/L)},\mathcal{X}_{m(R/L)})\\
    =&-\frac{il}{8\pi G_{3}}\(n^3+ n\beta^2\(\frac{1-A^2l^2}{l^2}+\gamma\)\)\delta_{m+n}=-i\frac{c}{12}\(n^3+ n\beta^2\(\frac{1-A^2l^2}{l^2}+\gamma\)\)\delta_{m+n},
\end{aligned}
\end{equation}
with $c=\frac{3l}{2G_{3}}$.

As we can see above, the solutions of Weyl factor in C-metric are all $e^{\alpha+\mu\Xi+\gamma \Xi^2}$. The linear term in the index of the exponential function can be removed by shifting $\Xi$, thus the key point is the coefficient of the squared term. When $\gamma=0$, \eqref{Extended term in generic gauge of Class I}, \eqref{Extended term in generic gauge of Class II} and \eqref{Extended term in generic gauge of Class III} return to the results in \eqref{Extended term of three classes}
\bibliographystyle{unsrt}
\bibliography{reference_revised}
\end{document}